\documentclass[11pt]{article}
\usepackage[english]{babel}
\usepackage{amssymb,amsfonts,amsmath}

\newtheorem{proposition}{Proposition}

\newtheorem{lemma}{Lemma}
\newtheorem{theorem}{Theorem}
\newtheorem{corollary}{Corollary}
\newtheorem{remark}{Remark}

\newcommand{\coe}{\mathsf{c}}
\newcommand{\chara}{\mathtt{C}_U^{\phantom{x}}}

\newcommand{\tre}{\hspace{0.3mm}}
\newcommand{\quattro}{\hspace{0.4mm}}
\newcommand{\cinque}{\hspace{0.5mm}}
\newcommand{\sei}{\hspace{0.6mm}}
\newcommand{\sette}{\hspace{0.7mm}}

\newcommand{\fin}{\hspace{0.3mm}}

\newcommand{\spe}{\hspace{0.4mm}\mathrm{sp}\hspace{0.3mm}}

\newcommand{\preope}{\mathfrak{X}_U^{\circ}}
\newcommand{\ope}{\mathfrak{X}_U^{\phantom{2}}}
\newcommand{\opeq}{\mathfrak{X}_U^2}
\newcommand{\gu}{\gamma_U}
\newcommand{\ellekappa}{\hspace{0.3mm}\mathfrak{L}_{\hspace{-0.5mm}\hat{K}}}

\newcommand{\uj}{\mathfrak{U}_J^{\phantom{\ast}}}
\newcommand{\ujast}{\mathfrak{U}_J^\ast}
\newcommand{\prau}{\mathrm{P}_{\hspace{-1.4mm}\mathcal{A}_1}^{\phantom{x}}}
\newcommand{\ati}{\tilde{\mathcal{A}}}
\newcommand{\prati}{\mathrm{P}_{\hspace{-1.2mm}\ati}\hspace{0.7mm}}

\newcommand{\admu}{\mathsf{A}\hspace{0.3mm}(U)}
\newcommand{\au}{\mathcal{A}_U}
\newcommand{\ag}{\mathcal{L}(G)}

\newcommand{\ee}{\mathrm{e}}
\newcommand{\ut}{\widetilde{U}}

\newcommand{\mul}{\mu_\mathtt{L}}
\newcommand{\mur}{\mu_\mathtt{R}}

\newcommand{\gm}{\circledast}

\newcommand{\ammm}{{\mbox{\tiny $(\!-\!)$}}}
\newcommand{\ppp}{{\mbox{\tiny $(\!+\!)$}}}
\newcommand{\pmpm}{{\mbox{\tiny $(\!\pm\!)$}}}
\newcommand{\errep}{\mathbb{R}_{\hspace{0.3mm}\ast}^{\mbox{\tiny $+$}}}
\newcommand{\errem}{\mathbb{R}_{\hspace{0.3mm}\ast}^{\mbox{\tiny $-$}}}
\newcommand{\errast}{\mathbb{R}_{\hspace{0.3mm}\ast}}
\newcommand{\errepm}{\mathbb{R}_{\hspace{0.3mm}\ast}^{\mbox{\tiny $\pm$}}}
\newcommand{\charm}{\varepsilon_{\errem}}
\newcommand{\charp}{\varepsilon_{\errep}}
\newcommand{\charpm}{\varepsilon_{\errepm}}
\newcommand{\funpm}{\Lambda^{\hspace{-0.6mm}\pmpm}_n}
\newcommand{\lur}{\mathrm{L}^1(\mathbb{R})}
\newcommand{\lurpm}{\mathrm{L}^1(\errepm)}
\newcommand{\kerint}{\varsigma^\pmpm_f}
\newcommand{\kerintf}{\varsigma^\pmpm_{\hspace{0.5mm}\mathsf{f}}}
\newcommand{\kerintfn}{\varsigma^\pmpm_{\hspace{0.5mm}\mathsf{f}_n}}

\newcommand{\aum}{\mathcal{A}_{\hspace{-0.1mm}\ammm}^{\phantom{\perp}}}
\newcommand{\aup}{\mathcal{A}_{\hspace{-0.1mm}\ppp}^{\phantom{\perp}}}

\newcommand{\rep}{{U\hspace{-0.5mm}\vee\hspace{-0.5mm} U}}
\newcommand{\repm}{{U^\ammm\hspace{-0.7mm}\vee\hspace{-0.4mm} U^\ammm}}
\newcommand{\repp}{{U^\ppp\hspace{-0.7mm}\vee\hspace{-0.4mm} U^\ppp}}

\newcommand{\hsm}{\mathcal{B}_2(\mathrm{L}^2(\errem))}
\newcommand{\hsp}{\mathcal{B}_2(\mathrm{L}^2(\errep))}
\newcommand{\hspm}{\mathcal{B}_2(\mathrm{L}^2(\errepm))}

\newcommand{\Hstar}{\mathrm{H}^\ast\hspace{-0.5mm}}
\newcommand{\isomo}{\hspace{0.3mm}\mathfrak{E}\hspace{0.2mm}}

\newcommand{\norhs}{\|_{\mathcal{B}_2}}
\newcommand{\ranglehs}{\rangle_{\mathcal{B}_2}}

\newcommand{\chG}{\check{G}}

\newcommand{\en}{\mathsf{N}}

\newcommand{\effe}{{\mathsf{f}}}
\newcommand{\uf}{\hat{U}(\effe)}

\newcommand{\ufn}{\hat{U}(\effe_n)}

\newcommand{\invo}{\hspace{0.4mm}\mathfrak{J}\hspace{0.5mm}}
\newcommand{\invom}{\hspace{0.4mm}\mathfrak{J}_\ammm}
\newcommand{\invop}{\hspace{0.4mm}\mathfrak{J}_\ppp}
\newcommand{\invol}{{\hspace{0.2mm}\mathsf{J}_{\mm}^{\phantom{\ast}}}}
\newcommand{\involu}{\hspace{0.2mm}\mathsf{J}_{\mm}}
\newcommand{\sinvo}{\hspace{0.2mm}\mathsf{J}\hspace{0.3mm}}

\newcommand{\mm}{{\hspace{0.3mm}\mathtt{m}\hspace{0.7mm}}}

\newcommand{\starp}{\stackrel{\hspace{0.4mm}U}{\star}}
\newcommand{\kstarp}{\stackrel{\hspace{0.4mm}U}{\star}_{\hspace{-5.75mm}\phantom{\frac{X}{\hat{X}}}_{\hat{K}}}}
\newcommand{\kcomp}{\hspace{0.8mm}{\circ}_{\hspace{-5.35mm}\phantom{\frac{X}{\hat{X}}}_{\hat{K}}}}
\newcommand{\starppm}{\stackrel{\hspace{0.15mm}\pmpm}{\star}}
\newcommand{\starpm}{\stackrel{\hspace{0.15mm}\ammm}{\star}}
\newcommand{\starpp}{\stackrel{\hspace{0.15mm}\ppp}{\star}}
\newcommand{\opkg}{\hat{K}_g}
\newcommand{\kgstarp}{\stackrel{\hspace{0.4mm}U}{\star}_{\hspace{-5.75mm}\phantom{\frac{X}{\hat{X}}}_{\opkg}}\hspace{-1.5mm}}

\newcommand{\de}{\mathrm{d}}
\newcommand{\dom}{\mathrm{Dom}}
\newcommand{\ran}{\mathrm{Ran}}
\newcommand{\Ker}{\mathrm{Ker}}

\newcommand{\ccc}{\mathbb{C}}
\newcommand{\hh}{\mathcal{H}}
\newcommand{\bh}{\mathcal{B}(\mathcal{H})}
\newcommand{\tch}{\mathcal{B}_1(\mathcal{H})}
\newcommand{\hs}{\mathcal{B}_2(\mathcal{H})}
\newcommand{\opa}{\hat{A}}
\newcommand{\opk}{\hat{K}}
\newcommand{\opt}{\hat{T}}
\newcommand{\optn}{\hat{T}_n}
\newcommand{\opf}{\hat{F}}
\newcommand{\ldg}{\mathrm{L}^2(G)}
\newcommand{\ld}{{\mathrm{L}^2}}

\newcommand{\lug}{\mathrm{L}^1(G)}

\newcommand{\ldx}{\mathrm{L}^2(X,\mu;\mathbb{C})}
\newcommand{\ranld}{\rangle_{\mathrm{L}^2}}
\newcommand{\norld}{\|_{\mathrm{L}^2}}
\newcommand{\norlu}{\|_{\mathrm{L}^1}}
\newcommand{\Bnorld}{\Big\|_{\mathrm{L}^2}}
\newcommand{\ldgu}{\ldg_{[U]}}

\newcommand{\tprochi}{\hspace{0.3mm}\mathfrak{R}_{\widehat{\chi}}}

\newcommand{\tprochin}{\hspace{0.3mm}\mathfrak{R}_{\widehat{\chi}_n}\hspace{-0.3mm}}
\newcommand{\chid}{\breve{\chi}}
\newcommand{\pic}{\hspace{0.3mm}\mathfrak{R}_{\breve{\chi}}}
\newcommand{\picn}{\hspace{0.3mm}\mathfrak{R}_{\breve{\chi}_n}\hspace{-0.3mm}}

\newcommand{\hfpa}{{\widehat{\phi_1\psi_1}}}
\newcommand{\hfpb}{{\widehat{\phi_2\psi_2}}}
\newcommand{\chichi}{{\widehat{\chi\chi}}}

\newcommand{\intG}{\int_G}
\newcommand{\mG}{{\mu_G}}
\newcommand{\modu}{\Delta_G}
\newcommand{\aam}{\forall_{\hspace{-0.5mm}\mG}\hspace{0.3mm}}
\newcommand{\aames}{\forall_{\hspace{-0.5mm}\mu}\hspace{0.3mm}}

\newcommand{\nn}{\mathcal{N}}
\newcommand{\nat}{\mathbb{N}}
\newcommand{\sun}{\sum_{n\in\mathcal{N}}}
\newcommand{\norsun}{\mbox{\tiny $\|\cdot\|_{\mathrm{L}^2}$}\hspace{-1.7mm}\sum_{n\in\mathcal{N}}\hspace{0.9mm}}
\newcommand{\norsunv}{\mbox{\tiny $\|\cdot\|_{\mathrm{L}^2}$}\hspace{-1mm}\sum_{n\in\mathcal{N}}}
\newcommand{\norsunat}{\mbox{\tiny $\|\cdot\|_{\mathrm{L}^2}$}\hspace{-1.7mm}\sum_{n\in\mathbb{N}}\hspace{0.9mm}}

\newcommand{\hsnorsun}{\mbox{\tiny $\|\cdot\|_{\mathcal{B}_2}$}\hspace{-1.5mm}\sum_{n\in\mathcal{N}}\hspace{0.9mm}}
\newcommand{\hsnorsunv}{\mbox{\tiny $\|\cdot\|_{\mathcal{B}_2}$}\hspace{-0.7mm}\sum_{n\in\mathcal{N}}}
\newcommand{\normsunv}{\mbox{\tiny $\|\cdot\|$}\hspace{-0.5mm}\sum_{n\in\mathcal{N}}}
\newcommand{\norlim}{\mbox{\tiny $\|\cdot\|_{\mathrm{L}^2}$}\hspace{-1.2mm}\lim_{n\rightarrow\infty}}

\newcommand{\hslim}{\mbox{\tiny $\|\cdot\|_{\mathcal{B}_2}$}\hspace{-0.4mm}\lim_{n\rightarrow\infty}}
\newcommand{\hsliml}{\mbox{\tiny $\|\cdot\|_{\mathcal{B}_2}$}\hspace{-0.4mm}\lim_{l\rightarrow\infty}}
\newcommand{\norliml}{\mbox{\tiny $\|\cdot\|_{\mathrm{L}^2}$}\hspace{-1.2mm}\lim_{l\rightarrow\infty}}
\newcommand{\sumu}{\mbox{\tiny $\|\cdot\|_{\mathrm{L}^2}$}\hspace{-1.7mm}\sum_{U\in\chG}}
\newcommand{\sumub}{\mbox{\tiny $\|\cdot\|_{\mathrm{L}^2}$}\hspace{-1mm}\sum_{U\in\chG}}

\newcommand{\dimu}{\delta(U)}

\newcommand{\wt}{\mathfrak{W}_{U}^{\hspace{0.3mm}\psi}}
\newcommand{\wta}{\mathfrak{W}_{U}^{\hspace{0.3mm}\psi\hspace{0.3mm}\ast}}
\newcommand{\wtt}{\mathfrak{W}_{V}^{\hspace{0.3mm}\eta}}
\newcommand{\wtta}{\mathfrak{W}_{V}^{\hspace{0.3mm}\eta\hspace{0.3mm}\ast}}
\newcommand{\wavchi}{\mathfrak{W}_{U}^{\hspace{0.3mm}\chi_{n}}}

\newcommand{\rem}{R_\mm\hspace{-0.3mm}}

\newcommand{\du}{\hat{D}_U}
\newcommand{\duu}{\hat{D}_U^{\phantom{1}}}
\newcommand{\ddu}{d_U^{\phantom{1}}}
\newcommand{\dupm}{\hat{D}_{\hspace{-0.3mm}\pmpm}}
\newcommand{\duupm}{\hat{D}_{\hspace{-0.3mm}\pmpm}^{\phantom{1}}}
\newcommand{\duum}{\hat{D}_{\hspace{-0.3mm}\ammm}^{\phantom{1}}}
\newcommand{\duup}{\hat{D}_{\hspace{-0.3mm}\ppp}^{\phantom{1}}}

\newcommand{\nuc}{\kappa_U}

\newcommand{\nucpm}{\kappa_\pmpm}
\newcommand{\chinpm}{\chi^\pmpm_n}
\newcommand{\bchinpm}{\breve{\chi}^\pmpm_n}
\newcommand{\keu}{\varkappa_U}

\newcommand{\fr}{\mathsf{FR}\hspace{0.3mm}(\mathcal{H})}
\newcommand{\fri}{\mathsf{FR}\hspace{0.3mm}(\mathcal{H};U)}

\newcommand{\frl}{\mathsf{FR}^{\mbox{\tiny $\langle |$}}\hspace{-0.3mm}(\mathcal{H};U)}

\newcommand{\frlr}{\mathsf{FR}^{\mbox{\tiny $|\rangle\langle |$}}\hspace{-0.3mm}(\mathcal{H};U)}

\newcommand{\frllr}{\mathsf{FR}^{\mbox{\tiny $\langle |\hspace{-0.3mm}|$}}\hspace{-0.3mm}(\mathcal{H};U)}

\newcommand{\wig}{\mathfrak{S}_U^{\phantom{x}}}
\newcommand{\wigg}{\mathfrak{S}_U}
\newcommand{\wigp}{\mathfrak{S}_{\hspace{-0.3mm}\ppp}^{\phantom{x}}}
\newcommand{\wigm}{\mathfrak{S}_{\hspace{-0.3mm}\ammm}^{\phantom{x}}}
\newcommand{\wigpm}{\mathfrak{S}_{\hspace{-0.3mm}\pmpm}^{\phantom{x}}}
\newcommand{\wiggm}{\mathfrak{S}_{\hspace{-0.3mm}\ammm}}
\newcommand{\wiggp}{\mathfrak{S}_{\hspace{-0.3mm}\ppp}}
\newcommand{\wiggpm}{\mathfrak{S}_{\hspace{-0.3mm}\pmpm}}
\newcommand{\twig}{\mathfrak{S}_{V}}

\newcommand{\ru}{\mathcal{R}_U^{\phantom{\perp}}}
\newcommand{\ruort}{\mathcal{R}_U^{\perp}}
\newcommand{\rup}{\mathcal{R}_{\hspace{-0.3mm}\ppp}^{\phantom{\perp}}}
\newcommand{\rum}{\mathcal{R}_{\hspace{-0.3mm}\ammm}^{\phantom{\perp}}}
\newcommand{\rupm}{\mathcal{R}_{\hspace{-0.3mm}\pmpm}^{\phantom{\perp}}}

\newcommand{\spa}{\hspace{-2mm}}

\newcommand{\intrr}{\int_{\mathbb{R}\times\mathbb{R}}}

\newcommand{\opb}{\hat{B}}
\newcommand{\tr}{{\hspace{0.3mm}\mathsf{tr}\hspace{0.3mm}}}
\newcommand{\lr}{\mathrm{L}^2(\mathbb{R})}
\newcommand{\lrr}{\mathrm{L}^2(\mathbb{R}\times\mathbb{R})}
\newcommand{\lurr}{\mathrm{L}^1(\mathbb{R}\times\mathbb{R})}

\newcommand{\rr}{{\mathbb{R}\times\mathbb{R}}}

\newcommand{\ima}{{\hspace{0.2mm}\mathrm{i}\hspace{0.4mm}}}

\newcommand{\hq}{{\hat{q}}}
\newcommand{\hp}{{\hat{p}}}

\newcommand{\hfp}{{\widehat{\phi\psi}}}
\newcommand{\qp}{{(q,p)}}
\newcommand{\disp}{\exp\!\left(\ima(p\hspace{0.5mm}\hq-q\hspace{0.3mm}\hp)\right)}

\newcommand{\intg}{\int_{G}}

\newcommand{\rrr}{\mathbb{R}}

\newcommand{\lx}{{\mathrm{L}^2(X)}}

\newcommand{\hrho}{\hat{\rho}}

\newcommand{\fs}{\mathcal{F}_{\hspace{-0.6mm}\mbox{\tiny sp}}^{\phantom{x}}}
\newcommand{\fsy}{\mathcal{F}_{\hspace{-0.6mm}\mbox{\tiny sp}}}
\newcommand{\two}{\mathcal{T}_{\mm}\hspace{-0.3mm}}
\newcommand{\mmm}{\overset{\leftrightarrow}{\mm}}

\newcommand{\proi}{{\mathrm{P}_{\hspace{-1.2mm}\mbox{\tiny $\ru$}}^{\phantom{x}}}}

\newcommand{\wigt}{\mathfrak{T}}

\newcommand{\sdp}{\hspace{-0.2mm}\rtimes}

\begin{document}

\title{Star products: a group-theoretical point of view}

\author{
Paolo Aniello  \vspace{2mm}\\
\small \it  Dipartimento di Scienze Fisiche dell'Universit\`a di
Napoli `Federico II', \\ \small \it Complesso Universitario di Monte
S.\ Angelo, via Cintia, 80126 Napoli, Italy  \\ \small \it and \\
\small \it Istituto Nazionale di Fisica Nucleare (INFN) -- Sezione
di Napoli
\\  \small \it and \\ \small \it  Facolt\`a di Scienze
Biotecnologiche, Universit\`a di Napoli `Federico II' \\
{\footnotesize E-mail: paolo.aniello@na.infn.it} }

\maketitle

\begin{abstract} \noindent
Adopting a purely group-theoretical point of view, we consider the
star product of functions which is associated, in a natural way, with a
square integrable (in general, projective) representation of a
locally compact group. Next, we show that for this (implicitly defined)
star product explicit formulae can be provided. Two significant examples are
studied in detail: the group of translations on phase space and the
one-dimensional affine group. The study of the first example leads to
the Gr\"onewold-Moyal star product. In the second example, the link
with wavelet analysis is clarified.
\end{abstract}

\section{Introduction}
\label{intro}

The concept of star product of functions is a remarkable achievement
of theoretical physics. The archetype --- and still nowadays, the
most important realization --- of this concept is the
Gr\"onewold-Moyal star product (see~{\cite{Zachos}} and references
therein). Although there is no unique general mathematical framework
encompassing all known star products, one can certainly single out a
simple leading idea to which the various possible definitions of
star products are more or less inspired: to replace the ordinary
pointwise product of ($\ccc$-valued) functions defined on a certain set (a
`phase space' endowed with some structures: a differentiable
manifold, a measure space etc.)\ with a suitable non-commutative,
associative product that mimics the typical non-commutative behavior
of linear operators.

We will make no attempt at surveying the rich and varied literature on star
products. We will content ourselves  with recalling that both
differential-geometric~{\cite{Cahen,Gutt,DeWilde}}
and algebraic~{\cite{Berezin}} approaches to the subject have been adopted,
also in view of different purposes and applications. It is also
worth mentioning the fact that the most important topics where
the formalism of star products plays a relevant role are,
probably, the construction of
quantum mechanics `on phase space' and the study of the classical
limit of quantum mechanics~{\cite{Zachos,Emch}}. Thus, one may regard
E.~Wigner~{\cite{Wigner}} and H.~Weyl~{\cite{Weyl}} as the
fathers of this formalism.

More recently, a general approach to star products based on the idea
of using suitable `quantizers' and `dequantizers' has been proposed
and developed by various authors~{\cite{Gracia-Annals,Gadella,
Gracia,Manko1,Manko2,Manko3,Aniello-compact}}. This approach is very
close to applications in quantum mechanics since the star products
of functions that one obtains are, by construction, nothing but the
`images' of the products of quantum-mechanical operators.

In our present contribution, we will adopt a purely
group-theoretical point of view which is conceptually similar to the
`quantizer-dequantizer' approach cited above. Indeed, rather than
trying to define a star product directly in a given space of
functions (as usual, for instance, in the differential-geometric
approach), we consider the star product (implicitly) induced by a
suitable group-theoretical quantization-dequantization scheme.
Clearly, at this point, the real problem is to find explicit
formulae for the implicitly defined star product.

Before illustrating the main points of our work,
it is worth mentioning that recently another group-theoretical approach
to star products --- in the context of a suitable quantization-dequantization scheme ---
has been elaborated; see ref.~{\cite{Aniello-framet}}. However, this approach, differently from
the approach adopted in the present paper,
relies on the concept of `frame transform' and it is not directly related
to the Groenewold-Moyal product.

Let us now briefly outline our method and our main results. First, we show that
by means of the quantization (Weyl) and dequantization (Wigner) maps
generated by a square integrable (in general, projective)
representation $U$ of a locally compact group $G$
--- see~{\cite{Aniello-framet,Ali1,Ali2}} --- it is possible to introduce, in a
natural way, a star product in the Hilbert space $\ldg$ of square
integrable $\ccc$-valued functions on $G$. The product of two
functions is obtained by quantizing them, by forming the product of
the two operators thus obtained and, finally, by dequantizing this
product. Endowed with the operation just described, $\ldg$ becomes a
$\Hstar$-algebra. We will then prove --- this is the main result of
the paper --- that the star product in $\ldg$ admits a simple
explicit formula. More precisely, we will show that with every
orthonormal basis in the Hilbert space of the representation $U$ is
associated a formula for the star product (however, all these
formulae share the same general form). This basic result can be
generalized or specialized in various ways. For instance, an
expression of the `$\opk$-deformed star product' ---
see~{\cite{Manko1,Manko2}}
--- which is an interesting generalization of the star product, can
also be obtained. On the other hand, in the case where $G$ is
unimodular, a particularly simple formula for the star product
--- a sort of `twisted convolution' \emph{\`a la}
Grossmann-Loupias-Stein~{\cite{GLP}} --- can be derived.

We believe that the point of view on star products adopted in this
paper is very close to the `original spirit' of the
Gr\"onewold-Moyal star product since it solely relies on
(generalized) Wigner and Weyl maps. In fact, `our' star product is
\emph{essentially} the Gr\"onewold-Moyal star product in the case
where the group $G$ is the group of translations on phase space;
i.e., the two products --- the twisted convolution and the Gr\"onewold-Moyal product ---
are related by the symplectic Fourier transform.

We stress that our approach relies on the \emph{existence} of a
square integrable representation $U$ of the locally compact group
$G$ for defining an associated star product in $\ldg$. This feature,
however, should not be regarded as a limit of this approach. As is
well known, when dealing with mathematics nothing is free: the
weaker are the assumptions, the poorer will be the results that one
is able to prove. Moreover, our group-theoretical point of view is
very natural having in mind applications to physics. If $G$ is
regarded as a `symmetry group' of a quantum system and $U$ as the
symmetry action of this group in the Hilbert space $\hh$ of the
system, then the associated star product in $\ldg$ is nothing but
the realization in terms of functions of the product of
quantum-mechanical operators (observables or states); moreover, it
turns out that the star product is `equivariant' with respect to the
natural action of the symmetry group. Namely, the natural action of
$G$ on operators in $\hh$ translates into (i.e.\ is intertwined by
the dequantization map with) a simple transformation of the
corresponding functions in $\ldg$, and the star product of two
\emph{transformed} functions coincides with the \emph{transformed}
product of the two \emph{untransformed} functions.

The paper is organized as follows. In Sect.~{\ref{known}}, we fix
the main notations and we briefly recall some mathematical notions;
in particular, we review some basic facts concerning square
integrable representations. Next, in
Sect.~{\ref{weyl-wigner}}, we define the dequantization (Wigner) and
quantization (Weyl) maps generated by a square integrable
representation, and we derive the relevant `intertwining properties'
of the Wigner map. On the basis of these definitions we then
introduce --- see Sect.~{\ref{defstar}} --- the notion of star
product associated with a square integrable representation, and we
study its main properties. The star product introduced in such a way
is, however, only \emph{implicitly} defined. As already mentioned,
it is a remarkable fact that it admits an \emph{explicit}
realization; furthermore, in the case of a unimodular group, a particularly simple formula
can be derived. These results
--- that form the \emph{core} of our paper --- are stated and proved in
Sect.~{\ref{main}}. In Sect.~{\ref{examples}}, we consider two
significant examples: the group of translations on phase space
--- which is related to the standard Gr\"onewold-Moyal star product --- and
the affine group, which plays a central role in wavelet analysis.
Eventually, in Sect.~{\ref{conclusions}}, a few conclusions are
drawn, with a glance at various possible developments of our work.

\section{Some known facts and notations}
\label{known}

In this section, we will recall some basic facts of the theory of
representations of topological groups; standard references on the
subject are~{\cite{Raja,Folland}}. We will also fix the main
notations that will be used in the following sections.

Let $G$ be a locally compact, second countable, Hausdorff
topological group (in short, l.c.s.c.\ group). We will denote by
$\mG$ and $\Delta_G$, respectively, a {\it left Haar measure} (of
course uniquely defined up to a multiplicative constant) and the
{\it modular function} on $G$. The symbol $e$ will indicate the unit
element in $G$.

For the scalar  product $\langle\cdot,\cdot\rangle$ in a separable
complex Hilbert space $\mathcal{H}$, we will always follow the
convention that it is linear in the \emph{second} argument. The
symbol $\mathcal{U}(\mathcal{H})$ will denote the {\it unitary
group} of $\mathcal{H}$ --- i.e.\ the group of all unitary operators
in $\mathcal{H}$, endowed with the strong operator topology ---
which is a metrizable, second countable, Hausdorff topological
group.

We will mean by the term {\it projective representation} of a
l.c.s.c.\ group $G$ a Borel projective representation of $G$ in a
separable complex Hilbert space $\mathcal{H}$ (see, for instance,
ref.~\cite{Raja}, chapter~{VII}), namely a map of $G$ into
$\mathcal{U}(\mathcal{H})$ such that
\begin{enumerate}
\item
$U$ is a weakly Borel map, i.e.\ $G\ni g\mapsto
\langle\phi,U(g)\,\psi\rangle\in\mathbb{C}$ is a Borel
function,\footnote{The terms {\it Borel function} (or map) and {\it
Borel measure} will be always used with reference to the natural
Borel structures on the topological spaces involved, namely to the
smallest $\sigma$-algebras containing all open subsets.} for any
pair of vectors $\phi,\psi\in\mathcal{H}$;
\item
$U(e)=I$, where $I$ is the identity operator in $\mathcal{H}$;
\item
denoting by $\mathbb{T}$ the circle group, namely the group of
complex numbers of modulus one, there exists a Borel function $\mm
\colon G\times G\rightarrow\mathbb{T}$ such that
\begin{equation}
U(gh)=\mm (g,h)\,U(g)\,U(h)\fin ,\ \ \ \forall\hspace{0.5mm} g,h\in
G \fin .
\end{equation}
\end{enumerate}
The function $\mm$ --- which is called the {\it multiplier
associated with} $U$ --- satisfies the following conditions:
\begin{equation}
\mm(g,e)=\mm(e,g)=1 \fin ,\ \ \ \ \forall\hspace{0.5mm} g\in G \fin
,
\end{equation}
and
\begin{equation} \label{multirelat}
\mm(g_1,g_2g_3)\, \mm(g_2,g_3)= \mm(g_1 g_2,g_3)\, \mm(g_1,g_2)\fin
,\ \ \ \ \forall\cinque g_1,g_2,g_3\in G \fin .
\end{equation}
It is, moreover, immediate to check that $\mm (g,g^{-1})= \mm (g^{-1},g)$.
Clearly, in the case where $\mm\equiv 1$, $U$ is a standard
unitary representation; in this case, according to a well known
result, the hypothesis that the map $U$ is weakly Borel implies that
it is, actually, strongly continuous. The notion of irreducibility
is defined for projective representations as for standard unitary
representations.

Let $\ut \colon G\rightarrow\mathcal{U}(\widetilde{\mathcal{H}})$ be
a projective representation of $G$ in a (separable complex) Hilbert
space $\widetilde{\mathcal{H}}$. We say that $\ut$ is
\emph{physically equivalent} to $U$ if there exist a Borel function
$\beta \colon G\rightarrow\mathbb{T}$, and a unitary or antiunitary
operator $W \colon \mathcal{H}\rightarrow\widetilde{\mathcal{H}}$,
such that
\begin{equation} \label{equirel}
\ut(g)=\beta(g)\, W\,U(g)\,W^\ast,\ \ \ \forall\hspace{0.4mm} g\in G
\fin .
\end{equation}
It is obvious that a projective
representation --- physically equivalent to an irreducible projective
representation --- is irreducible too. We will say that the representations $U$ and
$\ut$ are \emph{unitarily equivalent}
if, in relation~{(\ref{equirel})}, $\beta\equiv 1$ and $W$ is a unitary operator.

Observe that we can identify the unitary dual of $G$ with any (suitably topologized)
maximal set of mutually unitarily inequivalent, irreducible, unitary representations of $G$.
We will denote by $\chG$ such a set, and we will call it \emph{a realization of the unitary dual of} $G$.
It is well known that, if $G$ is compact,\footnote{We will include among the compact groups all the finite groups
(endowed with the discrete topology).}
then $\chG$ is a finite or countable set (endowed with the discrete topology);
moreover, $\chG$ consists of finite-dimensional representations.

Let $U$ be an \emph{irreducible} projective representation of the
l.c.s.c.\ group $G$ in the Hilbert space $\mathcal{H}$. Then, given
two vectors $\psi,\phi\in\mathcal{H}$, we define the function
(usually called `coefficient' of the representation $U$)
\begin{equation} \label{defcoe}
\coe_{\psi ,\phi}^U \colon G\ni g\mapsto \langle U(g)\,\psi
,\phi\rangle\in\ccc\fin ,
\end{equation}
and we consider the set (of `admissible vectors for $U$')
\begin{equation}
\admu :=\left\{\psi\in\mathcal{H}\,|\ \exists\phi\in\mathcal{H}:\,
\phi\neq 0,\, \coe_{\psi,\phi}^U \in \ldg\right\},
\end{equation}
where $\ldg\equiv\mathrm{L}^2(G,\mu_G;\ccc)$ (in the following, we
will denote by $\langle\cdot,\cdot\ranld$ and $\|\cdot\norld$ the
scalar product and the norm in $\ldg$). The representation $U$ is
said to be {\it square integrable} if $\admu\neq\{0\}. $ Square
integrable projective representations are characterized by the
following result --- see ref.~{\cite{Aniello}} --- which is a
generalization of a classical theorem of Duflo and
Moore~\cite{Duflo} concerning unitary representations:
\begin{theorem} \label{Duflo-Moore}
Let the projective representation $U\colon\,G\rightarrow
\mathcal{U}(\mathcal{H})$ be square integrable. Then, the set
$\admu$ is a dense linear span\footnote{Throughout the paper, we
call a nonempty subset of a vector space $\mathsf{V}$ a `linear
span' if it is a linear space itself (with respect to the operations
of $\mathsf{V}$), {\it with no extra requirement of closedness with
respect to any topology on} $\mathsf{V}$; we prefer to use the term
`(vector) subspace' of $\mathsf{V}$ for indicating a \emph{closed}
linear span (with respect to a given topology on $\mathsf{V}$).} in
$\mathcal{H}$, stable under the action of $U$, and, for any pair of
vectors $\phi\in\mathcal{H}$ and $\psi\in\admu$, the coefficient
$\coe_{\psi,\phi}^U$ is square integrable with respect to the left
Haar measure $\mu_G$ on $G$. Moreover, there exists a unique
positive selfadjoint, injective linear operator $\duu$ in
$\mathcal{H}$
--- {\em which we will call the `Duflo-Moore operator associated with} $U$' ---
such that
\begin{equation}
\admu =\mathrm{Dom}\big(\duu\big)
\end{equation}
and the following `orthogonality relations' hold:
\begin{equation} \label{orthrel}
\big\langle \coe_{\psi_1,\phi_1}^U,\hspace{0,5mm}
\coe_{\psi_2,\phi_2}^U\big\ranld = \langle\phi_1 ,\phi_2\rangle\,
\big\langle \duu\,\psi_2, \duu\,\psi_1\big\rangle  ,
\end{equation}
for all $\phi_1,\phi_2\in\mathcal{H}$ and all $\psi_1,\psi_2\in
\admu$. The Duflo-Moore operator $\duu$ is semi-invariant
--- with respect to $U$ --- with weight $\Delta_G^{1/2}$, i.e.
\begin{equation} \label{semi-invariance}
U(g)\,\duu = \Delta_G(g)^{\frac{1}{2}}\,\duu\, U(g)\fin ,\ \ \
\forall\hspace{0.4mm} g\in G\fin ;
\end{equation}
it is bounded if and only if $G$ is unimodular \emph{(i.e.\
$\Delta_G\equiv 1$)} and, in such case, it is a multiple of the
identity.
\end{theorem}
\begin{remark} \label{dufmoo}
{\rm If $U$ is square integrable, the associated Duflo-Moore
operator $\duu$, being injective and positive selfadjoint, has a
positive selfadjoint densely defined inverse. In the case where $U$
is a unitary representation, Duflo and Moore call the square of
$\du^{-1}$ the {\it formal degree} of the representation $U$. Note
that the operator $\duu$ is linked to the normalization of the Haar
measure $\mu_G$: if $\mu_G$ is rescaled by a positive constant, then
$\duu$ is rescaled by the square root of the same constant. We will
say, then, that $\duu$ is {\it normalized according to} $\mu_G$. On
the other hand, if a normalization of the left Haar measure on $G$
is not fixed, $\duu$ is defined up to a positive factor and we will
call a specific choice a {\it normalization of the Duflo-Moore
operator}. In particular, if $G$ is unimodular, then $\duu=I$ is a
normalization of the Duflo-Moore operator; the corresponding Haar
measure will be said to be \emph{normalized in agreement with the
representation} $U$. Moreover, observe that, as a consequence of
relation~{(\ref{semi-invariance})}, the dense linear span
$\mathrm{Dom}\big(\du^{-1}\big)=\mathrm{Ran}\big(\duu\big)$ --- like
the linear span $\admu =\mathrm{Dom}\big(\duu\big)$ --- is stable
under the action of $U$ and
\begin{equation} \label{semi-invariance-bis}
U(g)^{-1}\hspace{0.3mm}\du^{-1} =
\Delta_G(g)^{\frac{1}{2}}\,\du^{-1}\hspace{0.4mm} U(g)^{-1},\ \ \
\forall\hspace{0.4mm} g\in G\fin .
\end{equation}
From this relation, using the fact that $U(g)^{-1}=\mm(g,g^{-1})\,
U(g^{-1})$, we obtain:
\begin{equation} \label{useaaa}
U(g)\hspace{0.5mm}\du^{-1} =
\modu(g)^{-\frac{1}{2}}\,\du^{-1}\hspace{0.4mm} U(g)\fin ,\ \ \
\forall g\in G \fin ,
\end{equation}
We finally note that, in the case where $G$ is not unimodular, a
square integrable representation of $G$ cannot be finite-dimensional
(since the associated Duflo-Moore operator is
unbounded).~$\blacksquare$ }
\end{remark}

Let us list a few basic facts about square integrable
representations:
\begin{itemize}

\item The square-integrability of a representation is a property which extends to all its
physical equivalence class.

\item If the representation $U$ of $G$ is square integrable, then the
orthogonality relations~{(\ref{orthrel})} imply that, for every
nonzero admissible vector $\psi\in\admu$, one can define the linear
operator
\begin{equation}
\wt \colon \mathcal{H}\ni \phi\mapsto \big\|\duu\,\psi\big\|^{-1}\,
\coe_{\psi ,\phi}^U\in \ldg
\end{equation}
--- sometimes called {\it (generalized)
wavelet transform} generated by $U$, with {\it analyzing} or {\it
fiducial vector} $\psi$ --- which is an isometry.
For the adjoint
$\wta\colon\ldg\rightarrow\mathcal{H}$ of the isometry $\wt$ the
following weak integral `reconstruction formula' holds:
\begin{equation} \label{re-con-wt}
\wta\hspace{0.5mm} f = \big\| \duu\,\psi\big\|^{-1} \int_G
f(g)\hspace{0.5mm}\big(U(g)\,\psi\big)\,\de\mu_G(g)\fin ,\ \ \
\forall\hspace{0.4mm} f\in\ldg \fin .
\end{equation}
The ordinary wavelet transform arises in the special case where $G$
is the 1-dimensional affine group $\mathbb{R}\sdp\errep$
(see~\cite{Grossmann,Daubechies}); we will better clarify this point
in Sect.~{\ref{examples}}.

\item The isometry $\wt$ intertwines the square integrable
representation $U$ with the {\it left regular $\mm$-representation}
$R_\mm\hspace{-0.3mm}$ of $G$ in $\ldg$, see ref.~{\cite{Aniello}},
which is the projective representation (with multiplier $\mm$)
defined by:
\begin{equation} \label{defre}
\big(R_\mm\hspace{-0.3mm}(g) f\big)(g^\prime
)=\overset{\rightarrow}{\mm}(g,g^\prime)\, f(g^{-1}g^\prime )\fin ,\
\ \ g,g^\prime\in G\fin ,
\end{equation}
\begin{equation}
\overset{\rightarrow}{\mm}(g,g^\prime):=\mm(g,g^{-1})^\ast\,\mm(g^{-1},g^\prime)\fin
,
\end{equation}
for every $f\in \ldg$; namely:
\begin{equation} \label{first-intert}
\wt\, U(g) = R_\mm\hspace{-0.3mm}(g)\; \wt\fin ,\ \ \
\forall\hspace{0.4mm} g\in G \fin .
\end{equation}
Hence, $U$ is (unitarily) equivalent to a subrepresentation of
$R_\mm$. Note that, for $\mm\equiv 1$, $R\equiv
R_\mm\hspace{-0.3mm}$ is the standard left regular representation of
$G$.

\item Let the group $G$ be compact (hence, unimodular), and let $\chG$ be
a realization of the unitary dual of $G$. In this case, the \emph{unitary}
irreducible representations of $G$ are all finite-dimensional --- we
will denote by $\dimu$ the dimension of the Hilbert space of the
representation $U\in\chG$ --- and square integrable (since the Haar
measure on $G$ is finite and every coefficient of this
representation is a bounded function). They are ruled by the
Peter-Weyl theorem~{\cite{Folland,Simon}}. Precisely, the Hilbert
space $\ldg$ admits the following orthogonal sum decomposition
\begin{equation} \label{decldg}
\ldg = \bigoplus_{U\in\chG} \ldgu \fin ,
\end{equation}
where $\ldgu$ is a (closed) subspace of $\ldg$, characterized by the
following properties:
\begin{enumerate}
\item $\ldgu$ depends only on
the unitary equivalence class $[U]$ of $U$ and it is an invariant
subspace for the left regular representation $R$ of $G$;

\item for every orthonormal basis $\{\chi_n\}_{n=1}^{\dimu}$ in the Hilbert space of the
representation $U\in\chG$, we have that
\begin{equation} \label{decoldgu}
\ldgu = \bigoplus_{n=1}^{\dimu} \ran\big(\wavchi\big)
\end{equation}
--- hence: $\dim\big(\ldgu\big)=\dimu^2$; moreover,
$\ran\big(\wavchi\big)$ is an invariant subspace for the
regular representation $R$ and the restriction of $R$ to
$\ran\big(\wavchi\big)$ is unitarily equivalent to $U$;
therefore, `$U$ appears with multiplicity $\dimu$ in the left
regular representation $R$', namely, $R$ is unitarily equivalent to
the representation
\begin{equation}
\bigoplus_{U\in\chG} \overbrace{U \oplus \cdots \oplus U\hspace{0.2mm}}^{\dimu} \fin ;
\end{equation}

\item assuming that the Haar
measure $\mG$ is normalized as usual for compact groups --- i.e.\
$\mG(G)=1$ --- for any $\phi,\psi$ we have:
\begin{equation}
\dimu\int_G \langle\phi_1,U(g)\,\psi_1\rangle\, \langle
U(g)\,\psi_2,\phi_2 \rangle\ \de\mu_G (g)= \langle\phi_1
,\phi_2\rangle\, \langle \psi_2, \psi_1\rangle\fin ;
\end{equation}
hence, the Duflo-Moore operator associated with the unitary representation $U$
is of the form $\ddu\, I$, where $\ddu=\dimu^{-\frac{1}{2}}$.
\end{enumerate}

\item Let $\hat{q}$, $\hat{p}$ the standard position and momentum
operators in $\lr$. Then, the map
\begin{equation} \label{wey-sys}
U\colon\rr\ni(q,p)\mapsto\disp\in\mathcal{U}(\lr)
\end{equation}
is a projective representation of the (additive) group $\rr$. This
representation is square integrable and, fixing $(2\pi)^{-1}\de q\de
p$ as the Haar measure on $\rr$, we have that $\du=I$;
see~{\cite{Aniello-framet}}. Therefore, the Haar measure
$(2\pi)^{-1}\de q\de p$ is normalized in agreement with $U$. If
$\psi_0\in\lr$ is the ground state of the quantum harmonic
oscillator, then $\{ U(q,p)\,\psi_0\}_{q,p\in\mathbb{R}}$ is the
family of standard \emph{coherent
states}~{\cite{Perelomov,Klauder}}.

\end{itemize}

For the reader's convenience, we conclude this section fixing some further
notations and recalling a technical result that will be useful later on.

If $\hat{C}$ is a closable operator in $\hh$, the symbol
$\overline{\hat{C}}$ will indicate the closure of $\hat{C}$; a
\emph{core} for $\hat{C}$ is a linear span in $\hh$, contained in
the domain $\dom(\hat{C})$, such that the closure of the restriction
of $\hat{C}$ to this linear span coincides with the closure of
$\hat{C}$. For any pair of selfadjoint operators $\opa,\opb$ in
$\hh$, with a slight abuse of notation we will denote by
$\opa\otimes\opb$ the \emph{closure} of the ordinary tensor product
of $\opa$ by $\opb$, closure which is a selfadjoint operator. Given
a subspace $\mathcal{S}$ of $\hh$, we will denote by
$\mathcal{S}^\perp$ the orthogonal complement of $\mathcal{S}$ in
$\hh$. We will denote by $\bh$ the Banach space of bounded linear
operators in the Hilbert space $\hh$ and by $\|\cdot\|$ the
associated norm. We recall that the Hilbert space of Hilbert-Schmidt
operators $\hs$ in $\hh$ is a two-sided ideal in
$\bh$~{\cite{Reed}}; the associated scalar product and norm will be
denoted by $\langle\cdot,\cdot\ranglehs$ and $\|\cdot\norhs$,
respectively. Another two-sided ideal in $\bh$ is the Banach space
of trace class operators $\tch\subset\hs$. We will often use Dirac's
notation for rank-one operators: $|\phi\rangle\langle\psi| \sei \chi
:= \langle\psi,\chi\rangle \quattro \phi$, for any
$\phi,\psi,\chi\in\hh$.

Given a measure space $(X,\mu)$ the locution ``for $\mu$-almost all
$x$ in $X$'' will be usually substituted by the symbol $\aames x\in X$.
The following well known result will turn out to be very useful for
our purposes in Sect.~\ref{main}. Let the measure space $(X,\mu)$ be complete,
and let $\{f_n\}_{n\in \nat}$ be a sequence in $\ldx$
converging (in norm) to $f$. If there is a function $\tilde{f}\colon
X\rightarrow\ccc$ such that $\lim_{n\rightarrow\infty} f_n(x)
=\tilde{f}(x)$, $\aames x\in X$, then $\tilde{f}$ is
$\mu$-measurable and we have: $f=\tilde{f}$, the two functions being
regarded as elements of $\ldx$ (i.e.\ the two functions coincide
$\mu$-almost everywhere).\footnote{This result is a consequence of
the fact that the convergence with respect to the norm of $\ldx$
implies the convergence in $\mu$-measure.}

\section{Weyl-Wigner quantization-dequantization maps}
\label{weyl-wigner}

As we have recalled in the previous section, with every square
integrable representation of a l.c.s.c.\ group $G$ one can associate
an isometry --- the (generalized) wavelet transform --- mapping the
Hilbert space of the representation into the space $\ldg$. Beside
this map, one can define another important isometry. This isometry
maps the space of Hilbert-Schmidt operators --- acting in the
Hilbert space of the representation --- into $\ldg$. Since it
transforms operators into functions, it is called the \emph{Wigner
(dequantization) map}. Its adjoint, which transforms functions into
operators, is called the \emph{Weyl (quantization) map}.

Indeed, we recall that --- see~{\cite{Aniello-framet,Ali1,Ali2}} ---
given a square integrable projective representation $U \colon
G\rightarrow\mathcal{U}(\mathcal{H})$ (with multiplier $\mm$), with
every Hilbert-Schmidt operator $\opa\in\hs$ one can suitably
associate a function
\begin{equation}
G\ni g \mapsto \big(\wig \opa\big)(g)\in\ccc
\end{equation}
contained in $\ldg\equiv\mathrm{L}^2(G,\mu_G;\ccc)$. Denoting by
$\duu$, as in Sect.~{\ref{known}}, the Duflo-Moore operator
associated with $U$ (normalized according to a left Haar measure
$\mu_G$ on $G$), consider the following \emph{formal} definition: $
\big(\wig \opa\big)(g):=\tr\big(U(g)^\ast\opa\,\du^{-1}\big). $
Since the operator $U(g)^\ast\opa\,\du^{-1}$ (or, possibly, its
closure) is not, generally speaking, a trace class operator, the
given definition requires a rigorous interpretation. This can be
achieved by suitably restricting the class of Hilbert Schmidt
operators for which the definition makes sense, and then extending
`by density' the map obtained in such a way. To this aim, one can
exploit the fact that the \emph{finite rank operators} form a dense
linear span $\fr$ in the Hilbert space $\hs$.

Precisely, consider
those rank one operators in $\mathcal{H}$ that are of the type
\begin{equation}
\hfp=|\phi\rangle\langle\psi |,\ \ \ \phi\in\mathcal{H},\
\psi\in\mathrm{Dom}\big(\du^{-1}\big).
\end{equation}
The linear span generated by the operators of this form, namely, the
set
\begin{equation} \label{deffrl}
\frl := \big\{ \hat{F}\in\fr:\
\mathrm{Ran}(\hat{F}^\ast)=\Ker(\hat{F})^\perp\subset\mathrm{Dom}\big(\du^{-1}\big)\big\},
\end{equation}
is dense in $\fr$ and, hence, in $\hs$:
\begin{equation}
\overline{\frl}=\hs \fin .
\end{equation}
Explicitly, the elements of $\frl$ are those operators in $\fr$ that
admit a \emph{canonical decomposition} of the form
\begin{equation} \label{candeco}
\hat{F}=\sum_{k=1}^{\en} |\phi_k\rangle\langle \psi_k |\fin ,\ \ \
\en\in\nat\fin ,
\end{equation}
where $\{\phi_k\}_{k=1}^{\en}$, $\{\psi_k\}_{k=1}^{\en}$ are
linearly independent systems in $\hh$, with
$\{\psi_k\}_{k=1}^{\en}\subset \dom\big(\du^{-1}\big)$.
Later on, it will also turn out to be useful the definition of the following dense
linear span in $\hs$:
\begin{equation} \label{dlsp}
\frlr := \big\{ \hat{F}\in\fr:\ \mathrm{Ran}(\hat{F}),\tre
\mathrm{Ran}(\hat{F}^\ast)\subset\mathrm{Dom}\big(\du^{-1}\big)\big\}.
\end{equation}

Observe now
that, if we set
\begin{equation}
\big(\wig\,
\hfp\big)(g):=\tr\big(U(g)^\ast|\phi\rangle\langle\du^{-1}\tre\psi|\big)=\big\langle
U(g)\,\du^{-1}\tre\psi,\phi\big\rangle\fin ,\ \ \
\forall\hspace{0.4mm}\hfp\in\frl\fin ,
\end{equation}
then,  by virtue of the orthogonality relations~{(\ref{orthrel})},
for any $\hfpa\equiv |\phi_1\rangle\langle\psi_1 | ,\hfpb\in\frl$ we have:
\begin{eqnarray}
\int_G \big(\wig\, \hfpa\big)(g)^\ast\,\big(\wig\, \hfpb\big)(g)\;
\de\mu_G(g) \spa & = & \spa \int_G
\big\langle\phi_1,U(g)\,\du^{-1}\tre\psi_1\big\rangle\big\langle
U(g)\,\du^{-1}\tre\psi_2,\phi_2\big\rangle\; \de\mu_G(g) \nonumber\\
& = & \spa \langle\phi_1,\phi_2\rangle\,\langle\psi_2,\psi_1\rangle
= \big\langle \hfpa ,\hfpb\big\ranglehs\fin .
\end{eqnarray}
Therefore, extending the map $\wig$ to all $\fri$ by linearity, and
then to the whole Hilbert space $\hs$ by continuity, we obtain an
\emph{isometry}
\begin{equation}
\wig \colon\hs\rightarrow\ldg
\end{equation}
called the \emph{(generalized) Wigner map}, or \emph{Wigner
transform}, generated by the square integrable representation $U$.
We will denote by $\ru$ the range of the isometry $\wig$. It is easy
to check that $\ru$ depends only on the unitary equivalence class of
$U$. As the reader may prove, if the group $G$ is unimodular (hence:
$\duu=\ddu\hspace{0.4mm}I$, with $\ddu>0$), then for every trace
class operator $\hrho\in\mathcal{B}_1(\mathcal{H})$
--- in particular, for every density operator in $\mathcal{H}$ ---
we have simply:
\begin{equation} \label{non-formal}
\big(\wig\hspace{0.3mm}
\hrho\big)(g)=d_U^{-1}\,\tr(U(g)^\ast\hrho)\fin .
\end{equation}

\begin{remark} \label{ranus}
{\rm Suppose that $U$ is, in particular, a standard unitary
representation, and let $V$ be another square integrable unitary
representation of $G$ (acting in a Hilbert space $\mathcal{H}^\prime$), unitarily
\emph{inequivalent} to $U$. Then, it is easy to show that
\begin{equation} \label{rel-ranus}
\big(\ru\equiv\ran\big(\wig\big)\big)\perp\ran\big(\twig\big),
\end{equation}
where $\twig$ is the Wigner map generated by $V$. Indeed, let
$\wt$ and $\wtt$ be the wavelet transforms generated by $U$ and
$V$, with analyzing vectors $\psi\in\hh$ and $\eta\in\mathcal{H}^\prime$,
respectively. Relation~{(\ref{first-intert})} (with $\mm\equiv 1$)
implies that the bounded linear map
$\wtta\sette\wt\colon\hh\rightarrow\mathcal{H}^\prime$ intertwines the unitary
representation $U$ with $V$; hence, by Schur's lemma, it must be
identically zero. Therefore, we have that
\begin{equation}
0=\big\langle \wtta\sette\wt\quattro \phi,\xi\big\rangle =
\big\langle \wt\quattro \phi,\wtt\quattro\xi\big\rangle ,\ \ \
\forall\quattro\phi\in\hh \fin ,\ \forall\quattro\xi\in\mathcal{H}^\prime \fin ;
\end{equation}
i.e.\ $\ran\big(\wt\big)\perp\ran\big(\wtt\big)$. At this point,
relation~{(\ref{rel-ranus})} follows observing that
\begin{equation}
\ru=\overline{\mathrm{span}\big\{f\in\ran\big(\wt\big)\colon\
\psi\in\admu,\ \psi\neq 0 \big\}},
\end{equation}
and, of course, an analogous relation holds for the range of
$\twig$.~{$\blacksquare$}    }
\end{remark}

\begin{remark} \label{co-case}
{\rm Suppose that the group $G$ is compact  --- hence, unimodular
--- and $U$ is a (irreducible) \emph{unitary} representation. Then, by
relation~{(\ref{decoldgu})}, we have:
\begin{equation} \label{aa-bb}
\ldgu = \bigoplus_{n=1}^{\dimu} \ran\big(\wavchi\big)=
\mathrm{span}\left\{\coe_{\psi ,\phi}^U\colon \psi ,\phi\in\hh\right\}
=\ru\fin ,
\end{equation}
where the function $\coe_{\psi ,\phi}^U\in\ldg$ is the coefficient defined by~{(\ref{defcoe})}. Therefore,
by relation~{(\ref{decldg})}, we conclude that
\begin{equation} \label{cc-dd}
\ldg = \bigoplus_{U\in\chG} \ru \fin ,
\end{equation}
where we recall that the symbol $\chG$ denotes a realization of the unitary dual of $G$.~{$\blacksquare$}
}
\end{remark}

We will now explore the `intertwining properties' of the Wigner map
$\wig$ with respect to the natural action of the group $G$ in the Hilbert-Schmidt space $\hs$,
and to the standard complex conjugation in $\hs$. To this aim, let
us consider the map
\begin{equation}
\rep \colon G\rightarrow\mathcal{U}(\hs)
\end{equation}
defined by
\begin{equation} \label{definrep}
\rep(g)\hspace{0.3mm} \opa := U(g)\, \opa\, U(g)^\ast, \ \ \ \forall
\hspace{0.5mm}g\in G,\ \ \opa\in\hs\fin .
\end{equation}
The map $\rep$ is a (strongly continuous) \emph{unitary}
representation
--- even if, in general, the representation $U$ has only been
assumed to be \emph{projective}
---  which can be regarded as the standard action of the `symmetry
group' $G$ on the `quantum-mechanical operators' (`observables' or
`states'). Next, let us consider the map
\begin{equation}
\two \colon G\rightarrow \mathcal{U}(\ldg)
\end{equation}
defined by
\begin{equation} \label{two-sided}
\big(\two(g) f)(g^\prime) := \Delta_G(g)^{\frac{1}{2}}\hspace{0.9mm}
\mmm(g,g^\prime)\hspace{0.8mm} f(g^{-1}g^\prime g)\fin ,
\end{equation}
where the function $\mmm :G\times G\rightarrow \mathbb{T}$ has the
following expression:
\begin{equation} \label{demmm}
\mmm(g,g^\prime):= \mm(g,g^{-1}g^\prime)^\ast\hspace{0.5mm}
\mm(g^{-1}g^\prime,g)\fin .
\end{equation}
As the reader may check by means of a direct calculation involving
multipliers, the map $\two$ is a unitary representation; the
presence of the square root of the modular function $\Delta_G$ in
formula~{(\ref{two-sided})} takes into account the right action of
$G$ on itself. Notice that, for $\mm\equiv 1$, it coincides with the
restriction to the `diagonal subgroup' of the \emph{two-sided
regular representation} of the direct product group $G\times G$;
see~\cite{Folland,Segal}.

\begin{proposition}
The Wigner transform $\wig$ intertwines the representation $\rep$
with the representation $\two$; namely,
\begin{equation} \label{intertreps}
\wig\hspace{0.8mm} \rep (g)= \two (g)\, \wig\fin ,\ \ \ \forall
\hspace{0.5mm}g\in G\fin .
\end{equation}
Therefore, $\ru$ is an invariant subspace for the unitary
representation $\two$ and the representation $\rep$ is unitarily
equivalent to a subrepresentation of $\two$, i.e. to the restriction
of $\two$ to $\ru$.
\end{proposition}

\noindent {\bf Proof:} Let us first prove that $\wig\hspace{0.8mm}
\rep (g)\,\hfp = \two (g)\, \wig\,\hfp$  for any rank-one operator
$\hfp$ of the form
\begin{equation} \label{opeform}
\hfp\equiv|\phi\rangle\langle\psi |,\ \ \ \mbox{with}\ \phi\in\hh,\
\psi\in \mathrm{Dom}\big(\du^{-1}\big).
\end{equation}
Observe that, for every $g\in G$, we have:
\begin{eqnarray}
\big(\wig\hspace{0.8mm} \rep (g) \, \hfp \big)(g^\prime) =
\big(\wig\hspace{0.3mm}  |U(g)\,\phi\rangle\langle U(g)\,\psi |
\big)(g^\prime) \spa & = & \spa
\big\langle U(g^\prime)\, \du^{-1}\hspace{0.4mm} U(g)\,\psi, U(g)\,\phi\big\rangle \nonumber\\
& = & \spa \big\langle U(g)^{\ast}\hspace{0.4mm} U(g^\prime)\,
\du^{-1}\hspace{0.4mm} U(g)\,\psi, \phi\big\rangle  .
\end{eqnarray}
At this point, we can exploit relation~{(\ref{useaaa})} and the
standard properties of multipliers:
\begin{eqnarray}
\big(\wig\hspace{0.8mm} \rep (g) \, \hfp \big)(g^\prime) \spa & = &
\spa \Delta_G(g)^{\frac{1}{2}}\, \big\langle
U(g)^{\ast}\hspace{0.4mm} U(g^\prime)\,
U(g)\,\du^{-1}\hspace{0.3mm}\psi, \phi\big\rangle
\nonumber \\
& = & \spa \Delta_G(g)^{\frac{1}{2}}\, \mm (g,g^{-1})^{\ast}\,
\big\langle U(g^{-1})\, U(g^\prime)\,
U(g)\,\du^{-1}\hspace{0.3mm}\psi, \phi\big\rangle
\nonumber \\
& = & \spa \Delta_G(g)^{\frac{1}{2}}\, \mm
(g,g^{-1})^{\ast}\hspace{0.5mm}\mm(g^{-1},g^\prime)
\hspace{0.5mm}\mm(g^{-1}g^\prime,g)\, \big\langle U(g^{-1}g^\prime
g)\,\du^{-1}\hspace{0.3mm}\psi, \phi\big\rangle
\nonumber \\
& = & \spa \Delta_G(g)^{\frac{1}{2}}\, \mm(g,g^{-1}g^\prime)^\ast
\hspace{0.5mm} \mm(g^{-1}g^\prime,g) \, \big\langle U(g^{-1}g^\prime
g)\,\du^{-1}\hspace{0.3mm}\psi, \phi\big\rangle \nonumber \\
& = & \spa \big(\two (g)\, \wig\,\hfp\big)(g^\prime) \fin .
\end{eqnarray}
This relation extends to the linear span generated by the rank-one
operators of the form~{(\ref{opeform})}; i.e.\ to the dense linear
span $\frl$. Therefore, the bounded operators $\wig\hspace{0.8mm}
\rep(g)$ and $\two (g)\, \wig$ coincide on a dense linear span in
$\hs$; hence, they are equal.~{$\square$}
\begin{remark} {\rm
By a procedure analogous to the one adopted for proving relation~{(\ref{intertreps})}
one can check that
\begin{equation} \label{simpinter}
\wig\hspace{0.3mm} \big(U (g)\hspace{0.5mm}\opa\big) = \rem (g)\hspace{0.4mm}\big(\wig\opa\big),\ \ \ \forall
\hspace{0.3mm}\opa\in \hs \fin ,\ \forall
\hspace{0.4mm}g\in G \fin .
\end{equation}
This relation will be useful in Sect.~{\ref{main}}.~$\blacksquare$
}
\end{remark}

Let us consider, now, the antilinear map
$\invol\colon\ldg\rightarrow\ldg$ defined by
\begin{equation} \label{definvol}
\big(\invol f\big)(g) := \Delta_G(g)^{-\frac{1}{2}}\, \mm
(g,g^{-1})\, f(g^{-1})^\ast,\ \ \ \forall\hspace{0.4mm} f\in\ldg\fin
.
\end{equation}
We leave to the reader the easy task of verifying that the map
$\invol$ is (well defined and) a \emph{complex conjugation} in
$\ldg$: $\invol=\involu^{\hspace{0.3mm}\ast}$ and
$\involu^{\hspace{0.4mm}2}=I$ (i.e.\ $\invol$ is a selfadjoint
antiunitary map).

\begin{proposition} \label{propoinvol}
The isometry $\wig$ intertwines the standard complex conjugation
\begin{equation} \label{definvo}
\invo\colon\hs\ni \opa\mapsto\opa^\ast\in\hs
\end{equation}
in the Hilbert space $\hs$ with the complex conjugation $\invol$ in
$\ldg$; namely,
\begin{equation} \label{intcoco}
\wig\hspace{0.3mm} \invo =  \invol \wig\fin .
\end{equation}
Therefore, $\ru$ is an invariant subspace for the complex
conjugation $\invol$.
\end{proposition}

\noindent {\bf Proof:} We will first prove that $\wig\hspace{0.3mm}
\invo\,\hfp =  \invol \wig \,\hfp$ for any rank-one operator $\hfp$
of the form
\begin{equation} \label{opeform-bis}
\hfp\equiv|\phi\rangle\langle\psi |,\ \ \ \mbox{with}\ \phi, \psi\in
\mathrm{Dom}\big(\du^{-1}\big).
\end{equation}
Observe that we have:
\begin{eqnarray}
\big(\wig\hspace{0.3mm} \invo \, \hfp \big)(g) =
\big(\wig\hspace{0.3mm}  |\psi\rangle\langle \phi | \big)(g) =
\big\langle U(g)\, \du^{-1}\hspace{0.3mm} \phi, \psi\big\rangle \spa
& = & \spa
\big\langle
\du^{-1}\hspace{0.4mm}U(g)^\ast\hspace{0.3mm}\psi,\phi\big
\rangle^\ast \nonumber\\
& = & \spa \Delta_G(g)^{-\frac{1}{2}}\, \big\langle
U(g)^\ast\hspace{0.3mm}\du^{-1}\hspace{0.3mm}\psi,\phi\big\rangle^\ast,
\end{eqnarray}
where for obtaining the last equality we have used
relation~{(\ref{useaaa})}. Then, taking into account that
$U(g)^\ast=U(g)^{-1}=\mm(g,g^{-1})\, U(g^{-1})$, we find:
\begin{equation}
\big(\wig\hspace{0.3mm} \invo \, \hfp \big)(g) =
\Delta_G(g)^{-\frac{1}{2}}\, \mm(g,g^{-1})\, \langle
U(g^{-1})\hspace{0.8mm}\du^{-1}\hspace{0.4mm}\psi,\phi\rangle^\ast =
\big(\invol \wig \,\hfp\big)(g)\fin .
\end{equation}
Extending this relation to the linear span generated by the rank-one
operators of the form~{(\ref{opeform-bis})} --- i.e.\ to the dense
linear span $\frlr$  in $\hs$ --- and then, by continuity, to the
whole $\hs$ one completes the proof.~{$\square$}

Since the generalized Wigner transform $\wig$ is an isometry, the
adjoint map
\begin{equation}
\wigg^\ast \colon \ldg\rightarrow\hs
\end{equation}
is a partial isometry such that
\begin{equation}
\wigg^\ast\,\wig =I,\ \ \ \wig\,\wigg^\ast=\proi,
\end{equation}
where $\proi$ is the orthogonal projection onto the subspace
$\ru\equiv\mathrm{Ran}(\wig)=\mathrm{Ker}(\wigg^\ast)$ of $\ldg$.
Thus, the partial isometry $\wigg^\ast$ is the pseudo-inverse of
$\wig$ and we will call it \emph{(generalized) Weyl map} associated
with the representation $U$.

Let us provide an expression of the Weyl map. As is well known, the
weak integral
\begin{equation}
\uf :=\int_G \effe(g)\, U(g)\; \de\mu_G(g)\fin ,\ \ \
\forall\hspace{0.5mm}\effe\in\lug \fin ,
\end{equation}
defines a bounded operator in $\hh$ (here the square-integrability
of $U$ does not play any role). Then, one can easily prove the
following result:
\begin{proposition}
For every $\effe\in\lug\cap\ldg$, the densely defined operator $\uf\hspace{0.4mm}\du^{-1}$ extends
to a Hilbert-Schmidt operator and
\begin{equation}
\overline{\uf\hspace{0.4mm}\du^{-1}}= \wigg^\ast\hspace{0.5mm}
\effe\fin .
\end{equation}
Therefore, for every function $f\in\ldg$ --- given a sequence
$\{\effe_n\}_{n\in\nat}$ in $\ldg$, contained in the dense linear
span $\lug\cap\ldg$, such that $\displaystyle \norlim \effe_n =f$
--- we have:
\begin{equation} \label{forwey}
\wigg^\ast\hspace{0.5mm} f = \hslim \wigg^\ast\hspace{0.5mm} \effe_n = \hslim \overline{\ufn\hspace{0.4mm}\du^{-1}}.
\end{equation}
In the case where the group $G$ is unimodular, the following weak integral formula holds:
\begin{equation}
\wigg^\ast\hspace{0.5mm} f=d_{U}^{-1}\int_G f(g)\, U(g)\;
\de\mu_G(g),\ \ \ \forall\hspace{0.3mm}f\in\ldg\fin .
\end{equation}
\end{proposition}
We will now prove a result that will be useful in
Sect.~{\ref{examples}}.
\begin{proposition} \label{propouf}
Suppose that the Hilbert space $\hh$ where the representation $U$
acts is a space $\lx\equiv\ldx$ of square integrable functions on a
$\sigma$-finite measure space $(X,\mu)$. Then, for every
$\effe\in\lug$ and every $\phi\in\lx$, the function $G\ni g\mapsto
\effe(g)\, \big(U(g)\,\phi\big)(x)\in\ccc$ belongs to $\lug$ for
$\mu$-a.a.\ $x\in X$, and the following relation holds:
\begin{equation} \label{weakint}
\big(\uf\hspace{0.5mm} \phi\big)(x)= \int_G \effe(g)\,
\big(U(g)\,\phi\big)(x)\; \de\mu_G(g)\fin ,\ \ \ \aames x\in X\fin .
\end{equation}
Therefore, for every $\effe\in\lug\cap\ldg$ and every
$\varphi\in\dom\big(\du^{-1}\big)\subset\lx$, we have:
\begin{equation} \label{forwey-bis}
\Big(\big(\wigg^\ast\hspace{0.5mm}
\effe\big)\hspace{0.5mm}\varphi\Big)(x) = \int_G \effe(g)\,
\big(U(g)\,\du^{-1}\tre\varphi\big)(x)\; \de\mu_G(g)\fin ,\ \ \
\aames x\in X\fin .
\end{equation}
\end{proposition}

\noindent {\bf Proof:} By definition of the operator $\uf$ (which
involves a weak integral), we have that, for every $\effe\in\lug$
and any $\phi,\psi\in\lx$,
\begin{eqnarray}\hspace{-6mm}
\langle\psi,\uf\hspace{0.5mm} \phi\rangle= \langle \psi, \intG
\effe(g)\, U(g)\; \de\mu_G(g)\hspace{1mm} \phi\rangle \spa & = &
\spa \intG \effe(g)\, \langle\psi, U(g)\,\phi\rangle\; \de\mu_G(g)
\nonumber\\   \label{lame} & = & \spa \intG \de\mu_G(g) \int_X
\de\mu(x)\hspace{1mm} \effe(g)\,\psi(x)^\ast
\big(U(g)\,\phi\big)(x)\fin .
\end{eqnarray}
By the arbitrariness of $\psi\in\lx$, relation~{(\ref{weakint})} can be proved showing that
\begin{equation}
\langle \psi, \intG \effe(g)\, U(g)\; \de\mu_G(g)\hspace{1mm}
\phi\rangle = \int_X \de\mu(x)\hspace{1mm} \psi(x)^\ast  \intG
\de\mu_G(g) \hspace{1mm} \effe(g)\,\big(U(g)\,\phi\big)(x)\fin ,
\end{equation}
namely, that the iterated integrals in the last member of~{(\ref{lame})} can be permuted.
In fact, since $(X,\mu)$ (like $(G,\mG)$) is a $\sigma$-finite measure space,
we can use Tonelli's theorem; by this theorem, the Schwarz inequality and the fact that $\|U(g)\,\phi\|=\|\phi\|$,
we have:
\begin{eqnarray}
\int_{G\times X} \hspace{-0.5mm}\big|\psi(x)^\ast\,
\effe(g)\,\big(U(g)\,\phi\big)(x)\big|\;
\de\hspace{0.4mm}\mu_G\hspace{-0.4mm}\otimes\hspace{-0.4mm}\mu\hspace{0.5mm}(g,x)
\spa & = & \spa \intG \de\mu_G(g)\hspace{1mm} |\effe(g)|
\int_X \de\mu(x)\hspace{1mm} |\psi(x)|\hspace{0.7mm} \big|\big(U(g)\,\phi\big)(x)\big|
\nonumber\\
& \le & \spa  \|\effe\norlu \hspace{0.3mm} \|\psi\|\hspace{0.5mm}
\|\phi\|\fin .
\end{eqnarray}
Therefore, the function $G\times \hspace{-0.4mm}X\ni (g,x) \mapsto
\psi(x)^\ast\, \effe(g)\,\big(U(g)\,\phi\big)(x)$ belongs to
$\mathrm{L}^1(G\times \hspace{-0.4mm}X,
\mu_G\hspace{-0.4mm}\otimes\hspace{-0.4mm}\mu;\ccc)$, for any
$\phi,\psi\in\lx$. It follows by Fubini's theorem that the function
$G\ni g\mapsto \effe(g)\, \big(U(g)\,\phi\big)(x)$ belongs to
$\lug$, $\aames x\in X$,  and the iterated integrals in the last
member of~{(\ref{lame})} can indeed be permuted.~{$\square$}

\section{Star products from quantization-dequantization maps}
\label{defstar}

In this section, we will show that the quantization-dequantization
maps previously introduced induce, in a natural way, a `star product
of functions' enjoying remarkable properties. Let $U$ be a square
integrable projective representation  of the l.c.s.c.\ group $G$ in
the Hilbert space $\hh$, and let $\wig\colon \hs\rightarrow\ldg$ be
the associated Wigner map. Consider the following bilinear map from
$\ldg\times\ldg$ into $\ldg$:
\begin{equation} \label{defist}
(\cdot)\starp(\cdot)\colon \ldg\times\ldg\ni (f_1,f_2)\mapsto
\wig\Big(\big(\wigg^\ast\hspace{0.3mm} f_1\big)
\big(\wigg^\ast\hspace{0.3mm}
f_2\big)\Big)\hspace{-0.8mm}\in\ldg\fin ;
\end{equation}
i.e.\ $f_1\starp f_2$ is the function obtained dequantizing the
product (composition) of the two operators which are the `quantized
versions' of the functions $f_1$, $f_2$. We will call the bilinear
map~{(\ref{defist})} the \emph{star product associated with the
representation} $U$.

Before considering the properties of the star product associated
with $U$, it is worth fixing some terminology about algebras. By a
\emph{Banach algebra} we mean an associative algebra $\mathcal{A}$
which is a Banach space (with norm $\|\cdot\|_{\mathcal{A}}$) such
that
\begin{equation}
\|a\hspace{0.5mm} b\|_{\mathcal{A}}\le \|a\|_{\mathcal{A}}\, \|
b\|_{\mathcal{A}}\fin , \ \ \  \forall\hspace{0.4mm} a,b
\in\mathcal{A}\fin .
\end{equation}
Given Banach algebras $\mathcal{A}$ and $\mathcal{A}^\prime$, we
will say that a linear map
$\isomo\colon\mathcal{A}\rightarrow\mathcal{A}^\prime$ is an
(isometric) \emph{isomorphism of Banach algebras} if it is a
surjective isometry such that
$\isomo(a\hspace{0.5mm}b)=\isomo(a)\isomo(b)$, for all
$a,b\in\mathcal{A}$. A Banach algebra $\mathcal{A}$ --- endowed with
an involution\footnote{Let $\mathsf{V}$ be a vector space, and
$(\cdot,\cdot)\colon
\mathsf{V}\times\mathsf{V}\rightarrow\mathsf{V}$ a bilinear
operation in $\mathsf{V}$. We recall that an an \emph{involution} in
$\mathsf{V}$, with respect to the bilinear operation
$(\cdot,\cdot)$, is an antilinear map $\mathsf{V}\ni a\mapsto
a^\ast\in \mathsf{V}$ satisfying: $(a^\ast)^\ast=a$ and
$(a,b)^\ast=(b^\ast,a^\ast)$, $\forall\hspace{0.4mm}
a,b\in\mathsf{V}$.} ($a\mapsto a^\ast$) --- such that
\begin{equation} \label{consnorm}
\|a\|_{\mathcal{A}} = \|a^\ast\|_{\mathcal{A}}\fin , \ \ \
\forall\hspace{0.4mm} a \in\mathcal{A}\fin ,
\end{equation}
will be called a \emph{Banach $\ast$-algebra} (Banach star-algebra; of course, the `star' in $\ast$-algebra, which
refers to an involution, should not generate confusion with the `star' product).

A Banach $\ast$-algebra $\mathcal{A}$ is said to be a
\emph{$\Hstar$-algebra}~{\cite{Ambrose,Rickart}} if, in addition, it
is a (separable complex) Hilbert space (with
$\|a\|_{\mathcal{A}}=\sqrt{\langle
a,a\rangle_{\mathcal{A}}}\hspace{0.5mm}$) satisfying:
\begin{equation} \label{conscpr}
\langle a\hspace{0.5mm} b,c\rangle_{\mathcal{A}}=\langle b,a^\ast
c\rangle_{\mathcal{A}} \ \ \ \mbox{and}\ \ \ \langle a\hspace{0.5mm}
b,c\rangle_{\mathcal{A}}=\langle a, c\hspace{0.6mm}
b^\ast\rangle_{\mathcal{A}}\fin ,\ \ \ \forall\hspace{0.5mm}a,b,c
\in\mathcal{A}\fin ;
\end{equation}
Clearly, condition~{(\ref{consnorm})} now means that
the involution $\mathcal{A}\ni a\mapsto a^\ast\in\mathcal{A}$ is a complex conjugation
(an idempotent antiunitary operator).

\begin{remark} {\rm The definition of a $\Hstar$-algebra given above
may seem to be stricter than the usual definition. In fact, one
usually defines a $\Hstar$-algebra as a Banach algebra $\mathcal{A}$
which is a Hilbert space and satisfies the following condition: for
each $a\in\mathcal{A}$, there is an element
$a^\diamond\in\mathcal{A}$ (which need not be unique)
--- an \emph{adjoint} of $a$ --- such that
\begin{equation}
\langle a\hspace{0.5mm} b,c\rangle_{\mathcal{A}}=\langle
b,a^\diamond \hspace{0.3mm} c\rangle_{\mathcal{A}} \ \ \ \mbox{and}\
\ \ \langle a\hspace{0.5mm} b,c\rangle_{\mathcal{A}}=\langle a,
c\hspace{0.6mm} b^\diamond\rangle_{\mathcal{A}}\, ,\ \ \
\forall\hspace{0.5mm} b,c\in\mathcal{A}\, .
\end{equation}
Let us show that the two definitions are equivalent; i.e.\ that the
\emph{usual definition} implies the \emph{strict definition}. We
will use some terminology and results from~{\cite{Ambrose,Rickart}}.
Let $\mathcal{A}$ be a $\Hstar$-algebra, according to the usual
definition. For every element $x$ of a $\mathcal{A}$, the two
relations $x\hspace{0.3mm}\mathcal{A}=\{0\}$ and
$\mathcal{A}\hspace{0.5mm}x=\{0\}$ turn out to be equivalent. The
\emph{annihilator ideal} of $\mathcal{A}$ is the set defined by
\begin{equation}
\mathcal{A}_0 := \{x\in\mathcal{A}\colon\hspace{0.6mm}
x\hspace{0.3mm}\mathcal{A}=\{0\}\}=
\{x\in\mathcal{A}\colon\hspace{0.6mm}
\mathcal{A}\hspace{0.5mm}x=\{0\}\}.
\end{equation}
The annihilator ideal is a selfadjoint\footnote{A subset
$\mathcal{E}$ of $\mathcal{A}$ is said to be \emph{selfadjoint} if
the set $\mathcal{E}^\diamond$ consisting of all the adjoints of the
elements of $\mathcal{E}$ coincides with the set $\mathcal{E}$
itself.} closed two-sided ideal in $\mathcal{A}$; the set of all the
adjoints of any $x\in\mathcal{A}_0$ is $\mathcal{A}_0$ itself. A
$\Hstar$-algebra $\mathcal{A}$ is said to be \emph{proper} (or
\emph{semisimple}) if it satisfies the following two equivalent
conditions:
\begin{eqnarray}
\left(x\in\mathcal{A},\ \; x\hspace{0.3mm}\mathcal{A}=\{0\}\ \;
\Rightarrow  \; \  x=0\right) \ \ \ \mbox{and}\ \ \
\left(x\in\mathcal{A},\ \; \mathcal{A}\hspace{0.5mm}x=\{0\}\ \;
\Rightarrow  \ \;  x=0\right) ;
\end{eqnarray}
namely, if $\mathcal{A}_0=\{0\}$. Every element $y$ of a proper
$\Hstar$-algebra $\mathcal{A}$ has a \emph{unique} adjoint --- which
we denote by $y^\bullet$ --- and
$\|y\|_{\mathcal{A}}=\|y^\bullet\|_{\mathcal{A}}$; moreover, the map
$\mathcal{A}\ni y\mapsto y^\bullet\in\mathcal{A}$ is an involution.\\
A $\Hstar$-algebra $\mathcal{A}$ admits
an orthogonal sum decomposition of the following type:
\begin{equation} \label{candec}
\mathcal{A}= \mathcal{A}_0 \oplus \mathcal{A}_1 \fin ,
\end{equation}
where $\mathcal{A}_0$ is the annihilator ideal of $\mathcal{A}$, and
$\mathcal{A}_1$ is a closed two sided ideal which (endowed with the
restriction of the algebra operation of $\mathcal{A}$) is a proper
$\Hstar$-algebra. We will call $\mathcal{A}_1$ the \emph{canonical
ideal} of $\mathcal{A}$, and we will denote by $\prau$ the
orthogonal projection onto $\mathcal{A}_1$. The canonical ideal is
characterized by the relation
\begin{equation} \label{charcan}
a\hspace{0.5mm}b = \big(\prau a\big) \hspace{0.5mm} \big(\prau
b\big),\ \ \ \forall\hspace{0.5mm} a,b\in \mathcal{A}\fin ,
\end{equation}
in the following sense. Suppose that $\ati\subset \mathcal{A}$ is a
closed two-sided ideal, which is a proper $\Hstar$-algebra such that
$a\hspace{0.5mm}b = \big(\prati a\big) \hspace{0.5mm} \big(\prati
b\big)$, $\forall \cinque a,b\in \mathcal{A}$. Then, it is easy to
show that $\ati=\mathcal{A}_1$.\footnote{Indeed, for each
$a\in\ati^\perp$, it is clear that $a\hspace{0.5mm}b = 0$,
$\forall\cinque b\in\mathcal{A}$. Hence,
$\ati^\perp\subset\mathcal{A}_0$ and $\ati\supset\mathcal{A}_1$.
Now, let $c$ be a vector in $\ati$. Then, $c=c_0+c_1$, for some
$c_0\in\mathcal{A}_0$ and $c_1\in\mathcal{A}_1$, and, since
$\ati\supset\mathcal{A}_1$, $c_0=c-c_1\in\ati$. Therefore, $c_0=0$
as $\ati$ is a proper $\Hstar$-algebra. It follows that
$\ati=\mathcal{A}_1$.} Let
$J_0\colon\mathcal{A}_0\rightarrow\mathcal{A}_0$ be an arbitrary
complex conjugation in the annihilator ideal, and $J_1\colon
\mathcal{A}_1\rightarrow\mathcal{A}_1$ the involution defined by
$J_1\hspace{0.5mm} y = y^\bullet$, for any $y\in \mathcal{A}_1$
--- where $y^\bullet$ is the unique adjoint of $y$ belonging to
$\mathcal{A}_1$\footnote{It is clear that a generic adjoint of $y\in\mathcal{A}_1$
is of the form $y^\diamond=x+y^\bullet$, where $x$ is an arbitrary element of $\mathcal{A}_0$.} ---
which is a complex conjugation since $y$ and $y^\bullet$
satisfy: $\|y\|_{\mathcal{A}}=\|y^\bullet\|_{\mathcal{A}}$. We can
now define an antilinear map $a\mapsto a^\ast$ in $\mathcal{A}$ by
setting:
\begin{equation} \label{prototype}
a^\ast = (J_0 \oplus J_1)\hspace{0.8mm} a\fin ,\ \ \
\forall\hspace{0.4mm} a\in\mathcal{A}\fin .
\end{equation}
It is clear that this map is an involution that verifies both
conditions~{(\ref{consnorm})} and~{(\ref{conscpr})}. It is easy to
check that any involution $a\mapsto a^\ast$ in $\mathcal{A}$
satisfying~{(\ref{consnorm})} and~{(\ref{conscpr})} must be of the
form~{(\ref{prototype})}.~{$\blacksquare$} }
\end{remark}

A linear map $\isomo\colon\mathcal{A}\rightarrow\mathcal{A}^\prime$
--- where $\mathcal{A}$, $\mathcal{A}^\prime$ are $\Hstar$-algebras
--- is said to be an \emph{isomorphism of $\Hstar$-algebras} if it
is a unitary operator such that
\begin{equation}
\isomo(a\hspace{0.5mm}b)=\isomo(a)\isomo(b)\ \ \ \mbox{and}\ \ \
\isomo(a^\ast) =\isomo(a)^\ast,\ \ \ \forall\hspace{0.4mm} a,b
\in\mathcal{A}\fin .
\end{equation}

As is well known, the Hilbert space $\hs$ is a proper
$\Hstar$-algebra with respect to the ordinary composition of
operators (algebra operation) and to the standard complex
conjugation $\invo$ (involution), see~{(\ref{definvo})}.

The star product defined above is characterized by the following result:
\begin{proposition} \label{propost}
The bilinear map $(\cdot)\starp(\cdot)\colon
\ldg\times\ldg\rightarrow\ldg$ associated with the square integrable
projective representation $U$ enjoys the following properties:
\begin{enumerate}
\item the vector space $\ldg$, endowed with the operation $(\cdot)\starp(\cdot)$, is an associative algebra;

\item
the antilinear map $\invol$ is an involution in the vector space $\ldg$ with respect to the bilinear operation
$(\cdot)\starp(\cdot)$, i.e.
\begin{equation} \label{eqinvo}
\invol\big(\invol f\big)= f \ \ \ \mbox{and}\ \ \
\invol\Big(f_1\starp f_2\Big)= \big(\invol f_2\big)\starp
\big(\invol f_1\big),\ \ \ \forall\hspace{0.4mm} f,f_1,
f_2\in\ldg\fin ;
\end{equation}

\item
$\ldg$ --- regarded as a Banach space with respect to the norm $\|\cdot\norld$,
and endowed with the the star product associated with $U$
and with the involution $\invol$ --- is a Banach $\ast$-algebra; in particular, it satisfies the relation
\begin{equation} \label{disegua}
\Big\|f_1\starp f_2\Bnorld \le \|f_1\norld \|f_2\norld \fin ,\ \ \
\forall\hspace{0.4mm} f_1,f_2\in\ldg \fin ;
\end{equation}

\item $\au\equiv\big(\ldg, (\cdot)\starp(\cdot),\invol\big)$ is a
$\Hstar$-algebra; indeed, for all $f_1,f_2,f_3 \in\ldg$,
\begin{equation} \label{prophstar}
\langle f_1\starp f_2,f_3\rangle_{\ld}=\langle f_2, (\invol
f_1)\starp f_3\rangle_{\ld} \ \ \ \mbox{and}\ \ \ \langle f_1\starp
f_2,f_3\rangle_{\ld}=\langle f_1, f_3\starp(\invol f_2)\rangle_{\ld}
\fin ;
\end{equation}

\item for any $f_1,f_2\in\ldg$, we have that
\begin{equation} \label{contdue}
f_1\starp f_2\in\ru \fin ;
\end{equation}
therefore, the (closed) subspace $\ru\equiv\ran\big(\wig\big)$ of $\ldg$ is a
closed two-sided ideal in $\au$ and --- endowed with the restrictions
of the star product associated with $U$ and of the involution
$\invol$ {\rm ($\ru$ is an invariant subspace for $\invol$,
see Proposition~{\ref{propoinvol}})} --- is a $\Hstar$-algebra;

\item the $\Hstar$-algebra $\ru$ is proper and, for any $f_1,f_2\in\ldg$, we have that
\begin{equation} \label{contuno}
f_1\starp f_2=\big(\proi\hspace{0.5mm} f_1\big)\starp \big(\proi\hspace{0.5mm} f_2\big);
\end{equation}
hence, $\ru$ and its orthogonal complement $\ruort$ are,
respectively, the canonical ideal and the annihilator ideal of $\au$,
and the $\Hstar$-algebra $\au$ is proper if and only if $\ru=\ldg$;

\item  the unitary operator
\begin{equation} \label{isostalg}
\hs\ni\opa\mapsto \wig\opa\in\ru
\end{equation}
is an isomorphism of (proper) $\Hstar$-algebras;

\item the canonical ideal $\ru$ is an invariant subspace for the
representation $\two$ --- {\rm (see~{(\ref{two-sided})})} --- and
the star product associated with $U$ is equivariant with respect to
this representation, i.e.
\begin{equation} \label{natact}
\two(g)\Big(f_1\starp f_2\Big)=\big(\two (g) f_1\big)\starp
\big(\two (g) f_2\big),\ \ \ \forall\hspace{0.4mm} f_1, f_2\in\ldg ,
\ \forall\hspace{0.4mm}g\in G \fin .
\end{equation}

\end{enumerate}
\end{proposition}

\noindent {\bf Proof:} Since the star product $(\cdot)\starp(\cdot)$
is by construction a bilinear map, the vector space $\ldg$ ---
endowed with this operation --- is an algebra. Let us prove that
this algebra is associative. Indeed, observe that --- using
definition~{(\ref{defist})}, and by virtue of the relation
$\wigg^\ast\hspace{0.5mm}\wig = I$ and of the associativity of the
$\Hstar$-algebra $\hs$ ---  for any $f_1,f_2,f_3\in\ldg$ we have:
\begin{eqnarray}
\Big(f_1\starp f_2\Big)\starp f_3 =
\wig\Big(\big(\wigg^\ast\hspace{0.3mm} f_1\big)
\big(\wigg^\ast\hspace{0.3mm} f_2\big)\Big)\starp f_3 \spa & = &
\spa
\wig\hspace{-0.5mm}\left(\left(\wigg^\ast\hspace{0.5mm}\wig\Big(\big(\wigg^\ast\hspace{0.3mm}
f_1\big) \big(\wigg^\ast\hspace{0.3mm} f_2\big)\Big)\right)
\big(\wigg^\ast\hspace{0.3mm} f_3\big)\right)\nonumber\\ & = & \spa
\wig\hspace{-0.5mm}\left(\big(\wigg^\ast\hspace{0.3mm} f_1\big)
\big(\wigg^\ast\hspace{0.3mm} f_2\big) \big(\wigg^\ast\hspace{0.3mm}
f_3\big)\right).
\end{eqnarray}
From this relation, using again the fact that
$\wigg^\ast\hspace{0.5mm}\wig = I$ and definition~{(\ref{defist})},
we get
\begin{eqnarray}
\Big(f_1\starp f_2\Big)\starp f_3 & = & \spa
\wig\hspace{-0.5mm}\left(\big(\wigg^\ast\hspace{0.3mm} f_1\big)
\left(\wigg^\ast\hspace{0.5mm}\wig\Big(\big(\wigg^\ast\hspace{0.3mm}
f_1\big) \big(\wigg^\ast\hspace{0.3mm} f_2\big)\Big)\right) \right)
\nonumber\\ & = & \spa \wig\Big(\big(\wigg^\ast\hspace{0.3mm}
f_1\big) \Big(\wigg^\ast\Big( f_2\starp f_3\Big)\Big)\Big)\nonumber\\
& = & \spa f_1\starp \Big(f_2\starp f_3\Big),\ \ \
\forall\hspace{0.5mm} f_1,f_2,f_3\in\ldg\fin .
\end{eqnarray}

Let us now prove that the antilinear map $\invol$ is an involution
with respect to the bilinear operation $(\cdot)\starp(\cdot)$. The
first of relations~{(\ref{eqinvo})} is certainly satisfied (since
$\invol$ is a complex conjugation in $\ldg$). In order to prove the
second of relations~{(\ref{eqinvo})}, observe that the intertwining
relation~{(\ref{intcoco})} implies that
\begin{equation} \label{invols}
\invo\hspace{0.5mm}\wigg^\ast  =
\invo^{\hspace{-0.3mm}\ast}\hspace{0.3mm}\wigg^\ast =
\wigg^\ast\hspace{0.4mm} \involu^\ast = \wigg^\ast \hspace{0.4mm}
\invol \fin .
\end{equation}
Then --- using~{(\ref{intcoco})} and~{(\ref{invols})}, and the fact
that $\invo$ is an involution in the algebra $\hs$ --- we can argue
as follows:
\begin{eqnarray}
\invol\Big(f_1\starp f_2\Big) \hspace{-0.9mm} = \invol\wig\Big(\big(\wigg^\ast\hspace{0.3mm} f_1\big)
\big(\wigg^\ast\hspace{0.3mm} f_2\big)\Big)\hspace{-0.9mm} \spa & = &
\spa \wig\invo\Big(\big(\wigg^\ast\hspace{0.3mm} f_1\big)
\big(\wigg^\ast\hspace{0.3mm} f_2\big)\Big)\nonumber \\
& = & \spa \wig\Big(\big(\invo\wigg^\ast\hspace{0.3mm} f_2\big)
\big(\invo\wigg^\ast\hspace{0.3mm} f_1\big)\Big)
\nonumber \\
& = & \spa \wig\Big(\big(\wigg^\ast\invol f_2\big)
\big(\wigg^\ast\invol f_1\big)\Big) \nonumber \\
& = & \spa \big(\invol f_2\big)\starp \big(\invol f_1\big), \ \ \
\forall\hspace{0.5mm} f_1,f_2\in\ldg \fin . \label{procedure}
\end{eqnarray}

At this point, in order to show that $\au\equiv\big(\ldg,
(\cdot)\starp(\cdot), \invol\big)$ is a Banach $\ast$-algebra, it
remains to observe that
\begin{equation}
\|\invol f\norld\hspace{-1mm} = \| f\norld \fin ,\ \ \
\forall\hspace{0.4mm} f\in\ldg \fin ,
\end{equation}
($\invol$ is a antiunitary operator), and to prove
relation~{(\ref{disegua})}. In fact, we have:
\begin{eqnarray}
\Big\|f_1\starp f_2\Bnorld \hspace{-1.3mm} = \Big\|  \wig\Big(\big(\wigg^\ast\hspace{0.3mm} f_1\big)
\big(\wigg^\ast\hspace{0.3mm} f_2\big)\Big) \hspace{-0.3mm} \Bnorld \hspace{-1.3mm} \spa & = & \spa
\big\|  \big(\wigg^\ast\hspace{0.3mm} f_1\big)
\big(\wigg^\ast\hspace{0.3mm} f_2\big) \big\norhs \nonumber \\
& \le & \spa \big\|  \wigg^\ast\hspace{0.3mm} f_1 \big\norhs
\hspace{0.5mm} \big\|
\wigg^\ast\hspace{0.3mm} f_2 \big\norhs \nonumber \\
& \le & \spa \| f_1\norld \hspace{0.5mm}\| f_2\norld \fin , \ \ \
\forall\hspace{0.5mm} f_1,f_2\in\ldg \fin ,
\end{eqnarray}
where the last inequality is a consequence of the fact that $\wigg^\ast$ is a partial isometry.

We will now prove that $\au$ is a $\Hstar$-algebra. To this aim, let
us show that the first of relations~{(\ref{prophstar})} holds true.
In fact, for all $f_1,f_2,f_3 \in\ldg$, we have that
\begin{eqnarray}
\langle f_1\starp f_2,f_3\rangle_{\ld}=\left\langle
\big(\wigg^\ast\hspace{0.3mm} f_1\big) \big(\wigg^\ast\hspace{0.3mm}
f_2\big),\wigg^\ast\hspace{0.3mm} f_3\right\ranglehs\spa & = & \spa
\left\langle  \wigg^\ast\hspace{0.3mm} f_2
,\big(\wigg^\ast\hspace{0.3mm}
f_1\big)^{\hspace{-0.3mm}\ast}\hspace{0.3mm}\wigg^\ast\hspace{0.3mm}
f_3\right\ranglehs\nonumber\\ & = & \spa \left\langle f_2
,\wigg\big(\big(\invo\wigg^\ast\hspace{0.3mm}
f_1\big)\hspace{0.3mm}\wigg^\ast\hspace{0.3mm}
f_3\big)\right\rangle_{\ld}\nonumber\\ & = & \spa \left\langle f_2
,\wigg\big(\big(\wigg^\ast\invol
f_1\big)\hspace{0.3mm}\wigg^\ast\hspace{0.3mm}
f_3\big)\right\rangle_{\ld} \nonumber \\
& = & \spa \langle f_2, (\invol f_1)\starp f_3\rangle_{\ld}\,  .
\end{eqnarray}
The proof of the second of relations~{(\ref{prophstar})} is
analogous; we leave the details to the reader.

Properties~{(\ref{contdue})} and~{(\ref{contuno})}, and the fact
that the map~{(\ref{isostalg})} is an isomorphism of
$\Hstar$-algebras can be immediately verified by the definition of
the star product associated with $U$. Moreover, $\ru$ is a
\emph{proper} $\Hstar$-algebra, being isomorphic to $\hs$.

Eventually, we can prove the equivariance relation~{(\ref{natact})}
by a procedure similar to~{(\ref{procedure})}. In fact, exploiting
the intertwining relation~{(\ref{intertreps})}, for every
$f_1,f_2\in\ldg$, we have:
\begin{eqnarray}
\two(g)\Big(f_1\starp f_2\Big) \hspace{-0.8mm}  = \wig\hspace{0.8mm}
\rep (g) \Big(\big(\wigg^\ast\hspace{0.3mm} f_1\big)
\big(\wigg^\ast\hspace{0.3mm} f_2\big)\Big)\hspace{-0.9mm} \spa & =
& \spa \wig\Big(\big(\rep (g)\hspace{0.4mm}\wigg^\ast\hspace{0.3mm}
f_2\big) \big(\rep (g)\hspace{0.4mm}\wigg^\ast\hspace{0.3mm}
f_1\big)\Big)
\nonumber \\
& = & \spa \wig\Big(\big(\wigg^\ast\hspace{0.7mm} \two(g)\, f_1\big)
\big(\wigg^\ast \hspace{0.7mm} \two(g)\, f_2\big)\Big) \nonumber \\
& = & \spa \big(\two(g)\, f_1\big)\starp \big(\two(g)\, f_2\big)\fin
.
\end{eqnarray}

The proof is complete.~{$\square$}

It is interesting to note that the definition of the star
product~{(\ref{defist})} can be suitably generalized. In fact, since
$\hs$ is a two-sided ideal in $\bh$, with every bounded operator
$\opk\in\bh$ is associated a bilinear map
$(\cdot)\kcomp(\cdot)\colon \hs\times\hs\rightarrow\hs$
--- the \emph{$\opk$-product}\footnote{This notion has been considered for
`generic operators' in refs.~{\cite{Manko1,Manko2}}.} in $\hs$
--- defined by
\begin{equation} \label{defkcomp}
\opa\kcomp \opb := \opa\hspace{0.4mm}\opk\opb,\ \ \
\forall\hspace{0.4mm}\opa \fin ,\opb\in\hs \fin .
\end{equation}
Observe that $\hs$, endowed with the operation
$(\cdot)\kcomp(\cdot)$, is an associative algebra, and, if $\opk$ is
selfadjoint, then $\invo$ is an involution in $\hs$ with respect to
this operation. Moreover, since
\begin{equation}
\|\opa\hspace{0.4mm}\opk\opb\norhs \le \|\opa\norhs \hspace{0.5mm}
\|\opk\opb\norhs \le \|\opk\| \hspace{0.8mm} \|\opa\norhs
\hspace{0.5mm} \|\opb\norhs \fin ,
\end{equation}
it is clear that, if $\|\opk\|\le 1$, then $\big(\hs,
(\cdot)\kcomp(\cdot)\big)$ is a Banach algebra; if, furthermore,
$\opk$ is selfadjoint, then $\big(\hs,
(\cdot)\kcomp(\cdot),\invo\big)$ is a Banach $\ast$-algebra. The
operation~{(\ref{defkcomp})} allows us to introduce the following
bilinear map from $\ldg\times\ldg$ into $\ldg$:
\begin{equation} \label{defistk}
(\cdot)\kstarp(\cdot)\colon \ldg\times\ldg\ni (f_1,f_2)\mapsto
\wig\Big(\big(\wigg^\ast\hspace{0.3mm} f_1\big) \kcomp
\big(\wigg^\ast\hspace{0.3mm} f_2\big)\Big)\hspace{-0.8mm}\in\ldg
\fin .
\end{equation}
We will call the operation $(\cdot)\kstarp(\cdot)$
\emph{$\opk$-deformed star product associated with} $U$. Obviously,
the $\opk$-deformed star product coincides with the star product
defined by~{(\ref{defist})} in the case where $\opk=I$. By a
procedure analogous to the one adopted in the proof of
Proposition~{\ref{propost}}, one can derive the main properties of
the $\opk$-deformed star product:
\begin{proposition}
For every bounded operator $\opk\in\bh$, the bilinear map
$(\cdot)\kstarp(\cdot)$ enjoys the following properties:
\begin{enumerate}

\item the vector space $\ldg$, endowed with the operation $(\cdot)\kstarp(\cdot)$, is an associative algebra;

\item in the case where the operator $\opk$ is selfadjoint,
the antilinear map $\invol$ is an involution in the vector space $\ldg$ with respect to the bilinear operation
$(\cdot)\kstarp(\cdot)$, i.e.
\begin{equation}
\invol\big(\invol f\big)= f \ \ \ \mbox{and}\ \ \
\invol\Big(f_1\kstarp f_2\Big)= \big(\invol f_2\big)\kstarp
\big(\invol f_1\big), \ \ \ \forall\hspace{0.4mm} f,f_1, f_2\in\ldg
\fin ;
\end{equation}

\item if $\|\opk\|\le 1$, then $\ldg$ --- regarded as a Banach space with respect to the norm $\|\cdot\norld$,
and endowed with the $\opk$-deformed star product associated with
$U$
--- is a Banach algebra; in particular, it satisfies the relation
\begin{equation}
\Big\|f_1\kstarp f_2\Bnorld \le \|f_1\norld \|f_2\norld \fin ,\ \ \
\forall\hspace{0.4mm} f_1,f_2\in\ldg \fin ;
\end{equation}
if, furthermore, the operator $\opk$ is selfadjoint, then
$\big(\ldg,(\cdot)\kstarp(\cdot),\invol\big)$ is a Banach
$\ast$-algebra;

\item for any $f_1,f_2\in\ldg$, we have that
\begin{equation}
f_1\kstarp f_2\in\ru \fin ;
\end{equation}
therefore --- assuming that $\|\opk\|\le 1$ --- the (closed)
subspace $\ru$ of $\ldg$ is a closed two-sided ideal in the Banach
algebra $\big(\hs, (\cdot)\kcomp(\cdot)\big)$;

\item for any $f_1,f_2\in\ldg$, we have that
\begin{equation}
f_1\kstarp f_2=\big(\proi\hspace{0.5mm} f_1\big)\kstarp
\big(\proi\hspace{0.5mm} f_2\big);
\end{equation}

\item  assuming that $\|\opk\|\le 1$, the application
\begin{equation}
\hs\ni\opa\mapsto \wig\opa\in\ru
\end{equation}
is an isomorphism of the Banach algebras $\big(\hs,
(\cdot)\kcomp(\cdot)\big)$ and $\big(\ru,
(\cdot)\kstarp(\cdot)\big)$.

\item for any $f_1,f_2\in\ldg$, the following relation holds:
\begin{equation}
\two(g)\Big(f_1\kstarp f_2\Big)=\big(\two (g) f_1\big)\kgstarp
\big(\two (g) f_2\big),\ \ \opkg := U(g)\hspace{0.4mm}\opk
\hspace{0.6mm} U(g)^\ast,\ \ \ \forall\hspace{0.4mm} g\in G \fin .
\end{equation}

\end{enumerate}
\end{proposition}

It is a remarkable fact that the star product~{(\ref{defist})}, and
its generalization~{(\ref{defistk})}, which are implicitly defined
via the quantization-dequantization maps associated with a square
integrable representation, admit simple explicit formulae based on
certain integral kernels. The task of deriving such formulae will be
systematically pursued in the next section.

\section{Main results: explicit formulae for star products}
\label{main}

The aim of this section is to provide suitable expressions for the
star products associated with square integrable representations that
have been defined and characterized in Sect.~{\ref{defstar}}. For
the sake of clarity, we will split our presentation into a few
subsections. In particular, in Subsect.~{\ref{formulae}} we will
prove a simple formula for the star product $(\cdot)\starp(\cdot)$
--- see Theorem~{\ref{main-theorem}} ---
and from this formula we will derive various consequences, including
an expression for the $\opk$-deformed star product
(Corollary~{\ref{corkstar}}). We will then show in
Subsect.~{\ref{generalization}}
--- see Theorem~{\ref{general-theorem}} --- that
Theorem~{\ref{main-theorem}} can be actually generalized: there is a
`wide range' of possible realizations of the star product (in
general, of the $\opk$-deformed star product); as we will see, one
for each suitably characterized right approximate identity in the
$\Hstar$-algebra $\hs$. Of course, we could prove
Theorem~{\ref{general-theorem}} first, and regard
Theorem~{\ref{main-theorem}} just as a consequence. However, the
latter result can be obtained by means of a simpler procedure. So,
for the reader's convenience, we prefer to prove it first.

One can find various alternative routes for getting to the
main results of this section. We have tried to choose these routes
in such a way to allow the reader to achieve a certain insight
in `what is going on'.

\subsection{Assumptions and further notations}

In the following, we will always assume that $U$ is a square
integrable projective representation (with multiplier $\mm$) of the
l.c.s.c.\ group $G$ in the Hilbert space $\hh$. We will denote, as
usual, by $\duu$ the associated Duflo-Moore operator, normalized
according to a given left Haar measure $\mG$ on $G$. Recall that, if
$G$ is unimodular, then $\duu=\ddu I$, $\ddu > 0$; otherwise, $\duu$
is unbounded. We will use
--- often without any further explanation --- the notations and the
tools introduced in Sects.~{\ref{known}-\ref{defstar}}; in
particular, we will exploit the orthogonality relations for square
integrable representations and the result recalled at the end of
Sect.~{\ref{known}}.

Before starting our program, it is worth fixing a few additional
notations. We will denote by $\displaystyle \norlim$ the limit of a
sequence in $\ldg$ (converging with respect to the norm
$\|\cdot\norld$). Given a finite or countably infinite index set
$\nn$, we denote by $\norsunv$ either simply a finite sum in $\ldg$
($\nn$ finite) or an infinite sum in $\ldg$ (converging with respect
to the norm $\|\cdot\norld$). Clearly, analogous meanings will be
understood for the symbols $\displaystyle\hslim$ and $\hsnorsunv$
(of course, in this case the relevant space is $\hs$), or, in
general, $\normsunv$. Given a bounded operator $\opb$ in $\hh$, we
can define two natural bounded operators in the Hilbert-Schmidt
space $\hs$; i.e.\ the operators
\begin{equation}
\mathfrak{L}_{\opb}\colon\hs\ni\opa\mapsto\opb\hspace{0.4mm}\opa\in\hs,\
\ \ \mathfrak{R}_{\opb}\colon\hs\ni\opa\mapsto\opa\,\opb\in\hs \fin
.
\end{equation}
It is obvious that
$\mathfrak{L}_{\opb}\quattro\mathfrak{R}_{\opb^\prime}=\mathfrak{R}_{\opb^\prime}
\mathfrak{L}_{\opb}$. In particular, given a vector $\chi\in\hh$, we
will denote by $\tprochi$ the bounded linear operator in $\hs$
defined by
\begin{equation}
\tprochi\colon\hs\ni\opa\mapsto\opa\,\widehat{\chi}\in\hs \fin ,
\end{equation}
where we set: $\widehat{\chi}\equiv\chichi\equiv|\chi\rangle\langle
\chi |$. It is clear that --- for $\chi$ nonzero and normalized ---
$\tprochi$ is an orthogonal projector in the Hilbert space $\hs$.
\begin{remark} \label{remproj}
{\rm Let $J$ be a complex conjugation in $\hh$ (a selfadjoint
antiunitary operator). Then, the linear map
$\uj\colon\hh\otimes\hh\rightarrow\hs$, determined (in a consistent
way) by
\begin{equation} \label{detuj}
\uj\hspace{0.8mm} \phi\otimes\psi = |\phi\rangle\langle
J\hspace{0.4mm}\psi | \fin , \ \ \
\forall\hspace{0.5mm}\phi,\psi\in\hh \fin ,
\end{equation}
is a unitary operator (indeed, it is an isometry on the dense linear
span generated by the separable elements of $\hh\otimes\hh$, and the
image of this linear span is $\fr$, which is dense in $\hs$). It is
easy to check that
$\uj\hspace{0.3mm}(I\otimes\widehat{\chi}\hspace{0.4mm})\hspace{0.6mm}\ujast
= \mathfrak{R}_{\widehat{\chi}^{\hspace{0.4mm}\prime}}$, where
$\widehat{\chi}^{\hspace{0.4mm}\prime}=J\hspace{0.5mm}\widehat{\chi}\hspace{0.3mm}J=|J\hspace{0.4mm}\chi\rangle
\langle J\hspace{0.4mm}\chi|$. Let $\{\chi_n\}_{n\in\nn}$ be an
orthonormal basis in $\hh$. One can always choose the complex
conjugation $J$ in such a way that $J\hspace{0.4mm}\chi_n=\chi_n$,
for any $n\in\nn$; hence:
$\uj\hspace{0.3mm}(I\otimes\widehat{\chi}_n)\hspace{0.6mm}\ujast
=\tprochin$, with $\widehat{\chi}_n\equiv
|\chi_n\rangle\langle\chi_n|$. This choice of $J$ is convenient for
noting the fact that the relation $\normsunv
(I\otimes\widehat{\chi}_n\hspace{0.4mm}\Psi)=\Psi$,
$\forall\Psi\in\hh\otimes\hh$, is equivalent to $\hsnorsunv
\tprochin\hspace{0.3mm}\opa=\opa$, $\forall
\opa\in\hs$.~{$\blacksquare$} }
\end{remark}

Besides, given a vector $\chi$ contained in the dense linear span
$\dom\big(\du^{-1}\big)$, let $\chid$ be the linear operator in
$\hh$, of rank at most one, defined by
\begin{equation}
\chid:=|\chi\rangle\langle \du^{-1}\tre\chi | \fin .
\end{equation}
Then, we can consider the bounded linear operator
$\pic\colon\hs\ni\opa\mapsto\opa\,\chid\in\hs . $ Note that, if the
group $G$ is unimodular, we have: $\pic=d_U^{-1}\,\tprochi$.

Let us also introduce two integral kernels. Our formulae for star
products will be based on these kernels. First --- for any bounded
operator $\opk$ in $\hh$ and any vector $\chi\in\hh$, contained in
the dense linear span $\dom\big(\du^{-2}\big)$ --- consider the
integral kernel $\keu(\opk,\chi;\cdot,\cdot)\colon G\times
G\rightarrow\ccc$ defined by
\begin{equation} \label{defkeu}
\keu (\opk,\chi;g,h):=\big\langle U(g)\,\du^{-2}\tre\chi,
\opk\,U(h)\,\du^{-1}\tre\chi\big\rangle= \big\langle
\opk^\ast\hspace{0.4mm}
U(g)\,\du^{-2}\tre\chi,U(h)\,\du^{-1}\tre\chi\big\rangle.
\end{equation}
For notational convenience, we set $\keu (\chi;g,h)\equiv \keu
(I,\chi;g,h)=\big\langle U(g)\,\du^{-2}\tre\chi,
U(h)\,\du^{-1}\tre\chi\big\rangle$. Next,  again for every vector
$\chi$ contained in $\dom\big(\du^{-2}\big)$, let $\nuc(\chi;
\cdot,\cdot,\cdot) \colon G\times G \times G\rightarrow\ccc$ be the
integral kernel defined by\footnote{Recall that
$\ran\big(\du^{-1}\big)$ is a dense linear span in $\hh$, stable
under the action of the representation $U$; hence:
$\dom\big(\du^{-1}U(g)\,\du^{-1}\big)=\dom\big(\du^{-2}\big)$,
$\forall\hspace{0.5mm} g\in G$.}
\begin{equation} \label{defby}
\nuc(\chi; g,h,h^\prime)  :=  \big\langle
U(g)\,\du^{-1}\tre\chi,U(h)\,\du^{-1}\,U(h^\prime)\,\du^{-1}\tre\chi\big\rangle
.
\end{equation}
Exploiting relation~{(\ref{useaaa})} and the fact that
\begin{eqnarray} \label{usebb}
U(h^{-1}g)=\mm (h^{-1},g)\, U(h^{-1})\,U(g) \spa & = & \spa \mm
(h^{-1},g)\,\mm(h,h^{-1})^\ast\, U(h)^\ast\hspace{0.5mm}U(g)
\nonumber\\
& = & \spa \mm (h,h^{-1}g)^\ast\,U(h)^\ast\hspace{0.5mm}U(g) \fin ,
\end{eqnarray}
we find:
\begin{equation}\label{reladefi}
\nuc(\chi; g,h,h^\prime) =  \mm (h,h^{-1}g)^\ast \modu
(h^{-1}g)^{\frac{1}{2}}\, \keu (\chi;h^{-1} g,h^\prime)\fin ,\ \ \
\forall\hspace{0.5mm} g,h,h^\prime\in G \fin .
\end{equation}
Observe that --- since $\keu
(\opk,\chi;g,\cdot)=\wig\big(|\opk^\ast\hspace{0.6mm}U(g)\,\du^{-2}\tre\chi\rangle\langle\chi
|\big)^\ast$ --- for any $g\in G$, the function $G\ni h\mapsto\keu
(\opk,\chi;g,h)\in\ccc$ belongs to $\ldg$. Moreover, by
relation~{(\ref{reladefi})}, for any $g,h\in G$, the function $G\ni
h^\prime\mapsto\nuc(\chi; g,h,h^\prime)\in\ccc$ belongs to $\ldg$,
as well.

\subsection{Preliminary results}
\label{preliminary}

The following result will turn out to be fundamental for our purposes.
\begin{proposition} \label{proprirel}
For every bounded operator $\opk\in\bh$, for every function
$f\in\ldg$ and for every vector $\chi\in\dom\big(\du^{-2}\big)$, the
following formula holds:
\begin{equation} \label{prirel}
\big(\wig\pic\ellekappa\wigg^\ast f \big) (g)= \intg\de \mG (h)\,
\keu (\opk,\chi;g,h)\, f(h)\fin ,\ \ \ \aam g\in G\fin .
\end{equation}
\end{proposition}

\noindent {\bf Proof:} Indeed, for every $f\in\ldg$, we have:
\begin{eqnarray}
\intg\de \mG (h)\, \keu (\opk,\chi;g,h)\, f(h) \spa & = & \spa
\big\langle \wig\big(|\opk^\ast\hspace{0.6mm}
U(g)\,\du^{-2}\tre\chi\rangle\langle\chi|\big), f
\big\rangle_\ld
\nonumber\\
& = & \spa \big\langle \opk^\ast\hspace{0.6mm}
|U(g)\,\du^{-2}\tre\chi\rangle\langle\chi|, \wigg^\ast f
\big\ranglehs
\nonumber\\
& = & \spa  \tr\big( |\chi\rangle\langle U(g)\,\du^{-2}\tre
\chi|\,\opk\hspace{0.4mm}(\wigg^\ast f)\big)
\nonumber\\
& = & \spa \big\langle U(g)\,\du^{-2}\tre
\chi,\opk\hspace{0.4mm}(\wigg^\ast f)\hspace{0.5mm} \chi\big\rangle
,\ \ \ \aam g\in G\fin .
\end{eqnarray}
Hence, we conclude that
\begin{eqnarray}
\intg\de \mG (h)\, \keu (\opk,\chi;g,h)\, f(h) & = & \spa
\Big(\wig\big(\opk\hspace{0.4mm}(\wigg^\ast f)\hspace{0.5mm}
|\chi\rangle\langle \du^{-1}\tre \chi |\big)\Big)(g)
\nonumber\\
& = & \spa \big(\wig\pic\ellekappa\wigg^\ast f \big) (g)\fin ,\ \ \
\aam g\in G\fin .
\end{eqnarray}
The proof of formula~{(\ref{prirel})} is complete.~{$\square$}

At  this point, in order to prove the main theorem of this section,
we need to pass through three technical results. The first result
(Lemma~{\ref{lemrelare}}) will turn to be useful both in this
subsection and in Subsect.~{\ref{generalization}}. The third one
(Lemma~{\ref{lemrellem}}) `essentially contains' the expression of
the star product, already, but it requires a refinement (see
Proposition~{\ref{propneorela}} below) before getting to the main
theorem swiftly.

\begin{lemma} \label{lemrelare}
For every $f\in\ldg$ and for every $g\in G$, the following relation holds:
\begin{equation} \label{relare}
\big(\rem (g)\invol f\big) (h)^\ast = \mm (h,h^{-1}g)^\ast \modu
(h^{-1}g)^{\frac{1}{2}} \hspace{0.3mm} f(h^{-1}g) \fin ,
\end{equation}
$\aam h\in G$.
Therefore, for any $f_1,f_2\in\ldg$ and for every $g\in G$, the function
\begin{equation}
G\ni h\mapsto f_1(h)\, \mm (h,h^{-1}g)^\ast \modu
(h^{-1}g)^{\frac{1}{2}} \hspace{0.3mm} f_2(h^{-1}g)\in\ccc
\end{equation}
belongs to $\lug$ and
\begin{equation} \label{fortherefore}
\intG \de\mG (h)\, f_1(h)\, \mm (h,h^{-1}g)^\ast \modu
(h^{-1}g)^{\frac{1}{2}} \hspace{0.3mm} f_2(h^{-1}g)= \langle \rem
(g)\invol f_2,f_1\ranld \fin .
\end{equation}
\end{lemma}

\noindent {\bf Proof:} Recalling the definition of the
representation $\rem\colon G\rightarrow\mathcal{U}(\ldg)$
(see~{(\ref{defre})}) and of the complex conjugation $\invol\colon
\ldg\rightarrow\ldg$ (see~{(\ref{definvol})}), we have that, for
every $f\in\ldg$ and for every $g\in G$,
\begin{eqnarray}
\big(\rem (g)\invol f\big) (h) \spa & = & \spa \mm(g,g^{-1})^\ast\,
\mm (g^{-1},h)\hspace{0.6mm} \big(\invol f\big) (g^{-1}h)
\nonumber\\
& = & \spa \label{relarea} \mm(g,g^{-1})^\ast\, \mm
(g^{-1},h)\hspace{0.5mm} \mm(g^{-1} h,h^{-1}g)\hspace{0.6mm} \modu
(g^{-1}h)^{-\frac{1}{2}} \hspace{0.3mm} f(h^{-1}g)^\ast.
\end{eqnarray}
Observe now that --- identifying the group elements $g_1,g_2,g_3$ in
relation~{(\ref{multirelat})} with $g^{-1}$, $h$ and $h^{-1}g$,
respectively --- we have:
\begin{equation} \label{relareb}
\mm(g^{-1} h,h^{-1}g) = \mm (g^{-1},h)^\ast\,
\mm(g^{-1},g)\hspace{0.5mm} \mm(h,h^{-1}g)= \mm (g^{-1},h)^\ast\,
\mm(g,g^{-1})\hspace{0.5mm} \mm(h,h^{-1}g) \fin .
\end{equation}
From relations~{(\ref{relarea})} and~{(\ref{relareb})} one obtains
immediately formula~{(\ref{relare})}.~{$\square$}

\begin{lemma} \label{lemnovissimo}
For any $f_1,f_2\in\ldg$ and for every $\chi\in\dom\big(\du^{-2}\big)$, the following relation holds:
\begin{equation} \label{novissimo}
\intg\de \mG(h) \intg \de \mG(h^\prime)\, \nuc(\chi; g,h,h^\prime)\,
f_1(h) f_2(h^\prime)= \langle \rem (g)\invol \wig\pic\wigg^\ast
f_2,f_1\ranld \fin .
\end{equation}
\end{lemma}

\noindent {\bf Proof:}
Taking into account~{(\ref{reladefi})},
by relation~{(\ref{prirel})} --- with $\opk=I$ --- we obtain:
\begin{eqnarray}
\intg \de \mG(h^\prime)\, \nuc(\chi; g,h,h^\prime)\, f_2(h^\prime) \spa & = & \spa
\mm (h,h^{-1}g)^\ast \modu
(h^{-1}g)^{\frac{1}{2}}\intg \de \mG(h^\prime)\, \keu(\chi; h^{-1} g, h^\prime)\, f_2(h^\prime)
\nonumber\\
& = & \spa \mm (h,h^{-1}g)^\ast \modu
(h^{-1}g)^{\frac{1}{2}}\,\big(\wig\pic\wigg^\ast f_2\big) (h^{-1}g)
\fin .
\end{eqnarray}
At this point, relation~{(\ref{novissimo})} is a straightforward
consequence of Lemma~{\ref{lemrelare}}.~{$\square$}

\begin{lemma} \label{lemrellem}
Let $\chi$ be a vector belonging to $\dom\big(\du^{-2}\big)$. Then,
for every $\phi_1\in\hh$, and for any $\psi_1, \psi_2, \phi_2\in
\dom\big(\du^{-1}\big)$ --- setting, as usual,
$\widehat{\phi_j\tre\psi_j}\equiv |\phi_j\rangle\langle\psi_j|$, $j=1,2$ --- we have:
\begin{eqnarray}
\Big(\wig\tprochi\hspace{0.4mm}\big(\hfpa\hspace{0.4mm}\hfpb\big)\Big)(g)
\spa & = & \spa \intG\de\mG (h) \intG\de\mG (h^\prime)\,  \nuc
(\chi; g, h,h^\prime) \nonumber\\ \label{rellem} & \times & \spa
\big(\wig \hspace{0.3mm}\hfpa\big)(h)\, \big(\wig
\hspace{0.3mm}\hfpb\big)(h^\prime) \fin ,\ \ \ \aam g\in G \fin .
\end{eqnarray}
\end{lemma}

\noindent {\bf Proof:}
First observe that
\begin{eqnarray}
\intG\de\mG (h^\prime)\, \nuc (\chi; g, h,h^\prime)\, \big(\wig
\hspace{0.3mm} \hfpb\big)(h^\prime) \spa & = & \spa \intG\de\mG
(h^\prime)\; \big\langle\du^{-1}\,
U(h)^\ast\hspace{0.5mm}U(g)\,\du^{-1}\tre\chi,U(h^\prime)\,\du^{-1}\tre\chi\big\rangle
\nonumber\\
& \times & \spa \big\langle
U(h^{\prime})\hspace{0.4mm}\du^{-1}\tre\psi_2,\phi_2\big\rangle
\nonumber\\
& = & \spa  \big\langle \psi_2,\chi\rangle \, \langle \du^{-1}\,
U(h)^\ast\hspace{0.5mm}
U(g)\hspace{0.4mm}\du^{-1}\tre\chi,\phi_2\big\rangle \nonumber\\
& = & \spa \label{explo} \langle \psi_2,\chi\rangle \, \big\langle
U(g)\hspace{0.4mm}\du^{-1}\tre\chi,U(h)\hspace{0.4mm}\du^{-1}\tre\phi_2\big\rangle
,
\end{eqnarray}
$\forall\hspace{0.5mm}h,g\in G$, where we have used the fact that
$\phi_2$ is contained in $\dom\big(\du^{-1}\big)$. Then, exploiting
relation~{(\ref{explo})} and the fact that
\begin{equation}
\intG\de\mG (h)\, \big\langle
U(g)\hspace{0.4mm}\du^{-1}\tre\chi,U(h)\du^{-1}\tre\phi_2\big\rangle\,\big\langle
U(h)\hspace{0.4mm}\du^{-1}\tre\psi_1,\phi_1\big\rangle =  \langle
\psi_1,\phi_2\rangle\,\big\langle
U(g)\hspace{0.4mm}\du^{-1}\tre\chi,\phi_1\big\rangle
\end{equation}
--- note that $\big\langle U(h)\hspace{0.4mm}\du^{-1}\tre\psi_1,\phi_1\big\rangle
= \big(\wig\hspace{0.3mm} \hfpa\big)(h)$ --- we find:
\begin{eqnarray}
& & \spa \intG\de\mG (h) \intG\de\mG (h^\prime)\,  \nuc (\chi; g,
h,h^\prime) \, \big(\wig\hspace{0.3mm} \hfpa\big)(h)\, \big(\wig\hspace{0.3mm}
\hfpb\big)(h^\prime)
\nonumber\\
& = & \spa \langle \psi_2,\chi\rangle \intG\de\mG (h)\; \big\langle
U(g)\hspace{0.4mm}\du^{-1}\tre\chi,U(h)\du^{-1}\tre\phi_2\big\rangle\,
\big(\wig\hspace{0.3mm} \hfpa\big)(h)
\nonumber\\
& = & \spa \langle \psi_2,\chi\rangle\, \langle
\psi_1,\phi_2\rangle\, \big\langle
U(g)\hspace{0.4mm}\du^{-1}\tre\chi,\phi_1\big\rangle = \wig
\big(\hfpa\hspace{0.4mm}\hfpb\hspace{0.4mm}\widehat{\chi}\big)(g)
\fin .
\end{eqnarray}

The proof is complete.~{$\square$}

As anticipated, the following result can be regarded as a
generalization of Lemma~{\ref{lemrellem}}. It will allow us to prove
the main result of this section in a straightforward and transparent
way.
\begin{proposition} \label{propneorela}
Let $\chi$ be a  vector contained in $\dom\big(\du^{-2}\big)$. Then,
for any $f_1,f_2\in\ldg$, the following formula holds:
\begin{equation} \label{neorela}
\wig\tprochi\big((\wigg^\ast f_1)(\wigg^\ast f_2)\big) = \intG\de\mG
(h) \intG\de\mG (h^\prime)\,  \nuc (\chi; \cdot, h,h^\prime)\,
f_1(h) \, f_2(h^\prime) \fin .
\end{equation}
\end{proposition}

\noindent {\bf Proof:} By Lemma~{\ref{lemrellem}},
relation~{(\ref{neorela})} holds  for any pair of functions
$f_1,f_2$ belonging to the linear span $\wig(\frlr)$ (see~{(\ref{dlsp})}), which is dense
in $\ru$. Moreover
--- since $\Ker\big(\wigg^\ast\big)=\ruort$, and $\ru$ is an
invariant subspace for the complex conjugation $\invol$ and for the
representation $\rem$ --- for any pair of functions
$f_1,f_2\in\ldg$, of which at least one is contained in $\ruort$, we
have:
\begin{eqnarray}
\langle \rem (g)\invol \wig\pic\wigg^\ast f_2,f_1\ranld =0 \fin .
\end{eqnarray}
Thus, if $f_1$ and/or $f_2$ is contained in
$\ruort$, recalling relation~{(\ref{novissimo})} we conclude that
\begin{equation}
\intg\de \mG(h) \intg \de \mG(h^\prime)\, \nuc(\chi;
\cdot,h,h^\prime)\, f_1(h) f_2(h^\prime)=0 \fin .
\end{equation}
Therefore, relation~{(\ref{neorela})} is satisfied by $f_1,f_2$ in the dense
linear span $\wig(\frlr) + \ruort$.
In the case where the Hilbert space
$\hh$ is finite-dimensional (hence, $G$ is unimodular), this linear
span actually coincides with $\ldg$ itself and the proof is
complete.

Let us assume, instead, that $\dim(\hh)=\infty$, and let us prove
relation~{(\ref{neorela})} for a generic pair of functions in
$\ldg$. To this aim, consider first a pair of functions $f_1,f_2$ of
this kind: $f_1$ is an arbitrary function contained in the dense
linear span $\wig(\frlr) + \ruort$, and $f_2$ any function belonging
to $\ldg$. Next, take a sequence of functions
$\{f_{2;n}\}_{n\in\nat}\subset\ldg$, contained in $\wig(\frlr) +
\ruort$ and converging (with respect to the norm $\|\cdot\|_{\ld}$)
to $f_2$. Then, we have:
\begin{equation} \label{aaarel}
\norlim \wig\tprochi\big((\wigg^\ast f_1) (\wigg^\ast f_{2;n})\big)
= \wig\tprochi\big((\wigg^\ast f_1)(\wigg^\ast f_2)\big).
\end{equation}
On the other hand, by the first part of the proof and by
Lemma~{\ref{lemnovissimo}}, we have that
\begin{eqnarray}
\lim_{n\rightarrow\infty} \Big(\wig\tprochi\big((\wigg^\ast f_{1})
(\wigg^\ast f_{2;n})\big)\Big)(g) \spa & = & \spa
\lim_{n\rightarrow\infty} \intg\de \mG(h) \intg \de \mG(h^\prime)\,
\nuc(\chi; g,h,h^\prime)\, f_1(h) f_{2;n}(h^\prime)
\nonumber\\
& = & \spa \lim_{n\rightarrow\infty} \langle \rem
(g)\invol \wig\pic\wigg^\ast f_{2;n},f_{1}\ranld \nonumber\\
& = & \spa \langle \rem (g)\invol \wig\pic\wigg^\ast f_2,f_1\ranld
\nonumber\\
\label{lastmem} & = & \spa \intg\de \mG(h) \intg \de \mG(h^\prime)\,
\nuc(\chi; g,h,h^\prime)\, f_1(h) f_2(h^\prime)\fin .
\end{eqnarray}
From relations~{(\ref{aaarel})} and~{(\ref{lastmem})} it descends
that formula~{(\ref{neorela})} holds true for any pair of functions
$f_1\in \big(\wig(\frlr) + \ruort\big)$ and $f_2\in\ldg$. At this
point, using this result and a density argument analogous to the one
adopted for obtaining it, one proves relation~{(\ref{neorela})} for
a generic pair of functions in $\ldg$.~{$\square$}

\begin{remark} \label{permutint}
{\rm One can arrive at formula~{(\ref{neorela})} by various
alternative routes. For instance, one can derive it from Lemma~{\ref{lemnovissimo}}
using the intertwining relations~{(\ref{simpinter})} and~{(\ref{intcoco})}. This
way will be adopted for proving Theorem~{\ref{general-theorem}}. The above proof
offers the advantage of a direct computation.
Another way is to prove that, for every
$f\in\ldg$,
\begin{equation}
\intG\de\mG (h) \,  \nuc (\chi; g, h,(\cdot))\, f(h) = \wig\Big(
\big(\wigg^\ast f\big)^{\hspace{-0.3mm}\ast} \quattro
|U(g)\du^{-1}\tre\chi\rangle\langle\chi|\Big)^{\hspace{-0.5mm}\ast}
,
\end{equation}
and from this relation deduce that the function $\intG\de\mG
(h^\prime) \intG\de\mG (h)\, \nuc (\chi; \cdot, h,h^\prime)\, f_1(h)
\, f_2(h^\prime)$ is equal to the function on the l.h.s.\
of~{(\ref{neorela})}, for all $f_1,f_2\in\ldg$. Observe that this
shows, in particular, that \emph{the iterated integrals on the
r.h.s.\ of}~{(\ref{neorela})} \emph{can be
permuted}.~{$\blacksquare$} }
\end{remark}

\subsection{Formulae for star products}
\label{formulae}

We are now ready to prove the theorem that can be regarded as the
main result of this section. It provides a simple expression for the
star product associated with the square integrable projective
representation $U$.
\begin{theorem} \label{main-theorem}
Let $\{\chi_n\}_{n\in\nn}$ be an orthonormal basis in $\hh$,
contained in the dense linear span $\dom\big(\du^{-2}\big)$. Then,
for any $f_1,f_2\in\ldg$, the following formula holds:
\begin{equation} \label{fstpr}
f_1\starp f_2=\norsun \intg\de \mG(h) \intg \de \mG(h^\prime)\,
\nuc(\chi_n; \cdot,h,h^\prime)\, f_1(h) f_2(h^\prime)\fin ,
\end{equation}
where the integral kernel $\nuc(\chi_n; \cdot,\cdot,\cdot) \colon
G\times G \times G\rightarrow\ccc$ is defined by~{(\ref{defby})},
i.e.
\begin{equation}
\nuc(\chi_n; g,h,h^\prime)  :=  \big\langle
U(g)\,\du^{-1}\tre\chi_n,U(h)\,\du^{-1}\,U(h^\prime)\,\du^{-1}\tre\chi_n\big\rangle
.
\end{equation}
\end{theorem}

\noindent {\bf Proof:}
In order to prove formula~{(\ref{fstpr})} we can exploit relation~{(\ref{neorela})} and the fact that
\begin{equation} \label{limfor}
\hsnorsun \tprochin\hspace{0.3mm}\opa=\opa \fin ,\ \ \ \forall
\opa\in\hs \fin ,
\end{equation}
where $\widehat{\chi}_n\equiv |\chi_n\rangle\langle\chi_n|$;
see Remark~{\ref{remproj}}. Indeed, for any $f_1,f_2\in\ldg$, we
have:
\begin{eqnarray}
\norsun \intg\de \mG(h) \intg \de \mG(h^\prime)\, \nuc(\chi_n;
\cdot,h,h^\prime)\, f_1(h) f_2(h^\prime) \spa &  = & \spa \norsun
\wig\tprochin\big((\wigg^\ast f_1)(\wigg^\ast f_2)\big)
\nonumber\\
&  = & \spa \wig\,\hsnorsun \tprochin\big((\wigg^\ast
f_1)(\wigg^\ast f_2)\big)
\nonumber\\ \label{lastmemb} &  = & \spa
\wig \big((\wigg^\ast f_1)(\wigg^\ast f_2)\big).
\end{eqnarray}
By definition, the last member of~{(\ref{lastmemb})} is equal to
$f_1\starp f_2$.~{$\square$}

\begin{remark}
{\rm Taking into account the last assertion of
Remark~{\ref{permutint}}, we conclude that \emph{the iterated
integrals on the r.h.s.\ of formula}~{(\ref{fstpr})} \emph{can be
permuted}.~{$\blacksquare$}}
\end{remark}

\begin{remark}
{\rm One can readily derive from formula~{(\ref{fstpr})} various alternative
expressions for the star product; in particular:
\begin{eqnarray}
f_1\starp f_2 \spa & = & \spa \norsun \hspace{-0.8mm}\intG\hspace{-1.1mm} \de\mG (h)\, f_1(h)\, \mm (h,h^{-1}(\cdot))^\ast \modu
(h^{-1}(\cdot))^{\frac{1}{2}} \hspace{-1.5mm}\intG \hspace{-1.1mm}\de\mG (h^\prime)\, \keu(\chi_n; h^{-1}(\cdot),h^\prime)\, f_2(h^\prime)
\nonumber\\
&  = & \spa \norsun \hspace{-0.8mm}\intG\hspace{-1.1mm} \de\mG (h)\, f_1((\cdot)h)\, \mm ((\cdot)h,h^{-1})^\ast \modu
(h^{-1})^{\frac{1}{2}} \hspace{-1.5mm}\intG \hspace{-1.1mm}\de\mG (h^\prime)\, \keu(\chi_n; h^{-1},h^\prime)\, f_2(h^\prime)
\nonumber\\
&  = & \spa \norsun \hspace{-0.8mm}\intG\hspace{-1.1mm} \de\mG (h)\, f_1(h^{-1})\, \mm (h^{-1},h(\cdot))^\ast \modu
(h^{-1}(\cdot))^{\frac{1}{2}}
\nonumber\\
&  \times & \spa \label{alternative}
\intG \hspace{-1.1mm}\de\mG (h^\prime)\, \keu(\chi_n; h(\cdot),h^\prime)\, f_2(h^\prime)\fin ,
\ \ \ \forall\cinque f_1,f_2\in\ldg\fin .
\end{eqnarray}
The first expression is obtained by relation~{(\ref{reladefi})};
then, by the change of variables $h\mapsto gh$ and $h\mapsto h^{-1}$
(recall that $\intG \de\mG(h)\, f(h) = \intG \de\mG(h)\,\modu(h)^{-1} f(h^{-1})$), from
the first expression one obtains the other two.~$\blacksquare$
}
\end{remark}

\begin{remark}
{\rm It is rather boring, but straightforward, to check that, for every $\chi\in\dom\big(\du^{-2}\big)$,
\begin{eqnarray}
& & \spa \intg\de \mG(h) \intg \de \mG(h^\prime)\, \nuc(\chi;
\cdot,h,h^\prime)\, \big(\two (g)\, f_1\big)(h)\hspace{0.5mm}
\big(\two (g)\, f_2\big)(h^\prime) \nonumber\\ \label{confirm} & = &
\spa \two (g) \intg\de \mG(h) \intg \de \mG(h^\prime)\,
\nuc(U(g^{-1})\,\chi; \cdot,h,h^\prime)\, f_1(h) f_2(h^\prime) \fin
.
\end{eqnarray}
Since formula~{(\ref{fstpr})} does not depend on a specific choice
of the orthonormal basis $\{\chi_n\}_{n\in\nn}$ contained in
$\dom\big(\du^{-2}\big)$ (recall that this dense linear span is
stable with respect to $U$), relation~{(\ref{confirm})} confirms the fact that
$\big(\two (g)\, f_1\big)\starp \big(\two (g)\,
f_2\big)= \two (g) \, \Big(f_1 \starp f_2 \Big) $,
$\forall\hspace{0.4mm} f_1,f_2\in\ldg$ --- see~{(\ref{natact})}---
i.e.\ the equivariance of the star product with
respect to the representation $\two$.~{$\blacksquare$} }
\end{remark}

Theorem~{\ref{main-theorem}} has various implications. First of all, it is remarkable that, in the case
where $G$ is unimodular, the star product associated with the representation $U$ admits a simple alternative expression.

\begin{corollary} \label{unimod-case}
Suppose that the l.c.s.c.\ group $G$ is unimodular.
Then, for any $f_1,f_2\in\ldg$, we have:
\begin{eqnarray}
\Big(f_1\starp f_2\Big)(g) \spa & = & \spa d_U^{-1} \intG\de\mG (h)\, f_1(h)\, \mm
(h,h^{-1}g)^\ast  \big(\proi f_2\big)(h^{-1}g)
\nonumber\\
& = & \spa d_U^{-1} \intG\de\mG (h)\, \big(\proi f_1\big)(h)\, \mm
(h,h^{-1}g)^\ast  f_2(h^{-1}g) \nonumber\\ \label{threeeqs} & = &
\spa d_U^{-1} \intG\de\mG (h)\, \big(\proi f_1\big)(h)\, \mm
(h,h^{-1}g)^\ast  \big(\proi f_2\big)(h^{-1}g) \fin ,\ \ \ \aam g\in
G \fin .
\end{eqnarray}
Therefore, for any $f_1,f_2\in\ru$, the following formula holds:
\begin{equation} \label{fstpruni}
\Big(f_1\starp f_2\Big)(g) = d_U^{-1} \intG\de\mG (h)\, f_1(h)\, \mm
(h,h^{-1}g)^\ast  f_2(h^{-1}g) \fin ,\ \ \ \aam g\in G \fin .
\end{equation}

\end{corollary}

\noindent {\bf Proof:} Let $f_1,f_2$ be functions in $\ldg$. Then
--- using formula~{(\ref{fstpr})}, relation~{(\ref{novissimo})} and
the fact that $\pic=d_U^{-1}\,\tprochi$ (since $G$ is unimodular)
--- we have:
\begin{eqnarray}
f_1\starp f_2 \spa & = & \spa \norsun \intg\de \mG(h) \intg \de
\mG(h^\prime)\, \nuc(\chi_n; \cdot,h,h^\prime)\, f_1(h)
f_2(h^\prime) \nonumber\\ \label{pripri} & = & \spa \norsun \langle
\rem (\cdot)\invol \wig\picn\wigg^\ast f_2,f_1\ranld \nonumber\\ & =
& \spa d_U^{-1}\hspace{0.3mm}\norsun \langle \rem (\cdot)\invol
\wig\tprochin\wigg^\ast f_2,f_1\ranld \fin .
\end{eqnarray}
On the other hand --- by virtue of the continuity of the scalar
product in $\ldg$ and of the boundedness of the operators $\rem(g)$,
$\invol$ and $\wig$, and exploiting relations~{(\ref{limfor})} and,
then,~{(\ref{relare})} with ($\modu\equiv 1$) --- we also have that
\begin{eqnarray}
\sun \langle \rem (g)\invol \wig\tprochin\wigg^\ast f_2,f_1\ranld
\spa
& = & \spa \langle \rem (g)\invol \wig \wigg^\ast f_2,f_1\ranld
\nonumber\\
& = & \spa \big\langle \rem (g)\invol \proi f_2,f_1\big\ranld \nonumber\\
\label{secsec} & = & \spa \intG\de\mG (h)\, f_1(h)\, \mm
(h,h^{-1}g)^\ast \big(\proi f_2\big)(h^{-1}g) \fin .
\end{eqnarray}
Relations~{(\ref{pripri})} and~{(\ref{secsec})} imply that the first
of equations~{(\ref{threeeqs})} holds true; the other two are
obtained using the fact that $\proi$ is a projector satisfying $\rem
(g)\invol \proi=\proi\rem (g)\invol$.~{$\square$}

\begin{remark}
{\rm We stress that the particularly simple
formula~{(\ref{fstpruni})}
--- differently from formula~{(\ref{fstpr})} --- holds for any pair
of functions $f_1,f_2\in\ldg$ of which \emph{at least one} belongs
to the (closed) subspace $\ru$ of $\ldg$, which is the canonical
ideal of the $\Hstar$-algebra $\au$, see
Proposition~{\ref{propost}}. The r.h.s.\ of~{(\ref{fstpruni})} is a
`twisted convolution' generalizing the standard twisted
convolution~{\cite{GLP}} that appears in the case where $G$ is the
group of translations on phase space and $U$ is the projective
representation~{(\ref{wey-sys})} (we will examine this case in
Sect.~{\ref{examples}}).~{$\blacksquare$} }
\end{remark}

Let us derive another consequence of Theorem~{\ref{main-theorem}}.
In the case where the group $G$ is compact (hence, unimodular), there is a precise link between the convolution
product in $\ldg$~{\cite{Folland}} and the star products associated with a realization $\chG$ of the unitary dual
of $G$.
\begin{corollary} \label{cor-compa}
Suppose that the l.c.s.c.\ group $G$ is compact and that the Haar
measure $\mG$ is normalized as usual for compact groups, i.e.\ that
$\mG(G)=1$. Then, for any $f_1,f_2\in\ldg$, the following formula
holds:
\begin{equation} \label{fstpruni-comp}
\ldg\ni\intG\de\mG (h)\, f_1(h)\, f_2(h^{-1}(\cdot))=\sumu
\dimu^{-\frac{1}{2}}\, \big( f_1\starp  f_2\big) .
\end{equation}
\end{corollary}

\noindent {\bf Proof:} As is well known, since $G$ is compact, the
convolution of any pair of functions in $\ldg$ is again a function
belonging to $\ldg$. Moreover, from relation~{(\ref{cc-dd})} it
follows that $\sumub\proi f=f$, $\forall\quattro f\in\ldg$; hence
--- denoting by $R$ the left regular
representation of $G$ and by $\sinvo$ the complex conjugation
\begin{equation} \label{invsimp}
\ldg\ni f\mapsto f((\cdot)^{-1})^\ast\in\ldg
\end{equation}
--- for
any $f_1,f_2\in\ldg$ we have:
\begin{eqnarray}
\intG\de\mG
(h)\, f_1(h)\, f_2(h^{-1}g) \spa & = & \spa
\intG\de\mG (h)\, \Big(\sumu\proi f_1\Big)(h)\,
f_2(h^{-1}g)\nonumber\\
& = & \spa \big\langle R(g) \sinvo f_2,\tre \sumu \proi f_1
\big\ranld \nonumber\\ & = & \spa \sum_{U\in\chG}
\big\langle R(g) \sinvo f_2, \proi f_1 \big\ranld \nonumber\\
\label{factd} & = & \spa \sum_{U\in\chG} \intG\de\mG (h)\,
\big(\proi f_1\big)(h)\, f_2(h^{-1}g)  \fin ,
\end{eqnarray}
for all $g\in G$. On the other hand, by
Corollary~{\ref{unimod-case}} we have that
\begin{equation} \label{neos}
\intG\de\mG (h)\, \big(\proi f_1\big)(h)\, f_2(h^{-1}(\cdot)) =
\dimu^{-\frac{1}{2}}\, \big( f_1\starp  f_2\big),\ \ \
\forall\hspace{0.5mm} U\in\chG \fin ,
\end{equation}
where we recall that $\dimu^{-\frac{1}{2}}=\ddu$. Moreover, by
relations~{(\ref{contuno})} and~{(\ref{disegua})}, for any
$f_1,f_2\in\ldg$ we obtain the following estimate:
\begin{eqnarray}
\sum_{U\in\chG} \dimu^{-1} \, \big\| f_1\starp  f_2\big\|_{\ld}^2
\spa & = & \spa \sum_{U\in\chG} \dimu^{-1} \, \big\| \big(\proi
f_1\big)\starp  \big(\proi f_2\big)\big\|_{\ld}^2
\nonumber\\
& \le & \spa \sum_{U\in\chG} \dimu^{-1} \, \big\| \proi
f_1\big\|_{\ld}^2\sette \big\| \proi f_2\big\|_{\ld}^2
\nonumber\\
& \le & \spa \sum_{U\in\chG} \big\| \proi f_1\big\|_{\ld}^2\sette
\big\| \proi f_2\big\|_{\ld}^2 \le \| f_1\|_{\ld}^2 \sette \|
f_2\|_{\ld}^2 \fin  .
\end{eqnarray}
Hence, taking into account~{(\ref{contdue})}, we see that $\sumub
\dimu^{-\frac{1}{2}}\, \big( f_1\starp f_2\big)$ is a well defined
element of $\ldg$ and, by~{(\ref{neos})},
\begin{equation} \label{factu}
\sumu \intG\de\mG (h)\, \big(\proi f_1\big)(h)\,
f_2(h^{-1}(\cdot))=\sumu \dimu^{-\frac{1}{2}}\, \big( f_1\starp
f_2\big).
\end{equation}
At this point, relations~{(\ref{factd})} and~{(\ref{factu})} imply
that formula~{(\ref{fstpruni-comp})} holds true.~{$\square$}

We will now prove that it is possible to achieve a simple expression
of the $\opk$-deformed star product associated with the
representation $U$, for every bounded operator $\opk\in\bh$.
Although this result is more general than
Theorem~{\ref{main-theorem}} --- which corresponds to the case where
$\opk=I$ --- we will derive it as a consequence of
formula~{(\ref{fstpr})} for the star product. To this aim, it is
useful to observe that, by the definition of the $\opk$-deformed
star product and the fact that $\wigg^\ast\wig=I$, we have:
\begin{eqnarray}
f_1\kstarp f_2 \spa & := & \spa \wig \big(\wigg^\ast
f_1\,\opk\,\wigg^\ast f_2\big)
\nonumber\\ \label{obsk}
& = & \spa
\wig \Big(\wigg^\ast f_1\, \wigg^\ast\big(\wig(\opk\,\wigg^\ast
f_2)\big)\Big)=f_1\starp \big(\wig(\opk\,\wigg^\ast f_2)\big).
\end{eqnarray}
Moreover, for every bounded operator $\hat{K}$ in $\hh$ and for
every vector $\chi$ contained in $\dom\big(\du^{-2}\big)$, let us
define an integral kernel $\nuc(\hat{K},\chi; \cdot,\cdot,\cdot)
\colon G\times G \times G\rightarrow\ccc$ by setting:
\begin{eqnarray}
\nuc(\hat{K},\chi; g,h,h^\prime)  \spa & := & \spa  \langle \du^{-1}\,
U(h)^\ast\hspace{0.4mm}U(g)\,\du^{-1}\tre\chi,
\hat{K}\, U(h^\prime)\,\du^{-1}\tre\chi\rangle
\nonumber\\
& = & \spa \label{weset} \mm (h,h^{-1}g)^\ast \modu
(h^{-1}g)^{\frac{1}{2}}\, \keu (\opk,\chi;h^{-1} g,h^\prime) \fin .
\end{eqnarray}
Comparing this definition with~{(\ref{defby})}, it is clear that
$\nuc(\chi; g,h,h^\prime)\equiv\nuc(I,\chi; g,h,h^\prime)$.
\begin{corollary} \label{corkstar}
Let $\hat{K}$ be a bounded operator in $\hh$ and
$\{\chi_n\}_{n\in\nn}$ an orthonormal basis contained in the dense
linear span $\dom\big(\du^{-2}\big)$. Then, for any
$f_1,f_2\in\ldg$, the following formula holds:
\begin{equation} \label{fstpr-bis}
f_1\kstarp f_2=\norsun \intg\de \mG(h) \intg \de \mG(h^\prime)\,
\nuc(\hat{K},\chi_n; \cdot,h,h^\prime)\, f_1(h) f_2(h^\prime) \fin .
\end{equation}
\end{corollary}

\noindent {\bf Proof:} Taking into account relation~{(\ref{obsk})},
we can apply formula~{(\ref{fstpr})} for the (standard) star
product, and next we use relation~{(\ref{novissimo})}, thus getting
\begin{eqnarray}
f_1\kstarp f_2 \spa & = & \spa \norsun \intg\de \mG(h) \intg \de
\mG(h^\prime)\, \nuc(\chi_n; \cdot,h,h^\prime)\, f_1(h)
\big(\wig(\opk\,\wigg^\ast f_2)\big)(h^\prime)
\nonumber\\
& = & \spa \norsun \big\langle \rem (\cdot)\invol
\wig\picn\wigg^\ast
\big(\wig(\opk\,\wigg^\ast f_2)\big),f_1\big\ranld \nonumber\\
\label{eccoecco} & = & \spa \norsun \big\langle \rem
(\cdot)\invol\big( \wig\picn(\opk\,\wigg^\ast
f_2)\big),f_1\big\ranld \fin .
\end{eqnarray}
From~{(\ref{eccoecco})}, by virtue of
relations~{(\ref{fortherefore})} and~{(\ref{prirel})}, it follows
that
\begin{eqnarray}
f_1\kstarp f_2 \spa & = & \spa \norsun  \intG  \de\mG (h)\, f_1(h)\,
\mm (h,h^{-1}g)^\ast \modu (h^{-1}(\cdot))^{\frac{1}{2}}
\hspace{0.4mm} \big(\wig \picn \ellekappa\tre \wigg^\ast
f_2\big)(h^{-1}(\cdot))
\nonumber\\
& = & \spa \norsun  \intG \de\mG (h)\, f_1(h)\, \mm
(h,h^{-1}(\cdot))^\ast \modu (h^{-1}(\cdot))^{\frac{1}{2}}
\nonumber\\
& \times & \spa \intg \de \mG (h^\prime)\, \keu
(\opk,\chi_n;h^{-1}(\cdot),h^\prime)\, f_2(h^\prime)
\nonumber\\
& = & \spa \norsun  \intg  \de \mG(h) \intg \de \mG(h^\prime)\,
\nuc(\hat{K},\chi_n; \cdot,h,h^\prime)\, f_1(h) f_2(h^\prime) \fin ,
\end{eqnarray}
where the last member is obtained by~{(\ref{weset})}.

The proof is complete.~{$\square$}

Formula~{(\ref{fstpr-bis})} assumes a remarkably simple form in the
special case where the carrier Hilbert space $\hh$ of the
representation $U$ is finite-dimensional (so that the l.c.s.c.\
group $G$ must be unimodular; see the last assertion of
Remark~{\ref{dufmoo}}).
\begin{corollary} Suppose that the Hilbert space $\hh$, where the
square integrable representation $U$ acts, is finite-dimensional.
Then, for any pair of functions $f_1,f_2\in\ldg$, the following
formula holds:
\begin{equation} \label{fin-for}
f_1\kstarp f_2= d_U^{-3}\intg\de \mG(h) \intg \de \mG(h^\prime)\;
\tr(U(\cdot)^\ast\hspace{0.4mm}U(h)\,\opk\,U(h^\prime))\, f_1(h)
f_2(h^\prime)\fin .
\end{equation}
\end{corollary}

\noindent {\bf Proof:} If $\hh$ is finite-dimensional, then on the
r.h.s.\ of formula~{(\ref{fstpr-bis})} we have a finite sum and
$\du^{-1}= d_U^{-1} I$; therefore:
\begin{equation}
\Big(f_1\kstarp f_2\Big) (g)= \intg\de \mG(h) \intg \de
\mG(h^\prime)\sun \nuc(\hat{K},\chi_n; g,h,h^\prime)\, f_1(h)
f_2(h^\prime)\fin ,\ \ \ \aam g\in G\fin ,
\end{equation}
where
\begin{eqnarray}
\sun \nuc(\hat{K},\chi_n; g,h,h^\prime) \spa & = & \spa
d_U^{-3}\sun\langle \chi_n, U(g)^\ast\hspace{0.4mm}U(h)
\,\opk\,U(h^\prime)\,\chi_n\rangle
\nonumber\\
& = & \spa
d_U^{-3}\,\tr(U(g)^\ast\hspace{0.4mm}U(h)\,\opk\,U(h^\prime)) \fin ,
\end{eqnarray}
by definition of the trace.~{$\square$}

\begin{remark} {\rm
Assume that $G$ is a compact --- in particular, a finite --- group
and $U$ is a (irreducible) unitary representation. In this case,
formula~{(\ref{fin-for})} reads:
\begin{equation} \label{stpcp}
f_1\starp f_2 = \dimu^{\frac{3}{2}}\intg\de \mG(h) \intg \de
\mG(h^\prime)\; \chara\big((\cdot)^{-1}h h^\prime\big)\, f_1(h)
f_2(h^\prime)\fin .
\end{equation}
where the function $\chara\colon G\rightarrow\ccc$ is the character
of the finite-dimensional representation $U$; i.e.\
$\chara(g):=\tr(U(g))$. Then, since $\wig I=\dimu^{\frac{1}{2}}\,
\chara\big((\cdot)^{-1}\big)$, the obvious equation
\begin{equation}
\big(\wig I\big)\starp\big(\wig I\big)=\wig I
\end{equation}
translates into the following relation for the character $\chara$:
\begin{equation}
\chara(g) = \dimu^{2}\intg\de \mG(h) \intg \de \mG(h^\prime)\;
\chara\big(g h h^\prime\big)\, \chara\big(h^{-1}\big)\,
\chara\big((h^\prime)^{-1}\big)\fin .
\end{equation}
Thus, we recover results previously found in
ref.~{\cite{Aniello-compact}}.~{$\blacksquare$}}
\end{remark}

\subsection{A generalization of Theorem~{\ref{main-theorem}}}
\label{generalization}

As anticipated, Theorem~{\ref{main-theorem}} can be further
generalized. This generalization is based on the notion of
\emph{right approximate identity} in a Banach algebra --- in
particular, in a $\Hstar$-algebra. We will say that a sequence
$\{\optn\}_{n\in\nat}\subset\hs$  is a right approximate identity in
the $\Hstar$-algebra $\hs$ if $\displaystyle\hslim\opa\,\optn
=\opa$, for all $\opa\in\hs$; otherwise stated, if the sequence
$\big\{\mathfrak{R}_{\optn}\big\}_{n\in\nat}$ of bounded operators
in the Hilbert space $\hs$ is strongly convergent to the identity.
As is well known, the $\Hstar$-algebra $\hs$ admits an identity if
and only if $\hh$ is finite-dimensional, but it always admits a
right approximate identity. For instance, in the case where $\hh$ is
infinite-dimensional, for every orthonormal basis
$\{\chi_n\}_{n\in\nat}$ in $\hh$ the sequence $\{\sum_{k=1}^n
|\chi_k \rangle\langle \chi_k|\}_{n\in\nat}$ is a right (and left)
approximate identity in $\hs$. This example also provides the link
between Theorem~{\ref{main-theorem}} and its generalization, i.e.\
Theorem~{\ref{general-theorem}} below.

We will now define a positive selfadjoint operator in the Hilbert
space $\hs$ induced by the Duflo-Moore operator $\duu$.  In the
dense linear span $\frl\subset\hs$
--- see~{(\ref{deffrl})} --- we can define the linear operator
$\preope$ as follows:
\begin{equation}
\preope \opf := \sum_{k=1}^{\en} |\phi_k\rangle\langle
\du^{-1}\tre\psi_k |\fin , \ \ \ \opf\in\frl\fin ,
\end{equation}
where a canonical decomposition of $\opf$ is given
by~{(\ref{candeco})}. It is easy to check that $\preope$ is a
symmetric operator in the Hilbert space $\hs$; hence, it is
closable. Denoting by $\ope$ the closure of $\preope$ --- i.e.\
$\ope\equiv\overline{\preope}$
--- for every complex conjugation $J$ in $\hh$, we have:
\begin{equation}
\ope = \uj \big(I\otimes (J\hspace{0.5mm}\du^{-1}J)\big)\ujast \fin
,
\end{equation}
where $\uj$ is the unitary operator determined by~{(\ref{detuj})}.
$\ope$ is a positive selfadjoint operator (since it is unitarily
equivalent to $I\otimes (J\hspace{0.5mm}\du^{-1}J)$ and
$\spe\big(I\otimes
(J\hspace{0.5mm}\du^{-1}J)\big)=\spe\big(J\hspace{0.5mm}\du^{-1}J\big)=
\spe\big(\du^{-1}\big)$).

It is worth introducing the following dense linear span $\frllr$ in
$\hs$ (compare with definition~{(\ref{deffrl})}) defined by
\begin{equation}
\frllr := \big\{ \hat{F}\in\fr:\
\mathrm{Ran}(\hat{F}^\ast)\subset\mathrm{Dom}\big(\du^{-2}\big)\big\}\subset\frl\fin
.
\end{equation}
The elements of $\frllr$ are those finite rank operators in $\hh$
that admit a \emph{canonical decomposition} of the form
\begin{equation} \label{canf}
\hat{F}=\sum_{k=1}^{\en} |\phi_k\rangle\langle \chi_k | \fin ,\ \ \
\en\in\nat \fin ,
\end{equation}
where $\{\phi_k\}_{k=1}^{\en}$, $\{\chi_k\}_{k=1}^{\en}$ are
linearly independent systems in $\hh$, with
$\{\chi_k\}_{k=1}^{\en}\subset \dom\big(\du^{-2}\big)$. It is clear
that, if $\opf$ is positive selfadjoint, then one can set
$\phi_k=\chi_k$ in~{(\ref{canf})} (in particular, one can always
choose the vectors $\{\chi_k\}_{k=1}^{\en}$ mutually orthogonal).
Also note that, for every $\opf\in\frllr$, since the operator $\preope\opf$
belongs to $\frl$, we have:
\begin{equation} \label{clear}
\opeq\opf=
\ope(\ope\opf)=\ope(\preope\opf)=\preope(\preope\opf)=\sum_{k=1}^{\en}
|\phi_k\rangle\langle \du^{-2}\tre\chi_k |\fin ,
\end{equation}
where a canonical decomposition of $\opf$ is given
by~{(\ref{canf})}. Moreover --- since, for any complex conjugation
$J$ in $\hh$,
$\frllr=\uj\big(\hh\otimes\dom\big(J\hspace{0.5mm}\du^{-2}J\big)\big)$
and $\opeq = \uj \big(I\otimes
\big(J\hspace{0.5mm}\du^{-2}J\big)\big)\ujast$ --- the linear span
$\frllr$ is a core for the positive selfadjoint operator$\opeq$
(recall, indeed, that
$\hh\otimes\dom\big(J\hspace{0.5mm}\du^{-2}J\big)$ is a core for
$I\otimes \big(J\hspace{0.5mm}\du^{-2}J\big)$).

Let $\opk$ be a bounded operator in $\hh$ and $\opt$ an Hilbert-Schmidt operator contained in
the dense linear span $\dom\big(\opeq\big)$.
In association with these operators,
we can define the integral kernel $\gu(\opk,\opt;\cdot,\cdot)\colon G\times G\rightarrow\ccc$
as follows:
\begin{equation} \label{defnewintk}
\gu(\opk,\opt;g,h) := \Big(\wig\big(\opk^\ast\hspace{0.5mm}
U(g)\,(\opeq\hspace{0.3mm}\opt)^\ast\big)\Big)(h)^\ast .
\end{equation}
Observe that --- since, by virtue of the intertwining
relation~{(\ref{intcoco})},
$\big(\wig\opa\big)^\ast=\big(\invol\wig(\opa^\ast)\big)^\ast$,
$\forall\hspace{0.4mm}\opa\in\hs$
--- we have:
\begin{equation} \label{app-intcoco}
\gu(\opk,\opt;g,h) = \mm (h,h^{-1})^\ast \modu (h)^{-\frac{1}{2}}\,
\wig\big( (\opeq\hspace{0.3mm}\opt)\,
U(g)^\ast\hspace{0.4mm}\opk\big)(h^{-1}) \fin .
\end{equation}
In the case where $\opt\equiv\opf$ is a positive selfadjoint
belonging to $\frllr$ --- let $\sum_{k=1}^{\en}
|\chi_k\rangle\langle \chi_k |$ be a canonical decomposition of
$\opf$
--- the integral kernel~{(\ref{defnewintk})} has the following form (compare with~{(\ref{defkeu})}):
\begin{equation} \label{parcase}
\gu(\opk,\opf;g,h) = \sum_{k=1}^{\en} \big\langle
U(g)\,\du^{-2}\tre\chi_k, \opk\,U(h)\,\du^{-1}\tre\chi_k\big\rangle
\fin .
\end{equation}

The following result shows that with every \emph{suitable} right
approximate identity in $\hs$ is associated a formula for the
$\opk$-deformed star product $(\cdot)\kstarp (\cdot)$ in $\ldg$,
formula that (taking into account~{(\ref{parcase})}) extends the
first of the expressions~{(\ref{alternative})} for the star product.

\begin{theorem} \label{general-theorem}
Let $\hat{K}$ be a bounded operator in $\hh$, and let
$\{\optn\}_{n\in\nat}$ be a right approximate identity in
the $\Hstar$-algebra $\hs$ such that
$\{\optn\}_{n\in\nat}\subset\dom\big(\opeq\big)$. Then, for
any $f_1,f_2\in\ldg$, the following formula holds:
\begin{eqnarray} \label{fstpr-gen}
f_1\kstarp f_2 \spa & = & \spa\norlim \intg\de \mG(h)\;  f_1(h)\,
\mm (h,h^{-1}(\cdot))^\ast \modu
(h^{-1}(\cdot))^{\frac{1}{2}}   \nonumber \\
& \times & \spa \intg \de \mG(h^\prime)\; \gu(\opk,\optn;
h^{-1}(\cdot),h^\prime)\, f_2(h^\prime)\fin .
\end{eqnarray}

\end{theorem}

\noindent {\bf Proof:} Since the proof goes along lines similar to
those already traced in Subsects.~{\ref{preliminary}}
and~{\ref{formulae}}, we will be rather sketchy. For the sake of
clarity, we subdivide our argument into a few steps.

\begin{description}

\item[1)\hspace{3.6mm}] Let us first
show that, for every $\opf\in\frllr\subset\dom\big(\opeq\big)$ and
every $f\in\ldg$, we have:
\begin{equation} \label{anew}
\intG \de\mG (h)\; \gu(\opk,\opf;\cdot,h)\, f(h) = \wig\big(\opk(\wigg^\ast f)(\ope\opf)\big).
\end{equation}
In fact, if $\sum_{k=1}^{\en} |\phi_k\rangle\langle \chi_k |$ is a canonical decomposition of
$\opf$, taking into account relation~{(\ref{clear})} we find that
\begin{eqnarray}
\intG \de\mG (h)\; \gu(\opk,\opf;\cdot,h)\, f(h) \spa & = & \spa
\big\langle \wig\big(\opk^\ast\hspace{0.5mm} U(\cdot)\,(\opeq\opf)^\ast\big), f \big\rangle_\ld
\nonumber\\
& = & \spa \big\langle \opk^\ast\hspace{0.5mm} U(\cdot)\,(\opeq\opf)^\ast, \wigg^\ast f \big\ranglehs
\nonumber\\
& = & \spa \sum_{k=1}^{\en} \tr\big( |\phi_k\rangle\langle \du^{-2}\tre
\chi_k |\,U(\cdot)^\ast\hspace{0.5mm}\opk\hspace{0.4mm}(\wigg^\ast
f)\big)
\nonumber\\
& = & \spa \sum_{k=1}^{\en} \wig\big(\opk\hspace{0.4mm}(\wigg^\ast
f)\hspace{0.5mm} |\phi_k\rangle\langle \du^{-1}\tre
\chi_k |\big)
\nonumber\\
& = & \spa
\wig\big(\opk(\wigg^\ast f)(\ope\opf)\big).
\end{eqnarray}

\item[2)\hspace{3.6mm}] Let us observe the following fact. Let $\hat{C}$ be a selfadjoint operator
in a (complex separable) Hilbert space $\mathcal{S}$, and let
$\mathcal{S}_0$ be a dense linear span in $\mathcal{S}$ contained in
$\dom\big(\hat{C}^2\big)$; we suppose, furthermore, that
$\mathcal{S}_0$ is a core for the positive selfadjoint operator
$\hat{C}^2$. Then, for every vector
$\varphi\in\dom\big(\hat{C}^2\big)$, there exists a sequence
$\{\varphi_n\}_{n\in\nat}\subset\mathcal{S}_0$ such that both
$\lim_{n\rightarrow\infty}\|\varphi-\varphi_n\|=0$ and
$\lim_{n\rightarrow\infty}\|\hat{C}^2(\varphi-\varphi_n)\|=0$. These
two relations imply that
$\lim_{n\rightarrow\infty}\|\hat{C}(\varphi-\varphi_n)\|=0$, as
well. Indeed, this fact can be easily checked in the case where
$\mathcal{S}$ is a space of square integrable functions and
$\hat{C}$ is a multiplication operator by a measurable function
--- hint:
an unbounded multiplication operator can be written as the sum of a
(bounded) multiplication operator by a function of modulus not
larger than one and, possibly, of a multiplication operator by a
function of modulus larger than one, these two functions having
disjoint supports. Hence, by virtue of the spectral theorem in
`multiplication operator form' --- see~{\cite{Reed}} --- this
property holds true for a generic selfadjoint operator $\hat{C}$.

\item[3)\hspace{3.6mm}] It is possible to generalize formula~{(\ref{anew})};
i.e.\ one can prove that, for every $\opt\in\dom\big(\opeq\big)$ and
every $f\in\ldg$, the following relation holds:
\begin{equation} \label{bnew}
\intG \de\mG (h)\; \gu(\opk,\opt;\cdot,h)\, f(h) =
\wig\big(\opk(\wigg^\ast f)(\ope\hspace{0.3mm}\opt)\big).
\end{equation}
Indeed, let $\{\opf_l\}_{l\in\nat}$ be a sequence in $\frllr$ (which
is a core for $\opeq$) such that $\displaystyle\hsliml \opf_l =
\opt$ and $\displaystyle\hsliml \opeq \opf_l = \opeq\hspace{0.3mm}
\opt$. Then, we know that $\displaystyle\hsliml \ope \opf_l =
\ope\hspace{0.3mm} \opt$, as well. Observe now that by
relation~{(\ref{anew})} we have:
\begin{eqnarray} \hspace{-6mm}
\norliml \intG \de\mG (h)\; \gu(\opk,\opf_l;\cdot,h)\, f(h) \spa & = & \spa \norliml \wig\big(\opk(\wigg^\ast f)(\ope\opf_l)\big)
\nonumber\\
& = & \spa \wig\Big(\opk(\wigg^\ast f)\hspace{1mm}\hsliml(\ope\opf_l)\Big)
\nonumber\\
& = & \spa \label{cnew}
\wig\big(\opk(\wigg^\ast f)(\ope\hspace{0.3mm}\opt)\big),\ \ \ \forall\hspace{0.4mm}f\in\ldg \fin .
\end{eqnarray}
On the other hand, for every $f\in\ldg$ and every $g\in G$, we also have that
\begin{eqnarray}
\lim_{l\rightarrow\infty} \intG \de\mG (h)\; \gu(\opk,\opf_l;g,h)\, f(h) \spa & = & \spa
\lim_{l\rightarrow\infty} \big\langle\wig\big(\opk^\ast\hspace{0.5mm} U(g)\, (\opeq\opf_l)^\ast\big), f\big\rangle_\ld
\nonumber\\
& = & \spa \Big\langle\wig\big(\opk^\ast\hspace{0.5mm} U(g)\hspace{1mm} \hsliml (\opeq\opf_l)^\ast\big), f\Big\rangle_\ld
\nonumber\\
& = & \spa
\big\langle\wig\big(\opk^\ast\hspace{0.5mm} U(g)\,(\opeq\hspace{0.3mm}\opt)^\ast\big),f\big\rangle_\ld
\nonumber\\ \label{dnew}
& = & \spa \intG \de\mG (h)\; \gu(\opk,\opt;g,h)\, f(h)\fin .
\end{eqnarray}
From~{(\ref{cnew})} and~{(\ref{dnew})} it follows that relation~{(\ref{bnew})} holds true.

\item[4)\hspace{3.6mm}] Let us now prove that, for every $\opt\in\dom\big(\opeq\big)$ and any
pair of functions $f_1,f_2\in\ldg$,
\begin{eqnarray}
& & \spa \intg\de \mG(h)\;  f_1(h)\, \mm (h,h^{-1}(\cdot))^\ast
\modu (h^{-1}(\cdot))^{\frac{1}{2}} \intg \de \mG(h^\prime)\;
\gu(\opk,\opt; h^{-1}(\cdot),h^\prime)\, f_2(h^\prime) \nonumber \\
& = & \spa \label{enew} \langle
U(\cdot)\hspace{0.7mm}(\ope\hspace{0.3mm}\opt)^\ast(\wigg^\ast
f_2)^\ast\opk^\ast, \wigg^\ast f_1\ranglehs \fin .
\end{eqnarray}
Indeed, according to relations~{(\ref{bnew})}
and~{(\ref{fortherefore})}, for any $f_1,f_2\in\ldg$ we have:
\begin{eqnarray}
& & \hspace{-5mm} \spa \intg\de \mG(h)\;  f_1(h)\, \mm
(h,h^{-1}(\cdot))^\ast \modu (h^{-1}(\cdot))^{\frac{1}{2}} \intg \de
\mG(h^\prime)\; \gu(\opk,\opt; h^{-1}(\cdot),h^\prime)\,
f_2(h^\prime)
\nonumber\\
& = & \spa \intg\de \mG(h)\;  f_1(h)\, \mm (h,h^{-1}(\cdot))^\ast
\modu (h^{-1}(\cdot))^{\frac{1}{2}}\, \Big(\wig\big(\opk(\wigg^\ast
f_2)(\ope\hspace{0.3mm}\opt)\big)\Big)(h^{-1}(\cdot))
\nonumber\\
& = & \spa \big\langle \rem(\cdot)\invol \wig\big(\opk(\wigg^\ast
f_2)(\ope\hspace{0.3mm}\opt)\big), f_1\big\rangle_\ld \fin .
\end{eqnarray}
Then, using the intertwining relation~{(\ref{intcoco})} and
relation~{(\ref{simpinter})}, we find that
\begin{eqnarray}
& & \hspace{-5mm} \spa \intg\de \mG(h)\;  f_1(h)\, \mm
(h,h^{-1}(\cdot))^\ast \modu (h^{-1}(\cdot))^{\frac{1}{2}} \intg \de
\mG(h^\prime)\; \gu(\opk,\opt; h^{-1}(\cdot),h^\prime)\,
f_2(h^\prime)
\nonumber\\
& = & \spa \big\langle
\rem(\cdot)\hspace{0.7mm}\wig\big((\ope\hspace{0.3mm}\opt)^\ast(\wigg^\ast
f_2)^\ast\opk^\ast\big), f_1\big\rangle_\ld
\nonumber\\
& = & \spa \big\langle
\wig\big(U(\cdot)\hspace{0.7mm}(\ope\hspace{0.3mm}\opt)^\ast(\wigg^\ast
f_2)^\ast\opk^\ast\big), f_1\big\rangle_\ld
\nonumber\\
& = & \spa \langle
U(\cdot)\hspace{0.7mm}(\ope\hspace{0.3mm}\opt)^\ast(\wigg^\ast
f_2)^\ast\opk^\ast, \wigg^\ast f_1\ranglehs \fin .
\end{eqnarray}

\item[5)\hspace{3.6mm}] The next step is to show that,
for every $\opt\in\dom\big(\opeq\big)$ and any $f_1,f_2\in\ldg$,
\begin{equation} \label{enew-bis}
\langle
U(\cdot)\hspace{0.7mm}(\ope\hspace{0.3mm}\opt)^\ast(\wigg^\ast
f_2)^\ast\opk^\ast, \wigg^\ast f_1\ranglehs = \wig
\hspace{0.5mm}\mathfrak{R}_{\opt}\big((\wigg^\ast
f_1)\hspace{0.5mm}\opk\hspace{0.4mm}(\wigg^\ast f_2)\big).
\end{equation}
We will first give a proof of this relation in the case where
$\opt\equiv\opf\in\frllr$. In fact, considering a canonical
decomposition $\sum_{k=1}^\en |\phi_k\rangle\langle \chi_k |$ of
$\opf$, we get:
\begin{eqnarray}
\langle U(\cdot)\hspace{0.7mm}(\ope\opf)^\ast(\wigg^\ast
f_2)^\ast\opk^\ast, \wigg^\ast f_1\ranglehs \spa & = & \spa
\sum_{k=1}^\en \big\langle |U(\cdot)\,\du^{-1}\tre\chi_k
\rangle\langle \opk\hspace{0.4mm}(\wigg^\ast f_2)\,\phi_k|,
\wigg^\ast f_1\big\ranglehs
\nonumber \\
& = & \spa \sum_{k=1}^\en \tr\big( |\opk\hspace{0.4mm}(\wigg^\ast
f_2)\,\phi_k\rangle\,\langle U(\cdot)\,\du^{-1}\tre\chi_k
|\, \wigg^\ast f_1\big) \nonumber \\  & = & \spa \sum_{k=1}^\en
\wig\big((\wigg^\ast f_1)\hspace{0.5mm}\opk\hspace{0.4mm}(\wigg^\ast
f_2)\,
|\phi_k\rangle\langle\chi_k|\big) \nonumber\\
& = & \spa  \label{releffe}  \wig\big((\wigg^\ast
f_1)\hspace{0.5mm}\opk\hspace{0.4mm}(\wigg^\ast f_2)\,\opf\big).
\end{eqnarray}
Next, let $\opt$ be a generic Hilbert-Schmidt operator in
$\dom\big(\opeq\big)$, and let $\{\opf_l\}_{l\in\nat}$ be a sequence
in $\frllr$ such that $\displaystyle\hsliml \opf_l = \opt$ and
$\displaystyle\hsliml \opeq \opf_l = \opeq\hspace{0.3mm} \opt$
(hence: $\displaystyle\hsliml \ope \opf_l = \ope\hspace{0.3mm}
\opt$). Then, for any $f_1,f_2\in\ldg$, we have:
\begin{eqnarray}
\norliml \langle U(\cdot)\hspace{0.7mm}(\ope\opf_l)^\ast(\wigg^\ast
f_2)^\ast\opk^\ast, \wigg^\ast f_1\ranglehs \spa & = & \spa \norliml
\wig\hspace{0.5mm}\mathfrak{R}_{{\opf_l}}\big((\wigg^\ast
f_1)\hspace{0.5mm}\opk\hspace{0.4mm}(\wigg^\ast f_2)\big)
\nonumber\\
& = & \spa \label{fnew}
\wig\hspace{0.5mm}\mathfrak{R}_{\opt}\big((\wigg^\ast
f_1)\hspace{0.5mm}\opk\hspace{0.4mm}(\wigg^\ast f_2)\big),
\end{eqnarray}
where we have used relation~{(\ref{releffe})} and the fact that
$\displaystyle\hsliml\mathfrak{R}_{{\opf_l}}\opa=\mathfrak{R}_{\opt}\hspace{0.3mm}\opa$,
for every $\opa\in\hs$. Moreover, we also have that
\begin{equation} \label{gnew}
\lim_{l\rightarrow\infty}\langle
U(g)\hspace{0.7mm}(\ope\opf_l)^\ast(\wigg^\ast f_2)^\ast\opk^\ast,
\wigg^\ast f_1\ranglehs = \langle
U(g)\hspace{0.7mm}(\ope\hspace{0.3mm}\opt)^\ast(\wigg^\ast
f_2)^\ast\opk^\ast, \wigg^\ast f_1\ranglehs \fin ,
\end{equation}
for all $g\in G$. From~{(\ref{fnew})} and~{(\ref{gnew})} it follows
eventually that relation~{(\ref{enew})} holds true.

\item[6)\hspace{3.6mm}] Relations~{(\ref{enew})} and~{(\ref{enew-bis})}
imply that, for every $\opt\in\dom\big(\opeq\big)$ and any pair of
functions $f_1,f_2\in\ldg$,
\begin{eqnarray}
& & \spa \intg\de \mG(h)\;  f_1(h)\, \mm (h,h^{-1}(\cdot))^\ast
\modu (h^{-1}(\cdot))^{\frac{1}{2}} \intg \de \mG(h^\prime)\;
\gu(\opk,\opt; h^{-1}(\cdot),h^\prime)\, f_2(h^\prime) \nonumber \\
& = & \label{hnew} \spa \wig
\hspace{0.5mm}\mathfrak{R}_{\opt}\big((\wigg^\ast
f_1)\hspace{0.5mm}\opk\hspace{0.4mm}(\wigg^\ast f_2)\big).
\end{eqnarray}

\item[7)\hspace{3.6mm}]
Finally, as in the proof of Theorem~{\ref{main-theorem}}, one can
exploit relation~{(\ref{hnew})}, the boundedness of the Wigner map
and the fact that the sequence $\{\optn\}_{n\in\nat}$ is a right
approximate identity for obtaining formula~{(\ref{fstpr-gen})}.

\end{description}

The proof is complete.~$\square$

\section{Applications}
\label{examples}

In this section, we will consider two simple --- but extremely
significant
--- applications of the theory developed in
Sects.~{\ref{weyl-wigner}-\ref{main}}. We will first consider the
case of a square integrable
--- genuinely projective --- representation of a unimodular group; i.e., the group of translations on phase space.
The analysis of this case leads to the Gr\"onewold-Moyal star
product, i.e.\ the prototype of star product. Next, we will study a
case where square integrable unitary representations are involved of
a group which is \emph{not} unimodular; namely, the $1$-dimensional
affine group. As already mentioned, this group is at the base of
wavelet analysis.

\subsection{The group of translations on phase space}

Let us consider the group of translations on the $(1+1)$-dimensional
phase space, namely, the additive group $\rrr\times\rrr$ (the
extension to the $(n+n)$-dimensional case is straightforward). As is
well known (see, e.g., ref.~{\cite{Marmo}}), the map
\begin{equation}
\rrr\times\rrr\ni\qp\mapsto U\qp\in \mathcal{U}(\lr)\fin ,
\end{equation}
defined by
\begin{eqnarray}
U\qp \spa & := & \spa \disp \nonumber\\ \label{Weyl-sys}
& = & \spa
\ee^{-\frac{\ima}{2}\, qp}\, \exp(\ima p\hspace{0.5mm}\hq)\,
\exp(-\ima q\hspace{0.3mm}\hp) = \ee^{\frac{\ima}{2}\, qp}\,
\exp(-\ima q\hspace{0.3mm}\hp)\, \exp(\ima p\hspace{0.5mm}\hq) \fin
, \ \ q,p\in\mathbb{R} \fin ,
\end{eqnarray}
--- where $\hq$, $\hp$ are the standard position and momentum operators ---
is a projective representation of the unimodular group
$\rrr\times\rrr$, representation which we will call (with a slight abuse of
terminology) \emph{Weyl system}. The Weyl system is --- as already
observed in Sect.~{\ref{known}} --- a square integrable
representation. It `encodes' the canonical commutation relations of
quantum mechanics (in the integrated form), as shown by the last two
members of~{(\ref{Weyl-sys})}.

The (generalized) Wigner transform associated with the Weyl system
is not the standard Wigner transform but the so-called
\emph{Fourier-Wigner transform}~{\cite{Folland-bis}}. In fact, it
turns out that these maps are related by the \emph{symplectic
Fourier transform}, i.e.\ by the unitary operator $\fs \colon \lrr
\rightarrow\lrr$ determined by
\begin{equation}
\big(\fsy f\big)(q,p)=\frac{1}{2\pi}\intrr f(q^\prime,p^\prime)\,
\ee^{\ima (qp^\prime - pq^\prime)}\; \de q^\prime \de p^\prime,\ \ \
\forall\hspace{0.4mm} f\in\lurr\cap\lrr \fin .
\end{equation}
Recall that $\fsy$ enjoys the remarkable property of being both
unitary and selfadjoint:
\begin{equation}
\fs=\fsy^\ast \fin ,\ \ \ \fsy^2=I \fin .
\end{equation}

As already mentioned in Sect.~{\ref{known}}, $(2\pi)^{-1}\de q\de p$ is
the Haar measure on $\rrr\times\rrr$ normalized in agreement with the Weyl system
$U$. Then, in this case, the generalized Wigner transform $\wig$ is
the isometry from $\mathcal{B}_2(\lr)$ into
$\lrr\equiv\mathrm{L}^2\big(\rrr\times\rrr, (2\pi)^{-1}\de q\de
p\hspace{0.3mm};\ccc\big)$ determined by
\begin{equation} \label{deter}
\big(\wig\hspace{0.3mm} \hrho\big)\qp =\tr(U\qp^\ast\hrho)\fin ,\ \
\ \forall\hspace{0.4mm}\hrho\in\mathcal{B}_1(\lr) \fin .
\end{equation}
The multiplier
$\mm\colon(\rrr\times\rrr)\times(\rrr\times\rrr)\rightarrow\mathbb{T}$
associated with $U$ is of the form
\begin{equation}
\mm(q,p\hspace{0.6mm};q^\prime,p^\prime)=\exp\hspace{-0.5mm}\Big(\frac{\ima}{2}(qp^\prime-pq^\prime)\Big).
\end{equation}
Hence, for the function $\mmm\colon
(\rrr\times\rrr)\times(\rrr\times\rrr)\rightarrow\mathbb{T}$
(see~{(\ref{demmm})}) we find, in this case, the following
expression:
\begin{equation}
\mmm(q,p\hspace{0.6mm};q^\prime,p^\prime)=\mm(q,p\hspace{0.6mm};q^\prime-q,p^\prime-p)^\ast\,
\mm(q^\prime -q,p^\prime
-p\hspace{0.6mm};q,p)=\exp\hspace{-0.5mm}\big(\hspace{-0.7mm}-\ima(qp^\prime-pq^\prime)\big).
\end{equation}
Therefore, according to formula~{(\ref{two-sided})}, we have that the
generalized Wigner transform $\wig$ intertwines the unitary
representation
\begin{equation}
\rep \colon \rrr\times\rrr\rightarrow\mathcal{U}(\mathcal{B}_2(\lr))
\end{equation}
with the representation $\two \colon
\rrr\times\rrr\rightarrow\mathcal{U}(\lrr)$ defined by
\begin{equation}
\big(\two(q,p)\hspace{0.4mm} f\big)(q^\prime,p^\prime)=
\ee^{-\ima(qp^\prime - pq^\prime)}\hspace{0.8mm} f(q^\prime,
p^\prime) \fin ,\ \ \ \forall\hspace{0.3mm}f\in\lrr \fin .
\end{equation}
Moreover, $\wig$ intertwines the involution $\invo$ in $\hs$ with the complex
conjugation $\sinvo\equiv\invol$ that, in this case --- as the reader may readily check ---
takes the following form:
\begin{equation}
\big(\sinvo f\big)(q,p) = f(-q,-p)^\ast,\ \ \ \forall\hspace{0.4mm}
f\in\lrr \fin .
\end{equation}

As anticipated, the \emph{standard Wigner transform} --- we will
denote it by $\wigt$ --- is the isometry obtained composing the
isometry $\wig$, determined by~{(\ref{deter})}, with the symplectic
Fourier transform (see~{\cite{Aniello-framet}}):
\begin{equation}
\wigt := \fsy \,\wig\colon \mathcal{B}_2(\lr)\rightarrow \lrr \fin .
\end{equation}
It is clear that the isometry $\wigt$ intertwines the representation
$\rep$ with the unitary representation
$\mathcal{V}\colon\rrr\times\rrr\rightarrow\mathcal{U}(\lrr)$
defined by
\begin{equation}
\mathcal{V}\qp
:=\hspace{0.3mm}\fsy\hspace{0.6mm}\two\qp\hspace{0.7mm}\fsy \fin , \
\ \ \forall\hspace{0.5mm}\qp\in\rrr\times\rrr \fin ;
\end{equation}
as the reader may easily check, explicitly, we have:
\begin{equation}
\big(\mathcal{V}(q,p)\hspace{0.4mm} f\big)(q^\prime,p^\prime)=
 f(q^\prime-q, p^\prime-p) \fin ,\ \ \ \forall\hspace{0.4mm}f\in\lrr \fin .
\end{equation}
Thus, the representation $\mathcal{V}$ acts by simply translating
functions on phase space. It is also a remarkable fact --- see
ref.~{\cite{Pool}}
--- that
\begin{equation}
\mathrm{Ran}(\wigt)=\lrr \fin ;
\end{equation}
equivalently, $\ru\equiv\mathrm{Ran}\big(\wig\big)=\lrr$ (this fact can be verified
deducing the integral kernel of the Hilbert-Schmidt operator $\wigg^\ast f$,
for a generic $f\in\lrr$, and observing that $\Ker\big(\wigg^\ast\big)=\{0\}$).
Therefore, the standard Wigner transform $\wigt$ --- and its adjoint
$\wigt^\ast$, the \emph{standard Weyl map} --- are both unitary
operators.

Let us now study the star product in $\lrr$ induced by the Weyl
system $U$. Recalling Theorem~{\ref{unimod-case}}, and taking into
account the fact that, in this case, $\ru=\lrr$ (and $\ddu=1$), we
have:
\begin{eqnarray}
\Big(f_1\starp f_2\Big)(q,p) \spa & = & \spa \frac{1}{2\pi}\intrr
f_1(q^\prime,p^\prime)\hspace{0.8mm}
\mm(q,p\hspace{0.6mm};q-q^\prime,p-p^\prime)^\ast\, f_2(q-q^\prime,
p-p^\prime)\; \de q^\prime \de p^\prime \nonumber\\
& = & \spa \frac{1}{2\pi}\intrr f_1(q^\prime,p^\prime) \,
f_2(q-q^\prime,
p-p^\prime)\hspace{0.6mm}\exp\hspace{-0.5mm}\Big(\frac{\ima}{2}(qp^\prime-pq^\prime)\Big)
\, \de q^\prime \de p^\prime,
\end{eqnarray}
$\forall\hspace{0.5mm} f_1, f_2\in\lrr$. Thus, the star product
associated with the Weyl system is nothing but the \emph{twisted
convolution} of functions~{\cite{GLP,Folland-bis}}. According to the
results of Sect.~{\ref{defstar}}, $\big(\lrr, \starp, \sinvo\big)$
is a proper $\Hstar$-algebra and $\wig\colon\hs\rightarrow\lrr$ is
an isomorphism of $\Hstar$-algebras.

The unitary operators $\wigt$, $\wigt^\ast$ induce another star
product of functions
\begin{equation}
(\cdot)\gm(\cdot)\colon
\lrr\times\lrr\ni(f_1,f_2)\mapsto\hspace{0.3mm}\wigt\Big(\big(\wigt^\ast\hspace{0.3mm}
f_1\big) \big(\wigt^\ast\hspace{0.3mm} f_2\big)\Big)
\hspace{-0.4mm}\in\lrr \fin ,
\end{equation}
namely, the \emph{twisted product} (see~{\cite{GLP}}). Using the
fact that $\wigt = \fsy \,\wig$ and $\wigt^\ast = \wig\,\fsy$, we
obtain that
\begin{equation} \label{defist-bis}
f_1\gm f_2 = \hspace{0.3mm} \fs \Big(\big(\fs f_1\big)\starp
\big(\fs f_2\big)\Big).
\end{equation}
From this relation, by an explicit calculation, one finds that, for any pair of functions
$f_1,f_2$ in $\lurr\cap\lrr$,
\begin{equation} \label{twisted-pr}
\big(f_1\gm f_2\big) (q,p) = \frac{1}{\pi^2}\int_{\rr}
\hspace{-0.7mm} \de q^\prime \de p^\prime\,
\int_{\rr}\hspace{-0.7mm} \de q^{\prime\prime} \de p^{\prime\prime}\
\theta\big(q,p; q^\prime, p^\prime ; q^{\prime\prime},
p^{\prime\prime}\big)\, f_1(q^\prime, p^\prime)\,
f_2(q^{\prime\prime}, p^{\prime\prime}) \fin ,
\end{equation}
where we have set:
\begin{equation}
\theta\big(q,p; q^\prime, p^\prime ; q^{\prime\prime},
p^{\prime\prime}\big):=\exp\big(\ima 2(qp^\prime - pq^\prime +
q^\prime p^{\prime\prime} - p^\prime q^{\prime\prime} +
q^{\prime\prime} p - p^{\prime\prime} q)\big).
\end{equation}
The function $\theta\colon
(\rr)\times(\rr)\times(\rr)\rightarrow\mathbb{T}$ is the celebrated
\emph{Gr\"onewold-Moyal kernel}. The symplectic Fourier transform intertwines
the complex conjugation $\sinvo$ with the standard complex conjugation in $\lrr$:
$\fs \sinvo \hspace{0.8mm} \fs f =f^\ast$. Therefore, $\lrr$ endowed with the twisted
product and with the standard complex conjugation is again a proper $\Hstar$-algebra.
To the best of our knowledge, this fact has been noted for the first time by Pool~{\cite{Pool}}.

\subsection{The 1-dimensional affine group}

Let us consider, now, the {\it $1$-dimensional affine group},
namely, the  semi-direct product group $G=\mathbb{R}\sdp\errep$,
where $\errep$ is the subgroup of \emph{dilations}; i.e.\
$\errep$ is the group of strictly positive real numbers (we
will denote by $\errem$ the set of strictly negative real numbers)
which acts multiplicatively on $\mathbb{R}$.
Thus, $G$ consists of the
topological space $\mathbb{R}\times\errep$, endowed with the
composition law
\begin{equation}
(a,r)\,(a^\prime,r^\prime)=(a+r a^\prime, r r^\prime)\fin ,\ \ \
a\in\mathbb{R}\fin ,\ r\in \errep.
\end{equation}
This group is not unimodular. A pair $\mul$, $\mur$ of
--- left and right, respectively ---  conjugated Haar measures  on $G$ ($\intG
f(g)\,\de\mul(g)=\intG f(g^{-1})\,\de\mur(g)$) are given by
\begin{equation}
\de\mul (a,r)=r^{-2}\, \de a\,\de r \fin ,\ \ \, \de\mur
(a,r)=r^{-1}\, \de a\,\de r \fin , \ \ \ a\in\mathbb{R} \fin ,\
r\in\errep.
\end{equation}
Hence, the modular function $\Delta_G$ on $G$ is given by $\Delta_G
(a,r) = r^{-1}\!$, $\forall\quattro a\in\mathbb{R}$,
$\forall\quattro r\in\errep$. As already recalled in
Sect.~{\ref{known}}, this group is at the base of the theory of the
wavelet transform. For the sake of completeness, we will come back
to this point later on.
It is also worth
mentioning that the quantization-dequantization theory based on the
affine group has been studied by Aslaksen and Klauder~{\cite{Klauder2}},
who obtained the Wigner and Weyl maps associated with the
representations this group. However, they did not
consider the concept of star product.

Using Mackey's little group method for classifying the irreducible
representations of semi-direct products with abelian normal factors (see~{\cite{Raja}}),
and the results of ref.~{\cite{Aniello-sdp}} on the characterization of
square integrable representations of the groups of this type,
one finds out that $G$
admits a maximal set of (unitarily) inequivalent square integrable
irreducible representations consisting of two elements:
$\{U^\ammm\colon G\rightarrow\mathcal{U}(\ld(\errem)), U^\ppp\colon
G\rightarrow\mathcal{U}(\ld(\errep))\}$. These two unitary
representations are defined by:
\begin{eqnarray}
\big(U^\ammm (a,r)\, \varphi^{\ammm}\big)(x )\spa & := & \spa
r^{\frac{1}{2}}\, \ee^{\ima ax} \varphi^{\ammm}(r x) \fin ,\ \ \
a\in\mathbb{R} \fin ,\ r\in \errep,\ x\in\errem,
\ \,  \varphi^{\ammm}\in \ld(\errem) \fin ,\\
\big( U^\ppp (a,r)\, \varphi^{\ppp}\big) (x ) \spa & := & \spa
r^{\frac{1}{2}}\, \ee^{\ima ax }\varphi^{\ppp}(r x)\fin ,\ \ \
a\in\mathbb{R} \fin ,\ r\in \errep,\ x\in\errep,\ \,
\varphi^{\ppp}\in \ld(\errep) \fin ,
\end{eqnarray}
where the Hilbert space $\ld(\errepm)$ is of course defined
considering the restriction to $\errepm$ of the Lebesgue measure on
$\rrr$. Moreover, by the results of ref.~{\cite{Aniello-sdp}}, the
Duflo-Moore operator $\duupm$ associated with the representation
$U^\pmpm$
--- and normalized according to $\mul$
--- is the unbounded multiplication operator (defined on its
natural domain) by the function
\begin{equation}
\errepm\ni x\mapsto
\left(\frac{2\pi}{|x|}\right)^{\hspace{-0.6mm}\frac{1}{2}}.
\end{equation}

The representations $U^\ammm$, $U^\ppp$ are unitarily inequivalent,
but they are intertwined by the aniunitary operator
$\mathfrak{Z}\colon \ld(\errem)\ni \varphi\mapsto
\varphi(-(\cdot))^\ast\in \ld(\errep)$. Hence, they are physically
equivalent. We will denote by $\wigm$ and $\wigp$, respectively, the
associated Wigner maps. These maps are isometries that intertwine
the unitary representations $\repm$ and $\repp$, respectively, with the
two-sided regular representation $\mathcal{T}$ of
$\mathbb{R}\sdp\errep$, representation which is defined by
\begin{equation}
\big(\mathcal{T}(a,r)\hspace{0.4mm} f\big)(a^\prime,r^\prime):=
r^{-\frac{1}{2}}\, f\big(r^{-1}(a^\prime -a +r^\prime
a),r^\prime\big),\ \ \
\forall\hspace{0.4mm}f\in\ld(\mathbb{R}\times\errep,\mul)\fin .
\end{equation}
The standard involutions $\invom$, $\invop$ in the Hilbert-Schmidt spaces
$\hsm$, $\hsp$ are intertwined by the Wigner maps $\wigm$ and $\wigp$, respectively,
with the map
\begin{equation}
\sinvo\colon
\ld(\mathbb{R}\times\errep,\mul)\rightarrow\ld(\mathbb{R}\times\errep,\mul)
\fin ,
\end{equation}
which is the complex conjugation defined by
\begin{equation}
\big(\sinvo f\big)(a,r) = r^{\frac{1}{2}}\,
f(-r^{-1}a,r^{-1})^\ast,\ \ \ \forall\hspace{0.4mm}
f\in\ld(\mathbb{R}\times\errep,\mul) \fin .
\end{equation}
The explicit form of the Weyl map
$\wiggpm^\ast\colon\ldg\rightarrow\hspm$ can be easily obtained
applying formula~{(\ref{forwey-bis})}. Indeed, for every function
$\effe\colon G\rightarrow\ccc$ in $\lug\cap\ldg$ and every vector
$\varphi^\pmpm$ in $\dom\big(\dupm^{-1}\big)$, we have:
\begin{eqnarray}
\Big(\big(\wiggpm^\ast\hspace{0.3mm}\effe\big)\,\varphi^\pmpm\Big)(x)
\spa & = & \spa \big(\uf\hspace{0.4mm}\dupm^{-1}\tre\varphi^\pmpm\big)(x)\nonumber\\
& = & \spa \intG \effe(a,r)\hspace{0.8mm} \big(U^\pmpm(a,r)\hspace{0.7mm}\dupm^{-1}\varphi^\pmpm\big)(x)
\hspace{1.2mm} \de\mul (a,r)
\nonumber\\
& = & \spa \intG \effe(a,r)\, \sqrt{r}\hspace{0.8mm} \ee^{\ima a x}
\sqrt{\frac{r|x|}{2\pi}}\hspace{1mm} \varphi^\pmpm(rx)\; \de\mul
(a,r)\fin ,\ \ \ \mbox{for a.a.\ } x\in\errepm .
\end{eqnarray}
Next, by virtue of Fubini's theorem and of a change of variables
($r\mapsto x^{-1}y$, with $x,y\in\errepm$), we get:
\begin{eqnarray}
\Big(\big(\wiggpm^\ast\hspace{0.3mm}\effe\big)\,\varphi^\pmpm\Big)(x)
\spa & = & \spa |x|^{\frac{1}{2}} \int_{\errepm}\hspace{-0.8mm}
|y|^{-1}\de y\hspace{1.3mm} \varphi^\pmpm(y) \int_\rrr\frac{\de
a}{\sqrt{2\pi}} \; \effe(a, x^{-1} y)\hspace{1mm} \ee^{\ima a x}
\nonumber\\
& = & \spa \int_{\errepm} \kerintf(x,y)\hspace{1.2mm}
\varphi^\pmpm(y)\hspace{1.2mm} \de y \fin ,\ \ \ \mbox{for a.a.\ }
x\in\errepm,
\end{eqnarray}
where, for every $f\in\ldg$, the integral kernel
$\kerint(\cdot,\cdot)\colon\errepm\times\errepm\rightarrow\ccc$ is
defined by
\begin{equation}
\kerint(x,y) :=
|x|^{\frac{1}{2}}\hspace{0.6mm}|y|^{-1}\hspace{0.5mm}
\big(\mathcal{F}_1 f\big)(-x, x^{-1} y) \fin ,\ \ \ x,y\in\errepm,
\end{equation}
with $\mathcal{F}_1$ denoting the Fourier transform with respect to the first variable. This result
--- by the essential uniqueness of the inducing kernel of a Hilbert-Schmidt operator (or, more in general, of
a Carleman operator; see, for instance, assertion~{(e)} of
Theorem~{6.13} of~{\cite{Weidmann}}) --- implies that
$\kerintf(\cdot,\cdot)$ is the integral kernel associated with the
Hilbert-Schmidt operator $\wiggpm^\ast\hspace{0.3mm}\effe$, for
every $\effe\in\lug\cap\ldg$; hence, we have that
\begin{eqnarray}\hspace{-8mm}
\|\wiggpm^\ast\hspace{0.3mm}\effe\norhs^2 =
\int_{\errepm}\hspace{-0.8mm} \de x \int_{\errepm} \hspace{-0.8mm}
\de y \hspace{1.2mm} \frac{|x|}{y^2} \hspace{0.7mm}
\big|\big(\mathcal{F}_1 \effe\big)(-x, x^{-1} y)\big|^2 \spa & = &
\spa \int_{\errepm}\hspace{-0.8mm} \de x \int_{\errep}
\hspace{-0.4mm} \frac{\de r}{r^2} \hspace{1.2mm}
\big|\big(\mathcal{F}_1 \effe\big)(-x, r)\big|^2
\nonumber\\
& \le & \spa \int_{\rrr}\hspace{-0.8mm} \de x \int_{\errep}
\hspace{-0.4mm} \frac{\de r}{r^2} \hspace{1.2mm}
\big|\big(\mathcal{F}_1 \effe\big)(-x, r)\big|^2 \nonumber \\
\label{argas} & = & \spa \intG |\effe(a,r)|^2\; r^{-2}\, \de a\,\de
r =\|\effe\norld^2 \fin .
\end{eqnarray}
Of course, what we have found  --- i.e.\
$\|\wiggpm^\ast\hspace{0.3mm}\effe\norhs^2\le\|\effe\norld^2$ --- is
coherent with the fact that the Weyl map $\wiggpm^\ast$ is a partial
isometry. Now, let $f$ be a \emph{generic} function in $\ldg$ and
$\{\effe_n\}_{n\in\nat}$ a sequence in the linear span
$\lug\cap\ldg$ such that $\lim_{n\rightarrow\infty}
\|f-\effe_n\norld=0$. Then, the sequence
$\{\wiggpm^\ast\hspace{0.3mm}\effe_n\}_{n\in\nat}\subset\hs$
converges to $\wiggpm^\ast\tre f$; equivalently, the sequence
$\{\kerintfn\}_{n\in\nat}$ converges in
$\mathrm{L}^2(\errepm\times\errepm)$ to the integral kernel of the
Hilbert-Schmidt operator $\wiggpm^\ast\hspace{0.3mm}f$, kernel which
for the moment is still `unknown'. But, arguing as
in~{(\ref{argas})}, we see that the function $\kerint$ belongs to
$\mathrm{L}^2(\errepm\times\errepm)$ and
\begin{equation}
\|\kerint -\kerintfn\|_{\mathrm{L}^2(\errepm\times\errepm)}^2 =
\int_{\errepm}\hspace{-0.8mm} \de x \int_{\errepm} \hspace{-0.8mm}
\de y \hspace{1.2mm} \frac{|x|}{y^2} \hspace{0.7mm}
\big|\big(\mathcal{F}_1 (f-\effe_n)\big)(-x, x^{-1} y)\big|^2 \le
\|f-\effe_n\norld^2 \fin .
\end{equation}
It follows that the integral kernel of $\wiggpm^\ast\hspace{0.3mm}f$
is $\kerint$ for \emph{every} $f\in\ldg$. Moreover, we have that
\begin{eqnarray}
\|\wiggm^\ast\hspace{0.3mm}f\norhs^2 +
\|\wiggp^\ast\hspace{0.3mm}f\norhs^2 \spa & = & \spa \int_{\errem}\hspace{-0.8mm} \de x
\int_{\errep} \hspace{-0.4mm} \frac{\de r}{r^2} \hspace{1.2mm}
\big|\big(\mathcal{F}_1 f\big)(-x, r)\big|^2 + \int_{\errep}\hspace{-0.8mm} \de x
\int_{\errep} \hspace{-0.4mm} \frac{\de r}{r^2} \hspace{1.2mm}
\big|\big(\mathcal{F}_1 f\big)(-x, r)\big|^2 \nonumber\\
& = & \spa \intG |f(a,r)|^2\; r^{-2}\, \de a\,\de r =\|f\norld^2
\fin ,\ \ \ \forall\hspace{0.4mm} f\in \ldg \fin .
\end{eqnarray}
Therefore, denoting by $\rupm$ the range of the Wigner map $\wigpm$
(we know that $\rum \hspace{-0.4mm} \perp \hspace{0.2mm}\rup$, see
Remark~{\ref{ranus}}) --- since
$\rupm=\Ker\big(\wiggpm^\ast\big)^\perp$
--- the following relation must hold:
\begin{equation}
\ldg = \rum \hspace{-0.4mm} \oplus \hspace{0.6mm}\rup \fin .
\end{equation}

Let us now consider the star products in $\ldg$ associated with the
square integrable representations $U^\ammm$ and $U^\ppp$. By
definition
--- see~{(\ref{defist})} --- we have
\begin{equation}
f_1 \starppm f_2 := \wigpm\Big(\big(\wiggpm^\ast\hspace{0.3mm}
f_1\big) \big(\wiggpm^\ast\hspace{0.3mm} f_2\big)\Big),\ \ \
\forall\hspace{0.4mm}f_1,f_2\in\ld(\mathbb{R}\times\errep,\mul) \fin
.
\end{equation}
Exploiting the results of Sect.~{\ref{main}} we can provide explicit
formulae for these star products. Let $\{\chinpm\}_{n\in\nat}$ be an orthonormal basis
in $\ld(\errepm)$ contained in $\dom\big(\dupm^{-2}\big)$; i.e., such that
\begin{equation}
\Big(\errepm\ni x\mapsto |x|\;\chinpm(x)\Big)\in\hspace{0.5mm}
\ld(\errepm)\fin .
\end{equation}
For instance, one can choose the Laguerre functions
\begin{equation}
\chinpm\colon \errepm\ni x\mapsto L_{n-1}(|x|)\,\ee^{-|x|/2},\ \ \ L_k(x):=\sum_{j=0}^k {k\choose j}
\frac{(-x)^j}{j\hspace{0.5mm}!}\,,\ k=0,1,2,\ldots\, ,
\end{equation}
where, of course, $L_k$ is the Laguerre polynomial of order $k$.
According to the main result of
Sect.~{\ref{main}} --- see Theorem~{\ref{main-theorem}} --- we have:
\begin{equation} \label{fstprod}
f_1\starppm f_2=\norsunat \intg\de\mul (a,r) \intg \de\mul
(a^\prime,r^\prime)\, \nucpm\big(\chinpm; \cdot,\cdot; a,r;
a^\prime,r^\prime\big)\, f_1(a,r) f_2(a^\prime,r^\prime) \fin ,
\end{equation}
where the integral kernel $\nucpm\big(\chinpm; \cdot,\cdot;
\cdot,\cdot; \cdot,\cdot\big)\colon G\times G \times
G\rightarrow\ccc$ is defined by
\begin{equation}
\nucpm\big(\chinpm; a_1,r_1; a_2,r_2; a_3,r_3\big)\hspace{-0.4mm} :=
\big\langle
U^\pmpm(a_1,r_1)\,\dupm^{-1}\,\chinpm,U^\pmpm(a_2,r_2)\,\dupm^{-1}
\,U^\pmpm(a_3,r_3)\,\dupm^{-1}\,\chinpm\big\rangle . \nonumber
\end{equation}
Recalling the explicit form of the the Duflo-Moore operators
$\duupm$, we have:
\begin{eqnarray}
\nucpm\big(\chinpm; a_1,r_1; a_2,r_2; a_3,r_3\big)\hspace{-0.8mm}
\spa & = & \spa \frac{r_2\sqrt{r_3}}{r_1}\hspace{0.5mm} \big\langle
\dupm^{-1}\,\chinpm,\dupm^{-2}
\,U^\pmpm(-(a_1-a_2-r_2a_3)/r_1,r_2r_3/r_1)\,\chinpm\big\rangle
\nonumber
\\
& = & \spa \hspace{-0.6mm}\left(\hspace{-0.5mm}\frac{r_2}{2\pi
r_1}\hspace{-0.5mm}\right)^{\hspace{-0.7mm}\frac{3}{2}}
\hspace{-0.3mm} r_3 \int_{\errepm}\hspace{-0.8mm}
|x|^{\frac{3}{2}}\, \ee^{-\ima (a_1-a_2-r_2a_3)x/r_1}\,
\chinpm(x)^\ast\, \chinpm(r_2r_3 x/r_1)\; \de x
\nonumber\\
& = & \spa \left(\hspace{-0.5mm}\frac{r_2}{
r_1}\hspace{-0.5mm}\right)^{\hspace{-0.7mm}\frac{3}{2}}\frac{r_3}{2\pi} \hspace{1.3mm}
\funpm\big((a_1-a_2-r_2a_3)/r_1,r_2r_3 /r_1\big),
\end{eqnarray}
where the function $\funpm\colon\rrr\times\errep\rightarrow \ccc$ is defined by
\begin{equation}
\funpm (\alpha,\varrho) := \mathcal{F}\big(|\cdot|^{\frac{3}{2}}\;
\bchinpm(\cdot)^\ast\, \bchinpm(\varrho(\cdot))\big)(\alpha) \fin ,
\ \ \ \alpha\in\rrr\fin ,\ \varrho\in\errep ,
\end{equation}
with $\mathcal{F}\colon\lr\rightarrow\lr$ denoting the Fourier
transform
($\big(\mathcal{F}\varphi\big)(a)=(2\pi)^{-1/2}\int_{-\infty}^{+\infty}\ee^{-\ima
ax}\varphi(x)\,\de x$, for $\varphi\in\lur$) and $\bchinpm\in\lr$
the function
\begin{equation}
\bchinpm (x)= \chinpm (x),\ \ \mbox{for}\ x\in\errepm ,\ \ \ \
\bchinpm (x)= 0 \fin ,\ \ \mbox{otherwise};
\end{equation}
i.e., $\bchinpm$ is the image of $\chinpm$ via the natural immersion of $\ld(\errepm)$ into $\lr$.
In conclusion, the triples
\begin{equation}
\aum\equiv\big(\ld(\mathbb{R}\times\errep,\mul),\starpm,\sinvo\big)\
\ \mbox{and}\ \
\aup\equiv\big(\ld(\mathbb{R}\times\errep,\mul),\starpp,\sinvo\big)
\end{equation}
are $\Hstar$-algebras. The mutually orthogonal subspaces $\rum$ and
$\rup$ of $\ld(\mathbb{R}\times\errep,\mul)$ are, respectively, the
canonical and the annihilator ideals in the standard decomposition of
the $\Hstar$-algebra $\aum$, while they are, respectively, the
annihilator and the canonical ideals for $\aup$. It is clear that one
may endow $\ld(\mathbb{R}\times\errep,\mul)$ with the structure of a
\emph{proper} $\Hstar$-algebra by considering the star product
\begin{equation} \label{sumstar}
f_1\star f_2 := \big(f_1\starpm f_2\big)  + \big(f_1\starpp f_2\big)
.
\end{equation}

Let us now clarify the link with the standard wavelet transform. To
this aim, let us consider the unitary representation $\ut\colon
G\rightarrow\mathcal{U}(\lr)$ defined as follows. Taking into
account the orthogonal sum decomposition $\lr =
\ld(\errem)\oplus\ld(\errep)$, we can consider the representation
$U^\ammm\hspace{-0.7mm}\oplus U^\ppp$ of $G$ in $\lr$; then, we set
\begin{equation}
\ut (a,r):= \mathcal{F}\Big(\big(U^\ammm\hspace{-0.7mm}\oplus
U^\ppp\big) (a,r)\Big)\mathcal{F}^\ast, \ \ \
\forall\hspace{0.5mm}(a,r)\in \mathbb{R}\sdp\errep.
\end{equation}
For every $\psi\in\lr$, we have:
\begin{equation}
\psi_{a,r}(a^\prime)\equiv\big(\ut(a,r)\,\psi\big)(a^\prime)=r^{-\frac{1}{2}}\,\psi((a^\prime-a)/r),\ \ \
a,a^\prime\in\rrr,\ r\in\errep.
\end{equation}
Observe that this is the typical dependence on the translation and
dilation parameters of a `wavelet frame' (see~{\cite{Daubechies}};
note that the symbols that we use here for these parameters are
non-standard). However, a function $\psi \in\lr$, in order to be a
`good mother wavelet' --- i.e.\ in order to verify the the
orthogonality relations
\begin{equation}
\int_G    \langle\phi,\psi_{a,r}\rangle\langle\psi_{a,r},\phi
\rangle\; \de\mul(a,r)= \langle \phi,\phi\rangle \fin , \ \ \
\forall\hspace{0.4mm}\phi\in\lr
\end{equation}
--- has to satisfy suitable conditions. Indeed, as the reader will easily understand, one has to require
that the following conditions hold:
\begin{enumerate}
\item the projection onto $\ld(\errepm)$ (regarded as a subspace of
$\lr$) of the Fourier transform of $\psi$ belongs to
$\dom\big(\duupm\big)$, i.e.
\begin{equation}
\Big(\errepm\ni x\mapsto
|x|^{-1}\,\big|\big(\mathcal{F}\,\psi\big)(x)\big|^2\Big)\in\hspace{0.5mm}
\lurpm \fin ;
\end{equation}

\item denoting by $\charpm$ the characteristic function of the
subset $\errepm$ of $\rrr$,\footnote{Observe that the orthogonal
projection of $\lr$ onto $\ld(\errepm)$ is just the multiplication operator by
$\charpm$.} the vectors
\begin{equation*}
\duum\hspace{0.3mm}
\big(\charm\hspace{0.3mm}(\mathcal{F}\hspace{0.4mm}\psi)\big)\in\ld(\errem)\
\ \mbox{and}\ \ \duup\hspace{0.3mm}
\big(\charp\hspace{0.3mm}(\mathcal{F}\hspace{0.4mm}\psi)\big)\in\ld(\errep)
\end{equation*}
are both
normalized, i.e.
\begin{equation}
2\pi\int_{\errem} |x|^{-1}\,
\big|\big(\mathcal{F}\,\psi\big)(x)\big|^2\; \de x =
2\pi\int_{\errep} |x|^{-1}\,
\big|\big(\mathcal{F}\,\psi\big)(x)\big|^2\; \de x =1 \fin .
\end{equation}

\end{enumerate}

One can obtain a variant of the scheme analyzed above by allowing
both positive and \emph{negative} dilations; this variant is widely
exploited in wavelet analysis, see~{\cite{Daubechies}}. It amounts
to considering the semidirect product $\mathbb{R}\sdp\errast$, with
$\errast$ denoting the group of nonzero real numbers (with respect
to multiplication). This semidirect product group admits a single
square integrable irreducible representation, up to unitary
equivalence; namely, the unitary representation $U\colon
G\rightarrow\mathcal{U}(\lr)$ defined by
\begin{equation}
\big(U (a,r)\, \varphi\big)(x ) :=  |r|^{\frac{1}{2}}\, \ee^{\ima
ax} \varphi(r x),\ \ \ a\in\mathbb{R},\ r\in \errast,\ x\in\rrr, \
\, \varphi\in \lr \fin .
\end{equation}
Of course, one can repeat for $\mathbb{R}\sdp\errast$ the same
analysis performed for the group $\mathbb{R}\sdp\errep$. We leave
this analysis as an exercise for the reader.

\section{Conclusions, final remarks and perspectives}
\label{conclusions}

In this paper we have considered star products from a purely
group-theoretical point of view. In particular, we have not assumed
to deal with Lie groups, but, in general, with locally compact
topological groups. Therefore, our treatment allows us to include in
a unified framework, for instance, all the finite groups (in the
paper regarded as compact groups). This feature is certainly
appealing in view of the increasing interest in realizing quantum
mechanics on discrete spaces (see~{\cite{Gibbons}} and references
therein). We think, in particular, that applying our results to a
formulation of quantum mechanics on finite groups would be extremely
interesting.

Let us briefly review the main points of our work.

We have first recalled --- see Sect.~{\ref{weyl-wigner}} --- that
with a square integrable (in general, projective) representation
$U\colon G \rightarrow \mathcal{U}(\mathcal{H})$ of a locally
compact group $G$ are naturally associated a dequantization (Wigner)
map $\wig$, which is an isometry, and its adjoint, the quantization
(Weyl) map $\wigg^\ast$. The standard Wigner and Weyl maps are
recovered in the case where the group under consideration is the
group of translations on phase space, up to a (symplectic) Fourier
transform. We stress that this Fourier transform does not play any
--- mathematically or conceptually --- relevant role; essentially, it
allows to obtain the usual quantization rule for the functions of
position and momentum.

Next, in Sect.~{\ref{defstar}}, we have observed that by means of
the quantization and dequantization maps associated with the
representation $U$ one can define a star product of functions
enjoying remarkable properties. Endowed with this product and with a
suitable involution, the Hilbert space $\ldg$ becomes a
$\Hstar$-algebra $\au$, and --- regarding $G$ as a `symmetry group'
of a quantum system --- the star product is, by construction,
equivariant with respect to the natural action of $G$ in $\au$,
i.e.\ the action with which the standard symmetry action of $G$ on
states or observables in the Hilbert space $\hh$ is intertwined via
the Wigner map. Observe that the star product associated with $U$ is
such that the canonical ideal of $\au$ --- ideal which coincides
with the range $\ru$ of $\wig$ --- is a \emph{simple}
$\Hstar$-algebra (see~{\cite{Ambrose,Rickart}}), isomorphic to
$\hs$. It is clear that the algebra $\au$ is commutative if and only
if $\dim(\hh)=1$ (in this case, the square-integrability of $U$
forces the group $G$ to be compact). Observe moreover that, in the
case where $G$ admits various (unitarily) inequivalent
\emph{unitary} representations, one can define more general star
products by forming suitable `orthogonal sums' of `simple' star
products; see, e.g., formula~{(\ref{sumstar})}. In
Sect.~{\ref{defstar}}, we have also considered an interesting
deformation of the star product associated with $U$, namely, the
$\opk$-deformed star product, and studied its main properties. We
will consider applications of this deformed product elsewhere.

At this point, our main task has been to derive explicit formulae
for the previously defined star products. This task has been
accomplished in Sect.~{\ref{main}}. We have shown that for every
orthonormal basis contained in the domain of the positive
selfadjoint operator $\du^{-2}$ (with $\duu$ denoting the
Duflo-Moore operator associated with $U$) one has a realization of
the star product, see Theorem~{\ref{main-theorem}}; more generally
--- see Theorem~{\ref{general-theorem}} --- for every suitable right
approximate identity in the $\Hstar$-algebra $\hs$ one can provide a
realization of the $\opk$-deformed star product. In the case where
the group $G$ is unimodular, the star product of two functions
\emph{belonging to the range of the Wigner map} $\wig$ assumes the
particularly simple form of a `twisted convolution', which reduces
to the standard convolution if $U$ is a unitary representation. It
is interesting to note, incidentally, that it is the Banach space
$\lug$ which is usually endowed with the structure of a Banach
$\ast$-algebra by means of convolution~{\cite{Folland}}, while in
$\ldg$ the convolution product is, in general, an `ill-posed'
operation. Namely, if the convolution product exists and belongs to
$\ldg$ \emph{for all pairs of functions} in $\ldg$, then the group
$G$ must be compact (recall, however, that by H\"older's inequality
the convolution of any pair of functions in $\ldg$ does exist, for
$G$ unimodular). This is a particular case ($p=2$) of the classical
`$\mathrm{L}^p$-conjecture' ($p> 1$), which has been finally proved
(in its general form) in~{1990} by Saeki~{\cite{Saeki}}. Therefore,
the \emph{whole} vector space $\ldg$ can be endowed with the
structure of an algebra by means of the convolution product if and
only if $G$ is compact.

Consider, now, the specific case where the group $G$ is compact. In
this case, one obtains a nice decomposition formula for the
convolution in $\ldg$ in terms of the star products associated with
a realization of the unitary dual $\chG$ of $G$; see
Corollary~{\ref{cor-compa}}. The Hilbert space $\ldg$, endowed with
the convolution product and with the involution~{(\ref{invsimp})},
is a $\Hstar$-algebra which we denote by $\ag$. The orthogonal sum
decomposition~{(\ref{cc-dd})}
--- complemented by formula~{(\ref{fstpruni-comp})} --- can be
regarded as the decomposition into minimal closed (two-sided) ideals
of $\ag$ prescribed by the `second Wedderburn structure theorem for
$\Hstar$-algebras'~{\cite{Ambrose,Rickart}}. Any of these ideals
--- say $\ru = \ldgu$ --- is a simple finite-dimensional $\Hstar$-algebra
which is embedded, in a natural way, in the $\Hstar$-algebra $\au$
determined by the star product~{(\ref{stpcp})} and by the
involution~{(\ref{invsimp})}; precisely, as already observed, $\ru$
is the canonical ideal of $\au$. It is actually the interest in the
algebra $\ag$ that motivated Ambrose's study of
$\Hstar$-algebras~{\cite{Ambrose}}. On our opinion, the formalism of
star products provides a concrete and conceptually clear framework
for Ambrose's ideas.

It is worth noting that --- differently from the
quantization-dequantization scheme which has been recently developed
in ref.~{\cite{Aniello-framet}} --- in the `Weyl-Wigner approach'
that is considered in the present contribution there is no canonical
way for representing a \emph{generic} quantum observable as a
suitable `phase space function' since, for $\hh$
infinite-dimensional, $\hs\subsetneq\bh$  (in the case of the
standard Weyl quantization, this problem has been studied, for
instance, in~{\cite{Daubechies-bis}). This feature, of course,
reflects into the fact that there is no standard way for
representing within the framework considered here the product of a
\emph{generic} quantum observable by a state as a star product of
functions. However, we believe that suitably extending the domain of
the first argument of the star product --- this time \emph{defined}
as the r.h.s.\ of~{(\ref{fstpr})} --- from $\ldg$ to some larger
space of functions (or distributions), and, possibly, restricting
the domain of the second argument, one should be able to generalize
the results obtained in the paper. This interesting topic will be
the object of further investigation.

One can, in principle,  elaborate several examples of star products
defined along the lines traced in the present paper that are
potentially relevant for applications. In addition to the case of
compact groups, for all groups admitting square integrable
projective representations it is possible to define star products of
functions. In Sect.~{\ref{examples}} we have considered the
significant examples of the group of translations on phase space and
of the affine group, but, of course, several other examples would
deserve attention. As an example, we mention the group
$\mathrm{SL}(2,\mathbb{R})$. According to classical results due to
Bargmann~{\cite{Bargmann}}, this group admits a (infinite) countable
set of mutually inequivalent square integrable unitary
representations
--- the `discrete series'
--- with carrier Hilbert spaces consisting of suitable holomorphic
functions on the upper half plane.

A wide class of groups with important applications in physics and
related research areas (in particular, signal analysis) is formed by
the semidirect products with an abelian normal factor. For these
groups square integrable representations can be suitably
characterized, see~{\cite{Aniello-sdp}}, and examples of such
groups, admitting square integrable representations and having
remarkable applications, can be found in refs.~{\cite{Ali1,Ali2}}.
From the point of view of signal analysis, the image through the
Weyl map of a function in $\ldg$ can be regarded as a
\emph{localization operator} of a different kind with respect to the
localization operators usually considered in wavelet and Gabor
analysis~{\cite{Daubechies}}. Thus, the star product provides a way
for characterizing the product of two localization operators.
Possible applications of our results to signal analysis is a further
topic that we plan to investigate in the future.



\begin{thebibliography}{99}


\bibitem{Zachos} C.K. Zachos, D.B. Fairlie and T.L. Curtright eds.,
{\it Quantum Mechanics in Phase Space}, World Scientific (2005).

\bibitem{Fronsdal} F. Bayen, M. Flato, C. Fronsdal, A. Lichnerowicz,
D. Sternheimer, ``Deformation theory and quantization. I.
Deformations of symplectic structures.'', {\it Ann. Phys.} {\bf 111}
(1978), 61-110; F. Bayen, M. Flato, C. Fronsdal, A. Lichnerowicz, D.
Sternheimer, ``Deformation theory and quantization. II. Physical
applications.'', {\it ibidem} {\bf 111} (1978), 111-151.

\bibitem{Cahen} M. Cahen, S. Gutt, ``Regular $\ast$-representations of Lie algebras'',
{\it Lett. Math. Phys.} {\bf 6} (1982), 395-404.

\bibitem{Gutt}  S. Gutt, ``An explicit $\ast$-product on the cotangent bundle of a Lie group'',
{\it Lett. Math. Phys.} {\bf 7} (1983), 249-258.

\bibitem{DeWilde}  M. De Wilde, P. Lecompte, ``Existence of star-products and of formal deformations
of the Poisson-Lie algebra of arbirary symplectic manifolds'',
{\it Lett. Math. Phys.} {\bf 7} (1983), 487-496.

\bibitem{Berezin} F.A. Berezin, ``General concept of quantization'',
{\it Comm. Math. Phys.} {\bf 40} (1975), 153-174.

\bibitem{Emch} G.G. Emch, ``Mathematical and Conceptual Foundations of 20th-Century Physics'',
North-Holland (1984).

\bibitem{Wigner} E. Wigner, ``On the quantum correction for
thermodynamic equilibrium'', {\it Phys. Rev.} {\bf 40} (1932),
749-759.

\bibitem{Weyl} H. Weyl, {\it The Theory of Groups and Quantum
Mechanics}, Dover (1950).

\bibitem{Gracia-Annals} J.C. V\'arilly, J.M. Gracia-Bond\'ia, ``The
Moyal representation for spin'', {\it Ann. Phys.} {\bf 190} (1989),
107-148.

\bibitem{Gadella} M. Gadella, M.M. Mart\'in, L.M. Nieto, M.A. del
Olmo, ``The Stratonovich-Weyl correspondence for one-dimensional
kinematical groups'', J. Math. Phys. {\bf 32} (1991), 1182-1192.

\bibitem{Gracia} J.M. Gracia-Bond\'ia, J.C. V\'arilly, H. Figueroa,
{\it Elements of Noncommutative Geometry}, Birkh\"auser, Boston
(2001).

\bibitem{Manko1} O.V. Man'ko, V.I. Man'ko, G. Marmo, ``Alternative
commutation relations, star products and tomography'', {\it J. Phys.
A: Math. Gen.} {\bf 35} (2002), 699-719.

\bibitem{Manko2} V.I. Man'ko, G. Marmo, P. Vitale, ``Phase space distributions and a duality
symmetry for star products'', {\it Phys. Lett. A}
{\bf 334} (2005), 1-11.

\bibitem{Manko3} O.V. Man'ko, V.I. Man'ko, G. Marmo, P. Vitale, ``Star
products, duality and double Lie algebras'', {\it Phys. Lett. A}
{\bf 360} (2007), 522-532.

\bibitem{Aniello-compact} P. Aniello, A. Ibort, V.I. Man'ko, G.
Marmo, ``Remarks on the star product of functions on finite and
compact groups'', {\it Phys. Lett. A} {\bf 373} (2009), 401-408.

\bibitem{Aniello-framet} P. Aniello, V.I. Man'ko, G. Marmo, ``Frame transforms, star products
and quantum mechanics on phase space'', {\it J. Phys. A: Math.
Theor.} {\bf 41} (2008), 285304.

\bibitem{Ali1} S.T. Ali, J.P. Antoine, J.P. Gazeau, U.A. Mueller,
              ``Coherent states and their generalizations: a mathematical
                overview'', {\it Rev. Math. Phys.} {\bf 7} (1995),
                1013-1104.

\bibitem{Ali2} S.T. Ali, J.P. Antoine, J.P. Gazeau, {\it Coherent States,
               Wavelets and Their Generalizations}, Springer-Verlag (2000).

\bibitem{GLP} A. Grossmann, G. Loupias, E.M. Stein, ``An algebra of
pseudo-differential operators and quantum mechanics in phase
space'', {\it Ann. Inst. Fourier} {\bf 18} (1968), 343-368.


\bibitem{Raja} V.S. Varadarajan, {\it Geometry of Quantum
            Theory}, second edition, Springer (1985).

\bibitem{Folland} G.B. Folland, {\it A Course in Abstract
           Harmonic Analysis}, CRC Press (1995).

\bibitem{Aniello} P. Aniello, ``Square integrable projective
representations and square integrable representations modulo a
relatively central subgroup'', {\it Int. J. Geom. Meth. Mod. Phys.}
{\bf 3} (2006), 233-267.

\bibitem{Duflo}  M. Duflo, C.C. Moore, ``On the regular representation
                 of a nonunimodular locally compact group'',
                 {\it J. Funct. Anal.} {\bf 21} (1976), 209-243.

\bibitem{Grossmann} A. Grossmann, J. Morlet, T. Paul, ``Integral
           transforms associated to square integrable representations I.
           General results.'', {\it J. Math. Phys.} {\bf 26} (1985),
           2473-2479; A. Grossmann, J. Morlet, T. Paul, ``Integral
           transforms associated to square integrable representations II.
           Examples'', {\it Ann. Inst. H. Poincar\'e} {\bf 45}
           (1986), 293-309.

\bibitem{Daubechies} I. Daubechies, {\it Ten Lectures on Wavelets},
                     SIAM (1992).

\bibitem{Simon} B. Simon, {\it Representations of Finite and Compact Groups},
American Mathematical Society (1996).

\bibitem{Perelomov} A. Perelomov, {\it Generalized Coherent States and
Their Applications}, Springer-Verlag (1986).

\bibitem{Klauder} J.R. Klauder, E.C.G. Sudarshan, {\it Fundamentals of
Quantum Optics}, W.A. Benjamin (1968).

\bibitem{Reed} M. Reed, B Simon, {\it Methods of Modern Mathematical
Physics I: Functional Analysis}, Academic Press (1972).


\bibitem{Segal} I.E. Segal, ``The two-sided regular representation
of a unimodular locally compact group'', {\it Ann. Math.} {\bf 51}
(1950), 293-298.

\bibitem{Ambrose} W. Ambrose, ``Structure theorems for a certain
class of Banach algebras'', {\it Trans. Amer. Math Soc.} {\bf 57}
(1945), 364-386.

\bibitem{Rickart} C.E. Rickart, ``General Theory of Banach Algebras'',
Van Nostrand (1960).

\bibitem{Marmo} G. Esposito, G. Marmo, G. Sudarshan, {\it From
Classical to Quantum Mechanics}, Cambridge University Press (2004).

\bibitem{Folland-bis} G.B. Folland,  {\it Harmonic Analysis in Phase
Space}, Princeton University Press (1989).

\bibitem{Pool} J.C.T. Pool, ``Mathematical aspects of Weyl
correspondence'', {\it J. Math. Phys.} {\bf 7} (1966), 66-76.

\bibitem{Klauder2} E.W. Aslaksen, J.R. Klauder, ``Continuous representation
theory using the affine group'', {\it J. Math. Phys.} {\bf 10}
(1969), 2267-2275.

\bibitem{Aniello-sdp} P. Aniello, G. Cassinelli, E. De Vito,
           A. Levrero, ``Square-integrability of induced representations of
           semidirect products'', {\it Rev. Math. Phys.} {\bf 10}
           (1998), 301-313.

\bibitem{Weidmann} J. Weidmann, {\it Linear Operators in Hilbert Spaces},
Springer-Verlag (1980).

\bibitem{Gibbons} K.S. Gibbons, M.J. Hoffmann, W.K. Wootters,
``Discrete phase space based on finite fields'', {\it Phys. Rev. A}
{\bf 70} (2004), 062101.

\bibitem{Saeki} S. Saeki, ``The $\mathrm{L}^p$-conjecture and Young's
inequality'', {\it Illinois J. Math.} {\bf 34} (1990), 615-627.

\bibitem{Daubechies-bis} I. Daubechies, ``On the distributions
corresponding to bounded operators in the Weyl quantization'', {\it
Commun. Math. Phys.} {\bf 75} (1980), 229-238.

\bibitem{Bargmann} V. Bargmann, ``Irreducible representations of the
Lorentz group'', {\it Ann. of Math.} {\bf 48} (1947), 568-640.


\end{thebibliography}
\end{document}